\documentclass[10pt,authoryear]{article}

\usepackage[authoryear]{natbib}
\usepackage{lineno}
\usepackage{amsmath}
\usepackage{mathtools}
\usepackage{ae,aecompl}
\usepackage[]{graphicx}
\usepackage{amsfonts}
\usepackage{amssymb}
\usepackage{epstopdf}
\usepackage{epsfig}
\usepackage{float}
\def\lsim{\lower.5ex\hbox{$\; \buildrel < \over \sim \;$}}
\def\gsim{\lower.5ex\hbox{$\; \buildrel > \over \sim \;$}}

\providecommand{\keywords}[1]{\textit{Keywords:} #1}

\begin{document}

\title{Effect of matter 
geometry on low angular momentum black hole accretion in the 
Kerr metric}

\author{Pratik Tarafdar\\
	S. N. Bose National Centre For Basic Sciences\\ 
	Kolkata, India.\\
	pratikta16@gmail.com
	\and
	Deepika B. Ananda\\
	Nicolaus Copernicus Astronomical Centre\\
	Polish Academy of Sciences\\
	Warsaw, Poland.\\
	deepika@camk.edu.pl
	\and
	Sankhashubhra Nag\\
	Sarojini Naidu College for Women\\
	Kolkata, India.\\
	sankha@sncwgs.ac.in
	\and
	Tapas Kumar Das\\
	Harish-Chandra Research Institute\\
	HBNI, Chhatnag Road, Jhunsi \\ 
	Allahabad, India.\\
	tapas@hri.res.in}
\date{}

\maketitle

\begin{abstract}
\noindent
This work illustrates how the formation of energy-preserving 
shocks for polytropic accretion and temperature-preserving shocks 
for isothermal accretion are influenced by various geometrical 
configurations of general relativistic, axisymmetric, 
low-angular-momentum flow in the Kerr metric. Relevant pre- and 
post-shock states of the accreting fluid, both dynamical and 
thermodynamic, have been studied comprehensively. 
Self-gravitational back-reaction on the metric has not been 
considered in the present context. An elegant eigenvalue-based 
analytical method has been introduced to provide qualitative 
descriptions of the phase-orbits corresponding to stationary 
transonic accretion solutions, without resorting to involved 
numerical schemes. Effort has been made to understand how the 
weakly-rotating flow behaves in close proximity of the event 
horizon and how such `quasi-terminal' quantities are influenced 
by the black hole spin for different matter geometries. Our main 
purpose is thus to mathematically demonstrate that for 
non-self-gravitating accretion, separate matter geometries, in 
addition to the corresponding space-time geometry, control 
various shock-induced phenomena observed within black hole 
accretion discs. This work is expected to reveal how such 
phenomena observed near the horizon depend on physical 
environment of the source harbouring a supermassive black hole at 
its centre. It is also expected to unfold correspondences between 
the dependence of accretion-related parameters on flow geometries 
and on black hole spin. Temperature-preserving shocks in 
isothermal accretion may appear bright as substantial amount of 
rest-mass energy of the infalling matter gets dissipated at the 
shock surface, and the prompt removal of such energy to maintain 
isothermality may power the X-ray/IR flares emitted from our 
Galactic centre.
\end{abstract}

\keywords{accretion, accretion discs, black hole physics, hydrodynamics, shock waves, gravitation}

\section{Introduction}
For axially symmetric accretion onto astrophysical black 
holes, the geometric configuration of infalling 
matter influences the dynamical and the thermodynamic 
behaviour of the accretion flow. In our previous paper 
(\cite{td18na}, hereafter $P1$), we demonstrated how the 
multitransonic post shock flow manifests the dependence 
of its physical properties on various geometric profiles 
of material accreting onto non-rotating black holes. 
Calculations presented in $P1$ were carried out within the 
framework of general relativistic Schwarzschild metric. 
In general, astrophysical black holes are believed to 
possess non-zero spin angular momentum 
(\cite{mrfng09apj}, 
\cite{dwrb10mnras}, \cite{kmtnm10mnras}, \cite{tnm10apj}, 
\cite{ziolkowski10}, 
\cite{bggnps11a}, \cite{daly11mnras}, \cite{mr11mnras}, 
\cite{mndgkoprs11cqg}, \cite{nckp11mnras}, 
\cite{rbltmrnf12aip}, \cite{tm12mnras}, 
\cite{brenneman13}, \cite{dcppv13apj}, \cite{mtb13s}, 
\cite{fpwmkrd14mnras}, \cite{hlz14arxiv}, 
\cite{jbs14arxiv}, \cite{nt14arxiv}, \cite{sbdr14arxiv}) 
and such spin angular 
momentum (the Kerr parameter `$a$') assumes a vital role 
in influencing the various characteristic features of 
accretion induced astrophysical phenomena 
(\cite{dnhbmcbwkn15na}). 
It is thus imperative to learn how the black hole spin 
dependence of accretion astrophysics depends on the matter 
geometry. This is precisely what we would like to 
explore in the present work. \\

We formulate and solve the general relativistic Euler 
equation and the continuity equation 
in the Kerr metric to obtain the stationary integral 
accretion solutions for the three different geometric 
configurations (see e.g., section 4.1 of \cite{bcdn14cqg}, 
and references therein, for the detailed classification 
of three such geometries) of low angular momentum axially 
symmetric advective flow onto spinning black holes. 
We then study the stationary phase portrait of multi-
transonic accretion and demonstrate how stationary shocks 
may form for such flow topologies. Then, we study the 
astrophysics of shock formation and demonstrate how 
Kerr parameter influences the location of the shock formed 
and other shock related quantities. We also report how 
such spin dependence varies from one type of flow geometry 
to the other. The overall process to accomplish such task 
has been performed in the following way. \\

We first try to address the issue, upto a certain extent, 
completely analytically. It can be shown that 
the stationary transonic solutions for inviscid accretion 
can be mapped as critical solutions on the phase portrait 
spanned by radial Mach number and the corresponding radial 
distance measured on the equatorial plane 
(\cite{gkrd07mnras}). 
For all three geometries, the Euler and the continuity 
equations are formulated in the Kerr metric. The time 
independent parts of these equations are considered to 
find out the corresponding critical/sonic points, 
borrowing certain methodology from the theory of dynamical 
systems. One can obtain more than one critical points as 
well. In such cases, one needs to classify what kind of 
critical points are those. We provide an eigenvalue based 
analytical method to find out the nature of the critical 
points and demonstrate that they are either 
centre type or saddle type. What can be done analytically 
is, once the nature of the critical 
points are analytically determined following the 
aforementioned procedures, the tentative nature of 
corresponding orbits on the phase portrait can roughly be 
anticipated and the overall multi-transonic 
phase portraits can be understood. The purview of the 
analytical regime is limited upto this point. The 
advantage of introducing the eigenvalue based analytical 
method is to qualitatively 
understand how the phase orbits corresponding to the 
stationary transonic solutions would look like, without 
incorporating any complicated numerical technique. \\

The exact shape of the phase orbits, however, can never be 
obtained analytically. One needs to numerically 
integrate the Euler and the continuity equations to obtain 
the stationary integral mono/multi-transonic 
solutions. In section 5, we provide the methodology for 
integrating the fluid equations 
to obtain the transonic solutions. \\

Multitransonic solutions require the presence of a 
stationary shock to join the integral solutions passing 
through the outer and the inner sonic points, 
respectively. In section 6, we discuss the shock formation 
phenomena in detail and demonstrate how the Kerr parameter 
influences the shock related quantities for three 
different matter geometries. In the subsequent sections, 
the concept of quasi-terminal values is introduced to 
understand how the weakly rotating accretion flow behaves 
at the extreme close proximity of the event horizon 
and how such behaviours, for three different matter 
geometries, are influenced by the black hole spin. \\

The entire study presented in our work has been divided 
into six sub-categories altogether. We study three 
different geometrical models of polytropic accretion, and 
then study the same models for isothermal accretion - and 
for each flow model, we analyze the black hole spin 
dependence of various flow properties.

\section{Polytropic flow structures for various matter geometries}

As shown in $P1$, one needs to obtain the expressions for 
two first integrals of motion - the conserved specific 
energy $\mathcal{E}$, which is obtained from integral 
solutions of the time independent part of the Euler 
equation, and remains invariant for all 
three matter geometries; and the mass accretion rate 
$\dot{M}$, which is obtained from the integral solutions 
of the time independent part of the continuity equation 
and varies with the geometric configuration of matter. We 
start with the constant height flow 
and then continue the same for two other flow geometries, 
i.e. wedge-shaped quasi-spherical or conical flow, and 
flow in hydrostatic equilibrium along the vertical 
direction. \\

For ideal fluid, the general relativistic Euler and the 
continuity equations are obtained through the co-variant 
differentiation of the corresponding energy momentum 
tensor \\
\begin{equation}
T^{\mu\nu} = (\epsilon+p) v^\mu v^\nu + pg^{\mu\nu}
\label{eqn1}
\end{equation}
where, \\
\begin{equation}
\epsilon = \rho + \frac{p}{\gamma-1}
\label{eqn1a}
\end{equation}
is the energy density (which includes the rest mass energy 
density and the internal energy density), 
$p$ is the fluid pressure, and $v^\mu$ is the velocity 
field. Using the Boyer-Lindquist co-ordinate, one can show 
(\cite{dnhbmcbwkn15na}) that the conserved specific energy 
as defined on the equatorial plane is expressed as \\
\begin{equation}
\mathcal{E} = \frac{\gamma-1}{\gamma-(1+c_s^2)} 
\sqrt{\frac{1}{1-u^2}\left[\frac{Ar^2\Delta}{A^2-4\lambda arA+\lambda^2 r^2(4a^2-r^2\Delta)}\right]}
\label{eqn2}
\end{equation}
In the above expression, $\gamma$ is the ratio of the two 
specific heat capacities $C_p$ and $C_v$, where an 
adiabatic equation of state of the form $p=K\rho^\gamma$ 
has been used $\rho$ being the matter density. 
$c_s$ denotes the position dependent adiabatic sound 
speed, defined as \\
\begin{equation}
c_s^2 = \left(\frac{\partial p}{\partial\epsilon}\right)_{\text{constant entropy}},
\label{eqn3}
\end{equation}
The advective velocity $u$ measured along the equatorial 
plane can be obtained by solving the following equation \\
\begin{equation}
v_t = \sqrt{\frac{g_{t\phi}^2-g_{tt}g_{\phi\phi}}{(1-\lambda\Omega)(1-u^2)(g_{\phi\phi}+\lambda g_{t\phi})}}
\label{eqn4}
\end{equation}
where $\left(g_{t\phi},g_{tt},g_{\phi\phi}\right)$ are the 
corresponding elements of the metric \\
\begin{equation}
ds^2 = g_{\mu\nu}dx^\mu dx^\nu = -\frac{r^2\Delta}{A}dt^2+\frac{r^2}{\Delta}dr^2+\frac{A}{r^2}(d\phi-\omega dt)+dz^2
\label{eqn5}
\end{equation}
where the line element has been expressed on the 
equatorial plane, using the Boyer-Lindquist co-ordinate, 
and \\
$\omega=\frac{2ar}{A}$, $\Delta=r^2-2r+a^2$, $A=r^4+r^2a^2+2ra^2$. \\
$\lambda$ and $\Omega$ are the specific angular momentum 
and the angular velocity repectively, as defined by \\
$\lambda=-\frac{v_\phi}{v_t}$, $\Omega=\frac{v^\phi}{v^t}=-\frac{g_{t\phi}+\lambda g_{tt}}{g_{\phi\phi}+\lambda g_{t\phi}}$. \\
It is to be noted that the expression for $\mathcal{E}$ 
has been obtained using the natural unit where the radial 
distance (measured along the equatorial plane) has been scaled by $GM_{BH}/c^2$ and the dynamical as well as the sound velocity 
have been scaled by the velocity of light in vacuum $c$, $M_{BH}$ being the mass of the black hole considered. We also 
normalize $G=c=M_{BH}=1$. \\

\subsection{Constant Height Flow}
The mass accretion rate may be obtained as \\
\begin{equation}
\dot{M}_{CH} = 4\pi\sqrt{\Delta}H\rho\sqrt{\frac{u^2}{1-u^2}}
\label{eqn6}
\end{equation}
where $H$ is the radius independent constant thickness of 
the accretion disc, and 
$\rho=\left[\frac{c_s^2(\gamma-1)}{\gamma K(\gamma-1-c_s^2)}\right]^\frac{1}{\gamma-1}$.
The corresponding entropy accretion rate may be obtained 
through the transformation 
$\dot{\Xi}=\dot{M}(K\gamma)^{\frac{1}{\gamma-1}}$ as, \\
\begin{equation}
\dot{\Xi}_{CH} = 4\pi\sqrt{\Delta}H \left[\frac{c_s^2(\gamma-1)}{\gamma-1-c_s^2}\right]^\frac{1}{\gamma-1} \sqrt{\frac{u^2}{1-u^2}}
\label{eqn7}
\end{equation}
The idea of entropy accretion rate was initially proposed 
by \cite{az81apj} and \cite{blaes87mnras} 
in order to calculate the stationary solutions for low 
angular momentum non-relativistic transonic accretion 
under 
the influence of \cite{pw80aa} pseudo-Newtonian potential 
onto a non-rotating black hole. \\
The space gradient of the acoustic velocity as well as the 
dynamical velocity can be computed as: \\
\begin{equation}
\frac{dc_s}{dr} = \frac{\mathcal{N}^{CH}_1}{\mathcal{D}^{CH}_1} 
\label{eqn8}
\end{equation}
\begin{equation}
\frac{du}{dr} = \frac{\mathcal{N}^{CH}_2}{\mathcal{D}^{CH}_2}
\label{eqn9}
\end{equation}
where $\mathcal{N}^{CH}_1=\frac{-2u}{2(1-u^2)}\frac{du}{dr}-\frac{f'}{2f}$, $\mathcal{D}^{CH}_1=\frac{2c_s}{\gamma-1-c_s^2}$,
$\mathcal{N}^{CH}_2=u(1-u^2)\left[\frac{r-1}{\Delta}c_s^2-\frac{f'}{2f}\right]$, $\mathcal{D}^{CH}_2=u^2-c_s^2$, 
$f=\frac{\Delta}{B}$, $B=g_{\phi\phi}+2\lambda g_{t\phi}+\lambda^2 g_{tt}$ and $f'$ denotes the space derivative of $f$, 
i.e., $\frac{df}{dr}$. Hereafter the sub/superscripts $CH$ will stand for 'Constant Height'. \\
The critical point conditions may be obtained as,
\begin{equation}
u^2|_{r_c}=c_s^2|_{r_c}=\frac{f'}{2f}|_{r_c}\frac{\Delta_c}{r_c-1}
\label{eqn10}
\end{equation}
Inserting the critical point conditions in the expression 
of the conserved specific energy, one can solve 
the corresponding algebraic equation for a specific set of 
values of $\left[\mathcal{E},\lambda,\gamma,a\right]$, 
to obtain the value of the critical point $r_c$. \\
The gradient of the sound speed and that of the dynamical 
velocity can also be evaluated at the critical points as, 
\begin{equation}
\frac{du}{dr}|_{r_c} = -\frac{\beta_{CH}}{2\alpha_{CH}} \pm \frac{1}{2\alpha_{CH}}\sqrt{\beta_{CH}^2-4\alpha_{CH}\Gamma_{CH}}
\label{eqn12}
\end{equation}
\begin{equation}
\frac{dc_s}{dr}|_{r_c} = \frac{\mathcal{N}_1}{\mathcal{D}_1}|_{r_c}
\label{eqn13}
\end{equation}
where, the co-efficients $\alpha_{CH}$, $\beta_{CH}$ and $\Gamma_{CH}$ are given by, \\
$\alpha_{CH}=\frac{\gamma-3 c_s^2+1}{\left(c_s^2-1\right)^2}|_{r_c}$, \\
$\beta_{CH}=\frac{2c_s(r-1)\left(c_s^2-\gamma+1\right)}{\left(c_s^2-1\right)\left(a^2+(r-2)r\right)}|_{r_c}$, \\
$\Gamma_{CH}=\frac{2\left(c_s^2-1\right)(r-1)^2}{\left(a^2+(r-2)r\right)^2}|_{r_c} -\frac{c_s^2-1}{a^2+(r-2)r}|_{r_c}
+\frac{c_s^2(r-1)^2\left(-c_s^2+\gamma-1\right)}{\left(a^2+(r-2)r\right)^2}|_{r_c} -$\\
\resizebox{0.5\textwidth}{!}{$\left[\frac{\splitfrac{-\frac{a^2\lambda^4\left(a^2(r+2)+r^3\right)}{\left(a^2(r+2)+\lambda^2r\right)^2}+\frac{2a^2\lambda^4(r-2)\left(a^2+\lambda^2\right) \left(a^2(r+2)+r^3\right)}{\left(a^2(r+2)+\lambda^2r\right)^3}-\frac{a^2\lambda^4(r-2)\left(a^2+3r^2\right)}{\left(a^2(r+2)+\lambda^2r\right)^2}}{+\frac{\lambda^4\left(a^2+\lambda^2\right)\left(r^3-a^2\left(r^2-8\right)\right)}{\left(a^2(r+2)+\lambda^2r\right)^2}+\frac{\lambda^4r\left(2a^2-3r\right)}{a^2(r+2)+\lambda^2r}-2a^2r+4a\lambda r+5r^4}}{r^4\left(-\frac{\lambda^4(r-2)\left(a^2(r+2)+r^3\right)}{r^3\left(a^2(r+2)+\lambda^2r\right)}+\frac{2a^2}{r}+a^2-\frac{4a\lambda}{r}+r^2\right)}\right]_{r_c}$} \\
$+\frac{4 \left(-\frac{a^2 \lambda ^4 (r-2) \left(a^2 (r+2)+r^3\right)}{\left(a^2 (r+2)+\lambda ^2 r\right)^2}-a^2 r^2+\frac{\lambda ^4 \left(a^2 \left(r^2-8\right)-r^3\right)}{a^2 (r+2)+\lambda ^2 r}+2 a \lambda  r^2+r^5\right)}{r^5 \left(-\frac{\lambda ^4 (r-2) \left(a^2 (r+2)+r^3\right)}{r^3 \left(a^2 (r+2)+\lambda ^2 r\right)}+\frac{2 a^2}{r}+a^2-\frac{4 a \lambda }{r}+r^2\right)}|_{r_c} +$\\
\resizebox{0.5\textwidth}{!}{$\left[\frac{\splitfrac{2\left(2 a^5 \lambda  r^2 (r+2)^2+4 a^3 \lambda ^3 r^3 (r+2)+2 a^2 \lambda ^2 r^3 (r+2) \left(r^3-a^2\right)+\lambda ^6 (-r) \left(r^3-a^2 \left(r^2-8\right)\right)\right.}{\left.+\lambda ^4 \left(a^4 (r-3) (r+2)^2-3 a^2 r^4+r^7\right)
+a^4 r^2 (r+2)^2 \left(r^3-a^2\right)+2 a \lambda ^5 r^4\right)^2}}{\splitfrac{r^2 \left(a^2 (r+2)+\lambda ^2 r\right)^2 \left(a^4 (r+2)^2 r^2-4 a^3 \lambda  (r+2) r^2\right.}{\left.+a^2 (r+2) \left(r^5+\lambda ^2 r^3-\lambda ^4 (r-2)\right)-4 a \lambda ^3 r^3+\lambda ^2 r^6-\lambda ^4 (r-2) r^3\right)^2}}\right]_{r_c}$} \\
\\
Using numerical techniques, eqns.(\ref{eqn8},\ref{eqn9},\ref{eqn10},\ref{eqn12},\ref{eqn13}) can simultaneously 
be solved to obtain the phase portrait corresponding to 
the transonic flow. 

\subsection{Conical Flow}
The corresponding expression for the mass and the entropy 
accretion rates for the conical flow come out to be \\
\begin{equation}
\dot{M}_{CF} = 4\pi\sqrt{\Delta}\Theta r \rho\sqrt{\frac{u^2}{1-u^2}}
\label{eqn14}
\end{equation}
\begin{equation}
\dot{\Xi}_{CF} = 4\pi\sqrt{\Delta}\Theta r\left[\frac{c_s^2(\gamma-1)}{\gamma-1-c_s^2}\right]^\frac{1}{\gamma-1} \sqrt{\frac{u^2}{1-u^2}}
\label{eqn15}
\end{equation}
where $\Theta$ is the solid angle subtended by the 
accretion disc at the horizon.
The space gradient of the sound speed and the flow 
velocity may be obtained as, \\
\begin{equation}
\frac{dc_s}{dr} = \frac{\mathcal{N}^{CF}_1}{\mathcal{D}^{CF}_1} 
\label{eqn16}
\end{equation}
\begin{equation}
\frac{du}{dr} = \frac{\mathcal{N}^{CF}_2}{\mathcal{D}^{CF}_2}
\label{eqn17}
\end{equation}
where $\mathcal{N}^{CF}_1=\mathcal{N}^{CH}_1$, $\mathcal{D}^{CF}_1=\mathcal{D}^{CH}_1$,
$\mathcal{N}^{CF}_2=u(1-u^2)\left[\frac{2r^2-3r+a^2}{\Delta r}c_s^2-\frac{f'}{2f}\right]$, $\mathcal{D}^{CF}_2=u^2-c_s^2$, 
$f=\frac{\Delta}{B}$, $B=g_{\phi\phi}+2\lambda g_{t\phi}+\lambda^2 g_{tt}$, where the sub/superscripts $CF$ stand 
for `Conical Flow'. \\
Hence, the corresponding critical point condition comes 
out to be \\
\begin{equation}
u^2|_{r_c}=c_s^2|_{r_c}=\frac{f'}{2f}|_{r_c}\frac{\Delta_c r_c}{2r_c^2-3r_c+a^2}
\label{eqn18}
\end{equation}
Substituting the critical point conditions in the 
expression of the conserved specific energy, one can solve 
the corresponding algebraic equation for a specific 
set of values of $\left[\mathcal{E},\lambda,\gamma,a\right]$, 
to obtain the value of the critical point $r_c$. \\
The space gradient of $c_s$ and $u$ at the critical points 
may be obtained as, \\
\begin{equation}
\frac{du}{dr}|_{r_c} = -\frac{\beta_{CF}}{2\alpha_{CF}} \pm \frac{1}{2\alpha_{CF}}\sqrt{\beta_{CF}^2-4\alpha_{CF}\Gamma_{CF}}
\label{eqn20}
\end{equation}
\begin{equation}
\frac{dc_s}{dr}|_{r_c} = \frac{\mathcal{N}^{CF}_1}{\mathcal{D}^{CF}_1}|_{r_c}
\label{eqn21}
\end{equation}
where, the co-efficients $\alpha_{CF}$, $\beta_{CF}$ and $\Gamma_{CF}$ are given by, \\
$\alpha_{CF}=\alpha_{CH}$, \\
$\beta_{CF}=\frac{2c_s(a^2+r(2r-3))\left(c_s^2-\gamma+1\right)}{\left(c_s^2-1\right)r\left(a^2+(r-2)r\right)}|_{r_c}$, \\
$\Gamma_{CF}=-\frac{c_s^2-1}{a^2+(r-2) r}|_{r_c}+\frac{2 \left(c_s^2-1\right) (r-1)^2}{\left(a^2+(r-2) r\right)^2}|_{r_c}
+\frac{c_s^2}{r^2}|_{r_c} 
+\frac{c_s^2 (r-1) \left(a^2+r (2 r-3)\right) \left(c_s^2-\gamma +1\right)}{\left(c_s^2-1\right) r \left(a^2+(r-2) r\right)^2}|_{r_c} \\
+\frac{c_s^2 \left(-c_s^2+\gamma -1\right)}{r^2}|_{r_c}+\frac{c_s^2 (r-1) \left(-c_s^2+\gamma -1\right)}{r \left(a^2+(r-2) r\right)}|_{r_c} 
+\frac{c_s^4 (r-1) \left(a^2+r (2 r-3)\right) \left(-c_s^2+\gamma -1\right)}{\left(c_s^2-1\right) r \left(a^2+(r-2) r\right)^2}|_{r_c}$ \\
\resizebox{0.5\textwidth}{!}{$-\left[\frac{\splitfrac{-\frac{a^2 \lambda ^4 \left(a^2 (r+2)+r^3\right)}{\left(a^2 (r+2)+\lambda ^2 r\right)^2}+\frac{2 a^2 \lambda ^4 (r-2) \left(a^2+\lambda ^2\right) \left(a^2 (r+2)+r^3\right)}{\left(a^2 (r+2)+\lambda ^2 r\right)^3}-\frac{a^2 \lambda ^4 (r-2) \left(a^2+3 r^2\right)}{\left(a^2 (r+2)+\lambda ^2 r\right)^2}}{+\frac{\lambda ^4 \left(a^2+\lambda ^2\right) \left(r^3-a^2 \left(r^2-8\right)\right)}{\left(a^2 (r+2)+\lambda ^2 r\right)^2}+\frac{\lambda ^4 r \left(2 a^2-3 r\right)}{a^2 (r+2)+\lambda ^2 r}-2 a^2 r+4 a \lambda  r+5 r^4}}{r^4 \left(-\frac{\lambda ^4 (r-2) \left(a^2 (r+2)+r^3\right)}{r^3 \left(a^2 (r+2)+\lambda ^2 r\right)}+\frac{2 a^2}{r}+a^2-\frac{4 a \lambda }{r}+r^2\right)}\right]_{r_c}$} \\
$+\frac{4 \left(-\frac{a^2 \lambda ^4 (r-2) \left(a^2 (r+2)+r^3\right)}{\left(a^2 (r+2)+\lambda ^2 r\right)^2}-a^2 r^2+\frac{\lambda ^4 \left(a^2 \left(r^2-8\right)-r^3\right)}{a^2 (r+2)+\lambda ^2 r}+2 a \lambda  r^2+r^5\right)}{r^5 \left(-\frac{\lambda ^4 (r-2) \left(a^2 (r+2)+r^3\right)}{r^3 \left(a^2 (r+2)+\lambda ^2 r\right)}+\frac{2 a^2}{r}+a^2-\frac{4 a \lambda }{r}+r^2\right)}|_{r_c}$ \\
\resizebox{0.5\textwidth}{!}{$+\left[\frac{\splitfrac{2 \left(2 a^5 \lambda  r^2 (r+2)^2+4 a^3 \lambda ^3 r^3 (r+2)+2 a^2 \lambda ^2 r^3 (r+2) \left(r^3-a^2\right)+\lambda ^6 (-r) \left(r^3-a^2 \left(r^2-8\right)\right)\right.}{\left.+\lambda ^4 \left(a^4 (r-3) (r+2)^2-3 a^2 r^4+r^7\right)+a^4 r^2 (r+2)^2 \left(r^3-a^2\right)+2 a \lambda ^5 r^4\right)^2}}{\splitfrac{r^2 \left(a^2 (r+2)+\lambda ^2 r\right)^2 \left(a^4 (r+2)^2 r^2-4 a^3 \lambda  (r+2) r^2\right.}{\left.+a^2 (r+2) \left(r^5+\lambda ^2 r^3-\lambda ^4 (r-2)\right)-4 a \lambda ^3 r^3+\lambda ^2 r^6-\lambda ^4 (r-2) r^3\right)^2}}\right]_{r_c}$}. \\
\\
Using numerical techniques, eqns.(\ref{eqn16},\ref{eqn17},\ref{eqn18},\ref{eqn20},\ref{eqn21}) 
may be solved to obtain the phase portrait of the 
transonic solutions.

\subsection{Flow in hydrostatic equilibrium along the vertical direction}
The mass accretion rate is found to be, \\
\begin{equation}
\dot{M}_{VE} = 4\pi\sqrt{\Delta}H(r)\rho\sqrt{\frac{u^2}{1-u^2}} 
\label{eqn22}
\end{equation}
The disc height can be calculated as, \\
\begin{equation}
H(r)=\sqrt{\frac{2}{\gamma}}r^2\left[\frac{c_s^2(\gamma-1)}{(\gamma-1-c_s^2)F}\right]^\frac{1}{2}
\label{eqn22a}
\end{equation}
with $v_t=\sqrt{\frac{f}{1-u^2}}$ and $F=\lambda^2 v_t^2-a^2(v_t-1)$. \\
The corresponding entropy accretion rate is given by, \\
\begin{equation}
\dot{\Xi}_{VE} = 4\pi ur^2\left[\frac{c_s^2(\gamma-1)}{\gamma-1-c_s^2}\right]^{\frac{\gamma+1}{2(\gamma-1)}}
\left[\frac{2\Delta}{\gamma(1-u^2)F}\right]^{\frac{1}{2}}
\label{eqn23}
\end{equation}
The space gradient of $c_s$ and $u$ can be obtained as, \\
\begin{equation}
\frac{dc_s}{dr} = \frac{\mathcal{N}^{VE}_1}{\mathcal{D}^{VE}_1}
\label{eqn24}
\end{equation}
\begin{equation}
\frac{du}{dr} = \frac{\mathcal{N}^{VE}_2}{\mathcal{D}^{VE}_2}
\label{eqn25}
\end{equation}
where, \\
$\mathcal{N}^{VE}_2=\frac{2 c_s^2}{\gamma +1}\left(-\frac{P1 v_t \left(2 \lambda ^2 v_t-a^2\right)}{4 F}+\frac{\Delta '}{2 \Delta }+\frac{2}{r}\right)-\frac{P1}{2}$, \\
$\mathcal{D}^{VE}_2=\frac{u}{1-u^2}-\frac{2 c_s^2}{\gamma +1}\frac{1}{\left(1-u^2\right)u}{\left(1-\frac{u^2 v_t \left(2 \lambda ^2 v_t-a^2\right)}{2 F}\right)}$, \\
$P1=\frac{\Delta'}{\Delta }+\frac{d\Omega}{dr}\frac{\lambda}{1-\Omega \lambda}-\frac{g_{\phi\phi}'+\lambda g_{t\phi}'}{g_{\phi\phi}+\lambda g_{t\phi}}$ and $\Omega=\frac{v^\phi}{v^t}$, where the sub/superscripts $VE$ stand for `Vertical Equilibrium'. \\
This provides the corresponding critical conditions as, \\
\begin{equation}
u^2|_{r_c}=\frac{\text{P1}}{\frac{\Delta '}{\Delta }+\frac{4}{r}}|_{r_c}
\label{eqn26}
\end{equation}
\begin{equation}
c_s^2|_{r_c}=\frac{(\gamma +1) \left(2 F u^2\right)}{2 \left(2 F-u^2 v_t \left(2 \lambda ^2 v_t-a^2\right)\right)}|_{r_c}
\label{eqn27}
\end{equation}
The corresponding space gradients of velocities at 
critical points are obtained as, \\
\begin{equation}
\frac{du}{dr}|_{r_c}=-\frac{\beta_{VE}}{2\alpha_{VE}} \pm \frac{1}{2\alpha_{VE}}\sqrt{\beta_{VE}^2-4\alpha_{VE}\Gamma_{VE}}
\label{eqn28}
\end{equation}
\begin{equation}
\frac{dc_s}{dr}|_{r_c} = \frac{\mathcal{N}^{VE}_1}{\mathcal{D}^{VE}_1}|_{r_c}
\label{eqn29}
\end{equation}
where the co-efficients $\alpha_{VE}$, $\beta_{VE}$ and $\Gamma_{VE}$ are given by, \\
$\alpha_{VE}=\frac{1+u^2}{\left(1-u^2\right)^2}-\frac{2nD_2D_6}{2n+1}$, $\beta_{VE} =\frac{2nD_2D_7}{2n+1}+\tau_4$, $\Gamma_{VE}=-\tau_3$, \\
$n=\frac{1}{\gamma -1}$, $D_2=\frac{c_s^2}{u\left(1-u^2\right)}\left(1-D_3\right)$, $D_6=\frac{3u^2-1}{u\left(1-u^2\right)}-\frac{D_5}{1-D_3}-\frac{\left(1-nc_s^2\right)u}{nc_s^2\left(1-u^2\right)}$, \\
$D_7=\frac{1-nc_s^2}{nc_s^2}\frac{P1}{2}+\frac{D_3D_4v_tP1}{2\left(1-D_3\right)}$, $\tau _3=\frac{2n}{2n+1}\left(c_s^2\tau
_2-\frac{v_tP1v_1}{2nv_t}\left(1-nc_s^2\right)-c_s^2v_5v_t\frac{P1}{2}\right)-\frac{P1'}{2}$, \\
$\tau _4=\frac{2n}{2n+1}\frac{v_tu}{1-u^2}\left(\frac{v_1}{nv_t}\left(1-nc_s^2\right)+c_s^2v_5\right)$, $v_1=\frac{\Delta'}{2\Delta
}+\frac{2}{r}-\left(2\lambda ^2v_t-a^2\right)v_t\frac{\text{P1}}{4F}$, \\
$D_3=\frac{u^2v_t\left(2\lambda^2v_t-a^2\right)}{2F}$, $D_4=\frac{1}{v_t}+\frac{2\lambda^2}{2\lambda^2v_t-a^2}-\frac{2\lambda^2v_t-a^2}{F}$,
$D_5=D_3\left(\frac{2}{u}+\frac{D_4v_tu}{1-u^2}\right)$, $\tau_2=\tau _1-\frac{v_t\left(2\lambda^2v_t-a^2\right)}{4F}P1'$, $v_5=\left(2\lambda^2v_t-a^2\right)\frac{P1}{4F}v_4$, \\
$\tau_1=\frac{1}{2}\left(\frac{\Delta''}{\Delta}-\frac{(\Delta')^2}{\Delta ^2}\right)-\frac{2}{r^2}$, 
$v_4=\frac{v_3}{\left(2\lambda^2v_t-a^2\right)F}$, $v_3=\left(4\lambda^2v_t-a^2\right)F-\left(2\lambda^2v_t-a^2\right)^2v_t$. \\
\\
Note that Mach number is not unity at the critical points. 
Hence, apparently the critical points and the sonic points 
are not isomorphic. This issue may be resolved in two 
different ways: \\
a) The time-dependent Euler equation and the continuity 
equation can be linearly perturbed to find out the 
corresponding 
wave equation which describes the propagation of the 
acoustic perturbation through the background fluid 
space-time. 
The speed of propagation of such perturbation can be taken 
as the effective adiabatic sound speed. The critical 
points become the sonic points for such effective sound 
speed. This treatment requires dealing with the time 
dependent perturbation techniques, which is beyond the 
scope of this present work. For related calculations, one 
may refer to \cite{bbd17na}, where such perturbation 
technique has been applied for accretion in the 
Schwarzschild metric. \\
b) The integral solutions may be numerically carried out 
starting from the critical point upto a certain radial 
distance where the Mach number becomes exactly equal to 
unity and the corresponding radial distance can be 
considered as the sonic point. 
In our present work, we shall follow this approach.

\section{Parameter Space for Polytropic Accretion}
One obtains (\cite{dnhbmcbwkn15na}) the limits for the 
four parameters governing the flow as 
$\left[ 0 < \mathcal{E} < 1, 0 < \lambda < 4, \frac{4}{3} < \gamma < \frac{5}{3}, -1 < a < 1 \right]$ 
For polytropic accretion in the Kerr metric, the parameter 
space is four dimensional. For our convenience, 
we deal with two dimensional parameter space. $^4C_2$ such 
spaces may be obtained. For the time being, we concentrate 
on $\left[\mathcal{E}-\lambda\right]$ parameter space for 
fixed values of $\left[\gamma,a\right]$. \\

\begin{figure}
\centering
\includegraphics[scale=0.7]{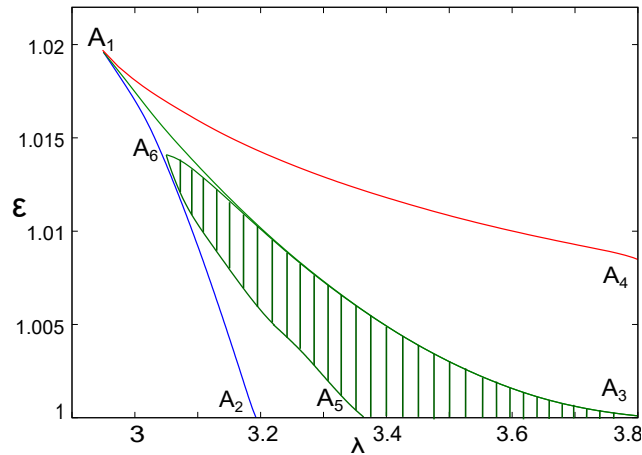}
\caption{$\mathcal{E}$-$\lambda$ plot for quasi-spherical disc geometry ($\gamma=1.35$ and $a=0.1$)}
\label{fig1}
\end{figure}

Figure \ref{fig1} shows the $\left[\mathcal{E}-\lambda\right]$ parameter space for adiabatic accretion in quasi-
spherical geometry for $\left[\gamma=1.35,a=0.1\right]$. 
Similar diagrams can be produced for the two other 
geometries as well. A$_1$A$_2$A$_3$A$_4$ represents the 
region of $\left[{\cal E},\lambda\right]$ for which the 
corresponding polynomial equation in $r_c$ along with the 
corresponding critical point conditions provides three 
real positive roots lying outside $r_+$, where 
$r_+=1+\sqrt{1-a^2}$, $a$ being the Kerr parameter. 
For region A$_1$A$_2$A$_3$, 
one finds ${\dot {\Xi}}_{\rm inner} > {\dot {\Xi}}_{\rm outer}$ and accretion is 
multi-critical. A$_3$A$_5$A$_6$ (shaded in green), which is a subspace of A$_1$A$_2$A$_3$ allows shock formation. Such a
subspace provides true multi-transonic accretion where the 
stationary transonic 
solution passing through the outer sonic point joins with 
the stationary transonic 
solution constructed through the inner sonic point through 
a discontinuous energy preserving shock
of Rankine-Hugoniot type. Such shocked multi-transonic 
solution contains two smooth transonic (from sub to super) 
transitions at two regular sonic points (of saddle type) 
and a discontinuous transition (from super to sub)
at the shock location.

On the other hand, the region A$_1$A$_3$A$_4$ represents 
the subset of 
$\left[{\cal E},\lambda,\gamma\right]_{\rm mc}$ (where 
`mc' stands for `multi-critical') for which 
${\dot {\Xi}}_{\rm inner} < {\dot {\Xi}}_{\rm outer}$ and 
hence incoming flow can have only one critical point of 
saddle type and the background flow
possesses one acoustic horizon at the inner saddle type 
sonic point. 
The boundary A$_1$A$_3$ between these two regions represents the value of $\left[{\cal E},\lambda,\gamma\right]$
for which multi-critical accretion is characterized by 
${\dot {\Xi}}_{\rm inner} = {\dot {\Xi}}_{\rm outer}$ and 
hence the transonic solutions passing through the inner 
and the outer sonic points are completely degenerate, 
leading to the formation of a 
heteroclinic orbit \footnote{Heteroclinic orbits are the 
trajectories defined on a phase
portrait which connects two different saddle type critical 
points. Integral solution configuration on phase portrait 
characterized by heteroclinic orbits are topologically 
unstable (\cite{js99}, \cite{strogatz01}).}
on the phase portrait. Such flow pattern 
may be subjected to instability and turbulence as well.\\

\begin{figure}
\centering
\includegraphics[scale=0.3]{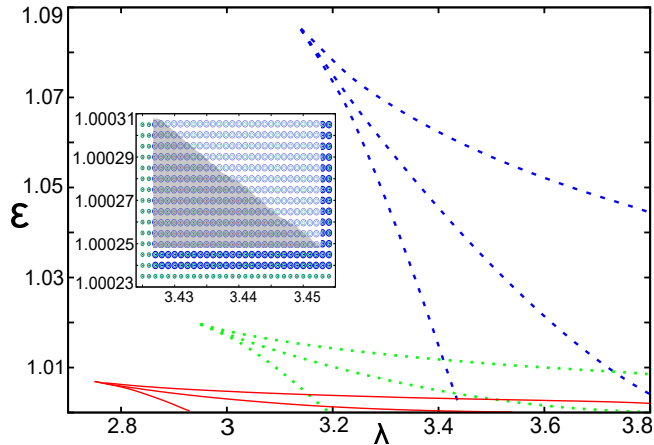}
\caption{Comparision of $\mathcal{E}$-$\lambda$ plot for three different flow geometries ($\gamma=1.35$ and $a=0.1$). 
Constant height disc, quasi-spherical flow and flow in vertical hydrostatic equilibrium represented by blue dashed lines, 
green dotted lines and red solid lines respectively. Shaded region in the inset depicts $\left[\mathcal{E},\lambda\right]$ space overlap 
with multicritical solutions for all three models.}
\label{fig2}
\end{figure}

In figure \ref{fig2}, for the same values of $\left[\gamma,a\right]$, we compare the parameter spaces 
for three different 
flow geometries. The common region for which multiple 
critical points are formed for all three flow geometries 
are shown in the inset.

\section{Classification of critical points for polytropic accretion}
In the previous section, we found that the transonic 
accretion may possess, depending on the initial boundary 
conditions defined by the values of $\left[\mathcal{E},\lambda,\gamma,a\right]$, one or three critical points. 
Since we consider inviscid, non-dissipative flow, the 
critical points are expected to be either of saddle type, 
or of centre type. No spiral (instead of the centre type) 
or nodal type points may be observed. The nature of 
the critical points, whose locations are obtained by 
substituting the critical point conditions for accretion 
flow in different geometries and solving for the equation 
of specific energy, cannot be determined from such 
solutions. A classification scheme has been developed 
(\cite{gkrd07mnras}) to accomplish such a task. Once the 
location of a critical point is identified, the linearized 
study of the space gradients of the square of the 
advective velocity in the close neighbourhood of such a 
point may be carried out to develop a complete and 
rigorous mathematical classification scheme to understand 
whether a critical point is of saddle type, or of 
centre type. Such methodology is based on a local 
classification scheme. The global understanding of the 
flow topology is not possible to accomplish using such 
scheme. For that purpose, study of the stationary integral 
flow solution is necessary, which can be accomplished only 
numerically. Such numerical scheme to obtain the global 
phase portraits will be discussed in detail in the 
subsequent sections. \\

Stationary axisymmetric accretion in the Kerr metric can be 
described by a first order autonomous differential 
equation (\cite{gkrd07mnras}) to apply the formalism 
borrowed from the dynamical systems theory and to 
find out the nature of the critical points using such 
formalism. Below, we generalize such analysis for 
polytropic accretion in three different models. \\

\subsection{Constant Height Flow}
The gradient of square of the sound speed and the dynamical 
flow velocity (the advective velocity) can be 
written as, \\
\begin{equation}
\frac{dc_s^2}{dr}=(\gamma-1-c_s^2)\left[\frac{-1}{2(1-u^2)}\frac{du^2}{dr}-\frac{f'}{2f}\right]
\label{eqn30}
\end{equation}
\begin{equation}
\frac{du^2}{dr}=\frac{2\left[\frac{r-1}{\Delta}c_s^2-\frac{f'}{2f}\right]}{\frac{1}{u^2}\left(\frac{1}{1-u^2}\right)(u^2-c_s^2)}
\label{eqn31}
\end{equation}
One can decompose the expression for $\frac{du^2}{dr}$ into 
two parametrized equations using a dummy 
mathematical parameter $\tau$ as \\
\begin{eqnarray}
\frac{du^2}{d\tau}=2\left[\frac{r-1}{\Delta}c_s^2-\frac{f'}{2f}\right] \nonumber \\
\frac{dr}{d\tau}=\frac{1}{u^2}\left(\frac{1}{1-u^2}\right)(u^2-c_s^2)
\label{eqn33}
\end{eqnarray}
The above equation is an autonomous equation and hence 
$\tau$ does not explicitly appear in their 
right hand sides. About the fixed values of the critical 
points, one uses a perturbation prescription 
of the following form, \\
\begin{equation}
u^2=u_c^2+\delta u^2
\label{eqn34}
\end{equation}
\begin{equation}
c_s^2={c_s}_c^2+\delta c_s^2
\label{eqn35}
\end{equation}
\begin{equation}
r=r_c+\delta r
\label{eqn36}
\end{equation}
and derives a set of two autonomous first order linear 
differential equations in the $\delta r-\delta u^2$ 
plane, by expressing $\delta c_s^2$ in terms of $\delta r$ 
and $\delta u^2$ as, \\
\begin{equation}
\frac{\delta c_s^2}{{c_s}_c^2}=(\gamma-1-{c_s}_c^2)\left[\frac{-1}{2u_c^2(1-u_c^2)}\delta u^2-\frac{r_c-1}{\Delta_c}\delta r\right]
\label{eqn37}
\end{equation}
This form of $\delta c_s^2$ has been derived using the 
modified form (in terms of $u^2$ instead of $u$) of the 
mass accretion rate (eqn.(\ref{eqn6})) and its 
corresponding expression for the entropy accretion rate 
(eqn.(\ref{eqn7})). 
Through this procedure, a set of coupled linear equations 
in $\delta r$ and $\delta u^2$ will be obtained as \\
\begin{equation}
\frac{d}{d\tau}(\delta u^2)=\mathcal{A}_{CH}\delta u^2+\mathcal{B}_{CH}\delta r
\label{eqn38}
\end{equation}
\begin{equation}
\frac{d}{d\tau}(\delta r)=\mathcal{C}_{CH}\delta u^2+\mathcal{D}_{CH}\delta r
\label{eqn39}
\end{equation}
where
\begin{eqnarray}
&{\cal A}_{CH} = \dfrac{(1-r_c)(\gamma - 1-c_{s_c}^2)}{\Delta_c (1-u_c^2)}\\
&{\cal B}_{CH} = 2\left[ \dfrac{c_{s_c}^2}{\Delta_c} -\dfrac{(r_c-1)^2c_{s_c}^2}{\Delta_c^2}(\gamma +1 -c_{s_c}^2)\right. \nonumber \\
&\left.-\dfrac{f''}{2f}+\dfrac{1}{2}\left(\dfrac{f'}{f}\right)^2\right]\\
&{\cal C}_{CH} = \left[1+\dfrac{(\gamma -1 -c_{s_c}^2)}{2(1-u_c^2)} \right] \dfrac{1}{u_c^2(1-u_c^2)}\\
&{\cal D}_{CH} = -{\cal A}_{CH}
\label{eqn39a}
\end{eqnarray}
Using trial solutions of the form $\delta u^2 \sim \exp(\Omega\tau)$ and $\delta r \sim \exp(\Omega\tau)$ 
($\Omega$, in this context, should not be confused with  
angular velocity of the flow in eqn.(\ref{eqn4})), the 
eigenvalues of the stability matrix 
can be expressed as, \\
\begin{equation}
\Omega_{CH}^2 \equiv \Omega_{{CH}_1}\Omega_{{CH}_2} = \mathcal{B}_{CH}\mathcal{C}_{CH}-\mathcal{A}_{CH}\mathcal{D}_{CH}
\label{eqn40}
\end{equation}
Once the numerical values corresponding to the location of the critical points are obtained, it is straightforward 
to calculate the numerical value corresponding to the expression for $\Omega^2$, since $\Omega^2$ is essentially 
a function of $r_c$. The accreting black hole system under consideration is a conservative system, hence either 
$\Omega^2 > 0$, which implies the critical points are saddle type, or one obtains $\Omega^2 < 0$, which implies 
the critical points are centre type. One thus understands the nature of the critical points (whether saddle type 
or centre type) once the values of $r_c$ is known. It has been observed that the single critical point solutions 
are always of saddle type. This is obvious, otherwise monotransonic 
solutions would not exist. It is also observed that for multi-critical flow, the middle critical point is of 
centre type and the inner and the outer critical points are of saddle type. This will be explicitly shown 
diagramatically in the subsequent sections.

\subsection{Conical Flow}
The gradient of square of the sound speed and the advective velocity are given by, \\
\begin{equation}
\dfrac{dc_s^2}{dr} = (\gamma -1-c_s^2)\left[\dfrac{-1}{2(1-u^2)}\dfrac{du^2}{dr}-\dfrac{f'}{2f}\right]
\label{eqn41}
\end{equation}
\begin{equation}
\dfrac{du^2}{dr} = \dfrac{2\left[\dfrac{(2r^2-3r+a^2)}{\Delta r}c_s^2 -\dfrac{f'}{2f} \right]}{\dfrac{1}{u^2}\left(\dfrac{1}{1-u^2}\right)(v^2-c_s^2)}
\label{eqn42}
\end{equation}
The parametrized form of the expression of $\frac{du^2}{dr}$ is given by the equations, \\
\begin{eqnarray}
\dfrac{du^2}{d \tau} = 2\left[\dfrac{(2r^2-3r+a^2)}{\Delta r}c_s^2 -\dfrac{f'}{2f} \right] \nonumber \\
\dfrac{dr}{d \tau} = (u^2-c_s^2)\dfrac{1}{u^2(1-u^2)}
\label{eqn43}
\end{eqnarray}
Using the perturbation scheme of eqns.(\ref{eqn34}), (\ref{eqn35}) and (\ref{eqn36}) we obtain, \\
\begin{eqnarray}
&\dfrac{\delta c_s^2}{c_{s_c}^2} = (\gamma -1 -c_{s_c}^2)\left[-\dfrac{1}{2u_c^2(1-u_c^2)}\delta u^2 \right. \nonumber \\
&\left.-\dfrac{(2r_c^2-3r_c+a^2)}{\Delta_c r_c}\delta r \right]
\label{eqn44}
\end{eqnarray}
where $\delta c_s^2$ has been derived using modified form of eqns.(\ref{eqn14}) and (\ref{eqn15}). 
The coupled linear equations in $\delta r$ and $\delta u^2$ are given by, \\
\begin{equation}
\dfrac{d}{d\tau}(\delta u^2) = {\cal A}_{CF} \delta u^2 + {\cal B}_{CF} \delta r
\label{eqn45}
\end{equation}
\begin{equation}
\dfrac{d}{d\tau}(\delta r) = {\cal C}_{CF} \delta u^2 + {\cal D}_{CF} \delta r
\label{eqn46}  
\end{equation}
where, \\
\begin{eqnarray}
&{\cal A}_{CF} = -\dfrac{(2r_c^2-3r_c+a^2)(\gamma - 1-c_{s_c}^2)}{\Delta_c r_c(1-u_c^2)}\\
&{\cal B}_{CF} = \dfrac{2(4r_c-3)c_{s_c}^2}{\Delta_c r_c} -\dfrac{f''}{f}+\left(\dfrac{f'}{f}\right)^2
-\dfrac{2 c_{s_c}^2}{\Delta_c^2 r_c^2}(2r_c^2-3r_c+a^2)\nonumber \\
&\left[ (3r_c^2-4r_c+a^2)+ (\gamma -1 -c_{s_c}^2)(2r_c^2-3r_c+a^2)\right]\\
&{\cal C}_{CF} = \left[1+\dfrac{(\gamma -1 -c_{s_c}^2)}{2(1-u_c^2)} \right] \dfrac{1}{u_c^2(1-u_c^2)}\\
&{\cal D}_{CF} = -{\cal A}_{CF}
\label{eqn46a}
\end{eqnarray}
Using the prescription mentioned in the previous subsection, eigenvalues of the stability matrix are obtained as, \\
\begin{equation}
\Omega_{CF}^2 \equiv \Omega_{{CF}_1}\Omega_{{CF}_2} = \mathcal{B}_{CF}\mathcal{C}_{CF}-\mathcal{A}_{CF}\mathcal{D}_{CF}
\label{eqn47}
\end{equation}

\subsection{Flow in hydrostatic equilibrium along the vertical direction}
The gradient of square of the advective velocity is given by, \\
\begin{equation}
\dfrac{du^2}{dr} = \frac{\beta^2 c_s^2\left[\dfrac{F_1'}{F_1}-\dfrac{1}{F}\dfrac{\partial F}{\partial r}\right]-\dfrac{f'}{f}}
{\left(1-\frac{\beta^2 c_s^2}{u^2}\right)\dfrac{1}{(1-u^2)}+\dfrac{\beta^2 c_s^2}{F}\left(\dfrac{\partial F}{\partial u^2}\right)}
\label{eqn49}
\end{equation}
where $F_1=\Delta r^4$ and $\beta=\sqrt{\frac{2}{\gamma+1}}$. \\
The parametrized form of the expression of $\frac{du^2}{dr}$ is given by the equations, \\
\begin{eqnarray}
\dfrac{du^2}{d\tau} = \beta^2 c_s^2\left[\dfrac{F_1'}{F_1}-\dfrac{1}{F}\dfrac{\partial F}{\partial r}\right]-\dfrac{f'}{f} \nonumber \\
\dfrac{dr}{d\tau} = \left(1-\frac{\beta^2 c_s^2}{u^2}\right)\dfrac{1}{(1-u^2)}+\dfrac{\beta^2 c_s^2}{F}\left(\dfrac{\partial F}{\partial u^2}\right)
\label{eqn50}
\end{eqnarray}
Using the perturbation scheme of eqns.(\ref{eqn34}), (\ref{eqn35}) and (\ref{eqn36}), and modified forms of 
eqns.(\ref{eqn22}) and (\ref{eqn23}) we obtain, \\
\begin{eqnarray}
\frac{\delta c_s^2}{c_{s_c}^2}=\mathcal{A}\delta u^2+\mathcal{B}\delta r
\label{eqn51}
\end{eqnarray}
where, \\
$\mathcal{A}=-\frac{\gamma-1-c_{s_c}^2}{\gamma +1}\left[\frac{1}{u_c^2\left(1-u_c^2\right)}
- \frac{1}{F_c}\left(\frac{\partial F}{\partial u^2}\right)\bigg\vert_c\right]$, \\
$\mathcal{B}=-\frac{\gamma-1-c_{s_c}^2}{\gamma +1}\left[\frac{F_1^\prime(r_c)}{F_1(r_c)}
- \frac{1}{F_c}\left(\frac{\partial F}{\partial r}\right)\bigg\vert_c\right]$. \\
The coupled linear equations in $\delta r$ and $\delta u^2$ are given by, \\
\begin{eqnarray}
\frac{\mathrm d}{{\mathrm d}\tau}(\delta u^2) &=& \beta^2 c_{\mathrm{sc}}^2
\left[\frac{{\mathcal A}F_1^{\prime}}{F_1} - \frac{\mathcal{AC}}{F}
+ \frac{\mathcal{CD}}{F^2} - \frac{\Delta_3}{F} \right]\, 
\delta u^2 
+ \left[\frac{\beta^2 c_{\mathrm{sc}}^2 F_1^{\prime}}{F_1} 
\left\{{\mathcal B} + \left(\frac{F_1^{\prime \prime}}{F_1^{\prime}} 
- \frac{F_1^{\prime}}{F_1}\right)\right\}\right. \nonumber \\
& & \left. - \frac{f^{\prime}}{f}
\left(\frac{f^{\prime\prime}}{f^\prime} - \frac{f^{\prime}}{f}\right)
-\frac{\beta^2 c_{\mathrm{sc}}^2 {\mathcal C}}{F} \left({\mathcal B}
- \frac{\mathcal C}{F} + \frac{\Delta_4}{\mathcal C} 
\right) \right] \delta r
\label{eqn52a}
\end{eqnarray}
\begin{eqnarray}
\frac{{\mathrm d}}{{\mathrm d}\tau}(\delta r) &=& 
\left[\frac{1}{\left(1-u_{\mathrm{c}}^2\right)^2} 
-\frac{\beta^2 c_{\mathrm{sc}}^2}{u_{\mathrm{c}}^2
\left(1-u_{\mathrm{c}}^2\right)} \left\{ {\mathcal A} 
+ \frac{2u_{\mathrm{c}}^2-1}{\left(1-u_{\mathrm{c}}^2\right)^2}\right\}
+ \frac{\beta^2 c_{\mathrm{sc}}^2 {\mathcal D}}{F}\left({\mathcal A}
- \frac{\mathcal D}{F} + \frac{\Delta_1}{\mathcal D}\right) 
\right] \delta u^2 \nonumber \\ 
& &+ \left[ -\frac{\beta^2 c_{\mathrm{sc}}^2 {\mathcal B}}
{u_{\mathrm{c}}^2\left(1-u_{\mathrm{c}}^2\right)}
+ \frac{\beta^2 c_{\mathrm{sc}}^2 {\mathcal D}}{F} \left({\mathcal B}
-\frac{\mathcal C}{F}+\frac{\Delta_2}{\mathcal D}\right)\right]\delta r 
\label{eqn52b}
\end{eqnarray}
where, \\
$\mathcal{C}=\left(\frac{\partial F}{\partial r}\right)\bigg\vert_c$, 
$\mathcal{D}=\left(\frac{\partial F}{\partial u^2}\right)\bigg\vert_c$, \\
$\Delta_1=\frac{\partial}{\partial u^2}\left(\frac{\partial F}{\partial u^2}\right)\bigg\vert_c$, 
$\Delta_2=\frac{\partial}{\partial r}\left(\frac{\partial F}{\partial u^2}\right)\bigg\vert_c$, 
$\Delta_3=\frac{\partial}{\partial u^2}\left(\frac{\partial F}{\partial r}\right)\bigg\vert_c$, 
$\Delta_4=\frac{\partial}{\partial r}\left(\frac{\partial F}{\partial r}\right)\bigg\vert_c$. \\
Using the prescription mentioned in the previous subsection, eigenvalues of the stability matrix are obtained as, \\
\begin{equation}
\Omega_{VE}^2=\beta^4 c_{s_c}^4 \chi^2 + \xi_1\xi_2
\label{eqn53}
\end{equation}
where,
$\chi=\left[\frac{F_1^{\prime}{\mathcal A}}{F_1}
-\frac{\mathcal{AC}}{F}+\frac{\mathcal{CD}}{F^2}
-\frac{\Delta_3}{F}\right]
=\left[\frac{\mathcal B}{u_{\mathrm c}^2\left(1-u_{\mathrm c}^2\right)}
-\frac{\mathcal{BD}}{F}
+\frac{\mathcal{CD}}{F^2}-\frac{\Delta_2}{F}\right]$, \\
$\xi_1=\frac{\beta^2 c_{\mathrm{sc}}^2 F^{\prime}_1}{F_1}
\left[{\mathcal B}+ \frac{F_1^{\prime\prime}}{F_1^{\prime}}
- \frac{F_1^{\prime}}{F_1} \right]
-\frac{f^{\prime}}{f}\left[\frac{f^{\prime\prime}}{f^{\prime}}
-\frac{f^{\prime}}{f}\right] 
- \frac{\beta^2 c_{\mathrm{sc}}^2 \mathcal{C}}{F}
\left[{\mathcal B}-\frac{\mathcal C}{F}+
\frac{\Delta_4}{\mathcal C}\right]$, and \\
$\xi_2=\frac{1}{\left(1-u_{\mathrm c}^2\right)^2}
- \frac{\beta^2c_{\mathrm{sc}}^2}{u_{\mathrm c}^2
\left(1-u_{\mathrm c}^2\right)}\left[{\mathcal A}+
\frac{2u_{\mathrm c}^2-1}{u_{\mathrm c}^2\left(1-u_{\mathrm c}^2\right)}
\right]+\frac{\beta^2 c_{\mathrm{sc}}^2 \mathcal{D}}{F}
\left[\mathcal{A}-\frac{\mathcal D}{F}+\frac{\Delta_1}{\mathcal D}\right]$.

\section{Dependence of $\Omega^2$ on flow and spin parameters for polytropic accretion with various matter geometries}

\begin{figure} [h!]
\centering
\begin{tabular}{cc}
\includegraphics[width=0.4\linewidth]{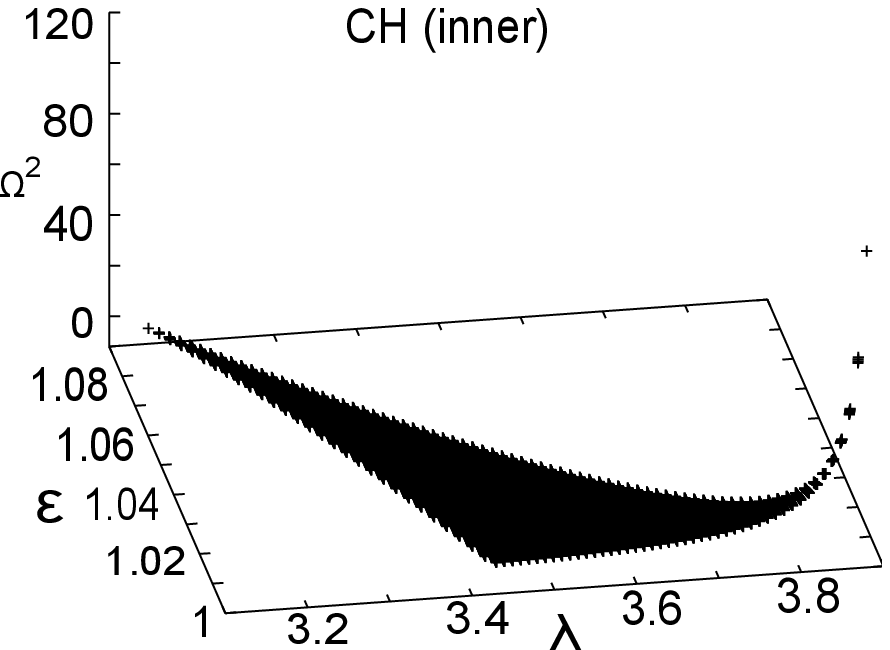} &
\includegraphics[width=0.4\linewidth]{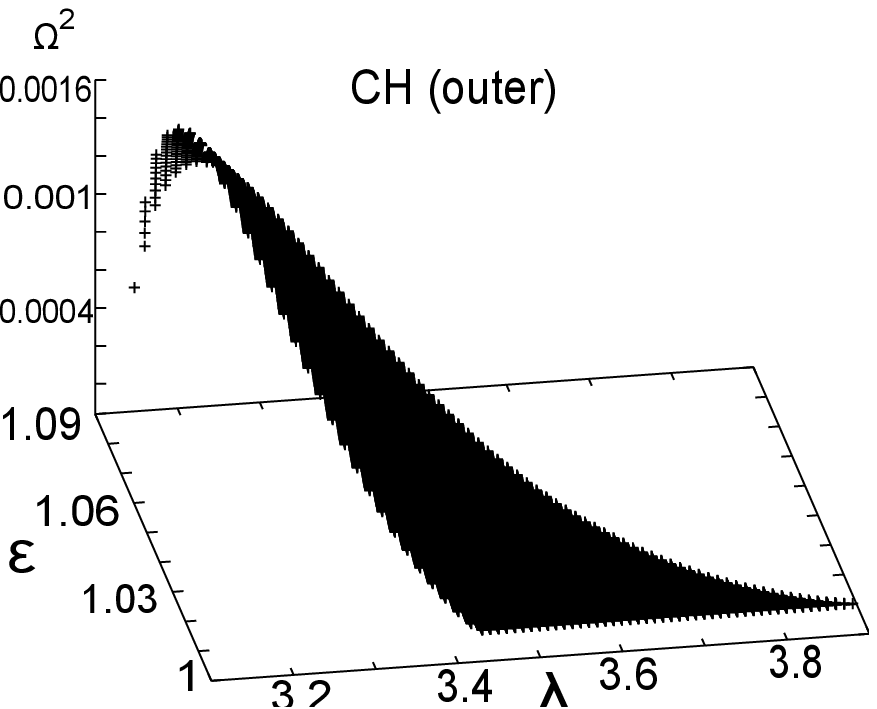} \\
\includegraphics[width=0.4\linewidth]{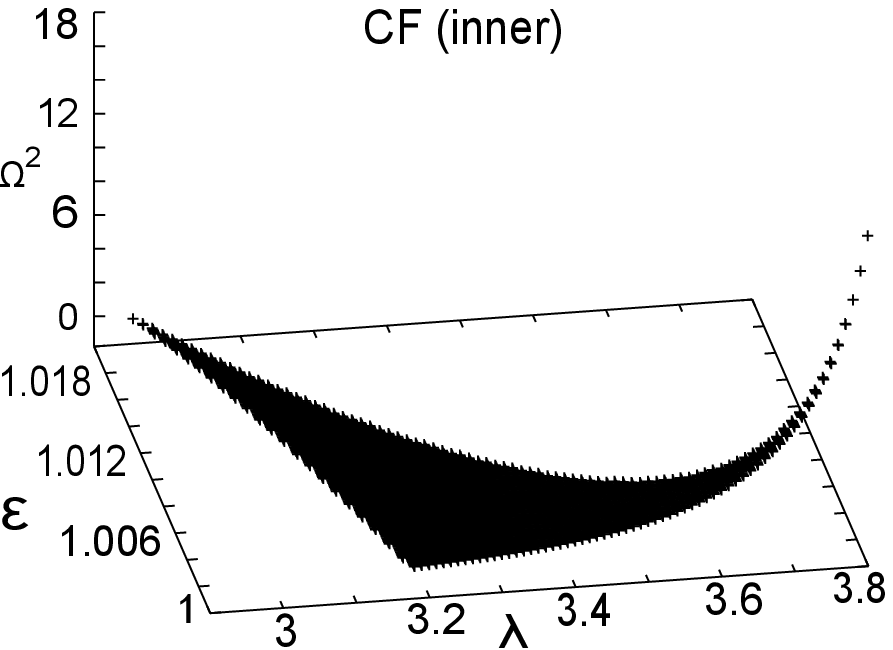} &
\includegraphics[width=0.4\linewidth]{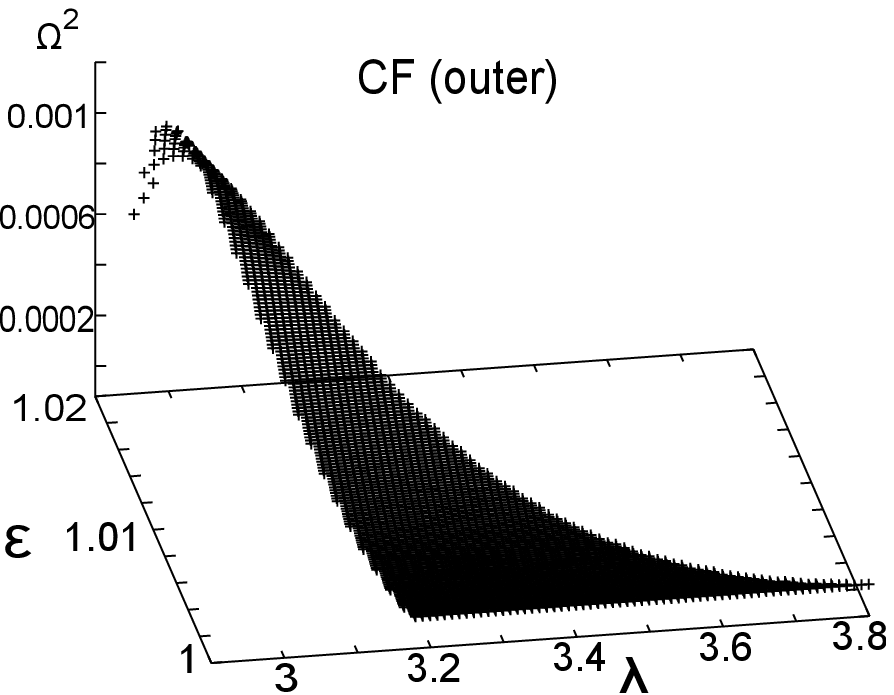} \\
\includegraphics[width=0.4\linewidth]{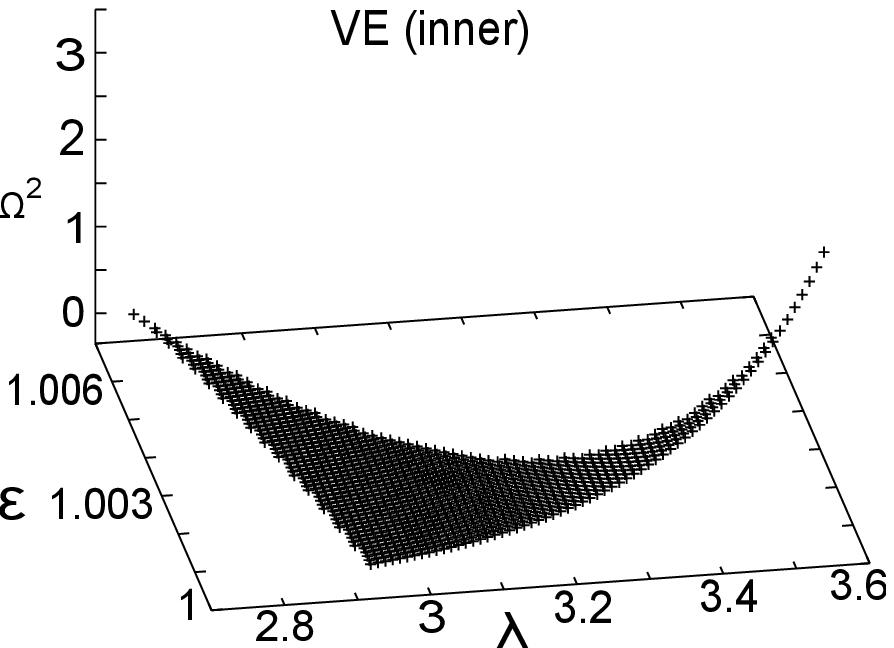} &
\includegraphics[width=0.4\linewidth]{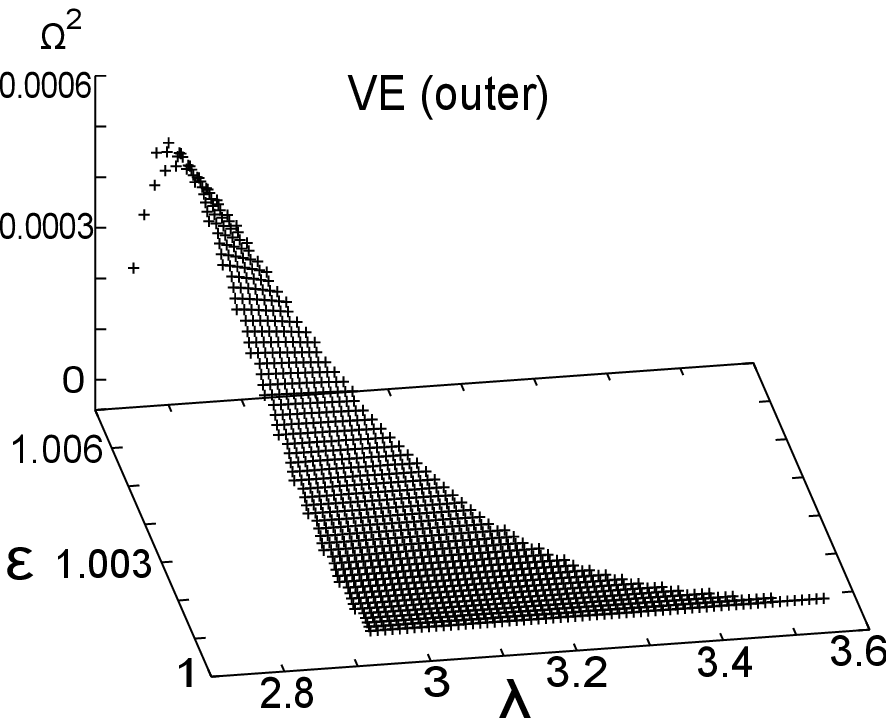} \\
\end{tabular}
\caption{Comparison of $\Omega^2$ vs. $\left[\mathcal{E},\lambda\right]$
of the inner and outer critical points for constant height flow (CH), 
quasi-spherical flow (CF) and flow in vertical hydrostatic equilibrium (VE) 
($\gamma=1.35$, $a=0.1$).}
\label{fig3}
\end{figure}

\begin{figure}[h!]
\centering
\begin{tabular}{cc}
\includegraphics[width=0.4\linewidth]{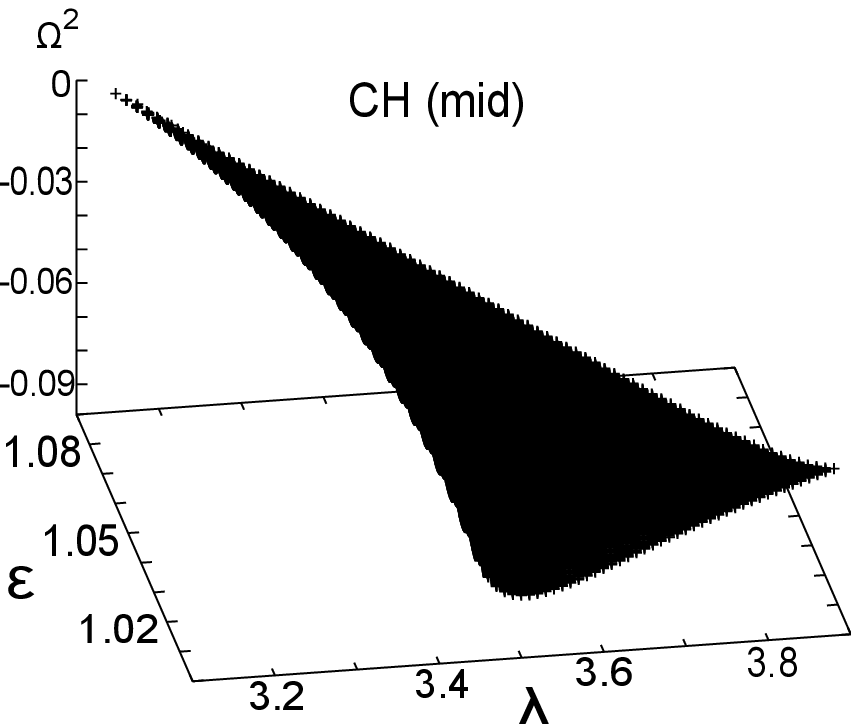} &
\includegraphics[width=0.4\linewidth]{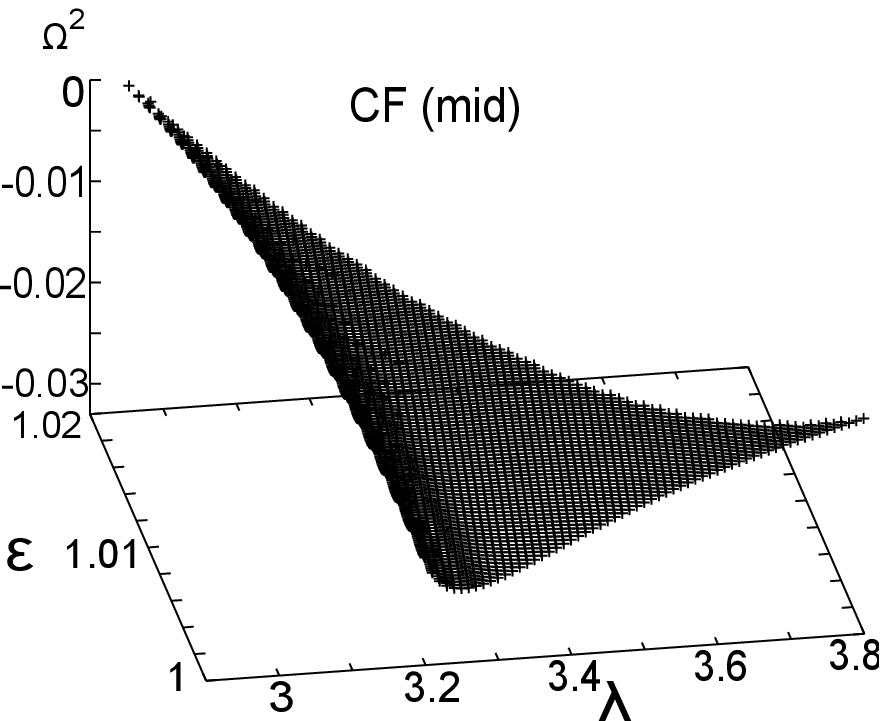} \\
\includegraphics[width=0.4\linewidth]{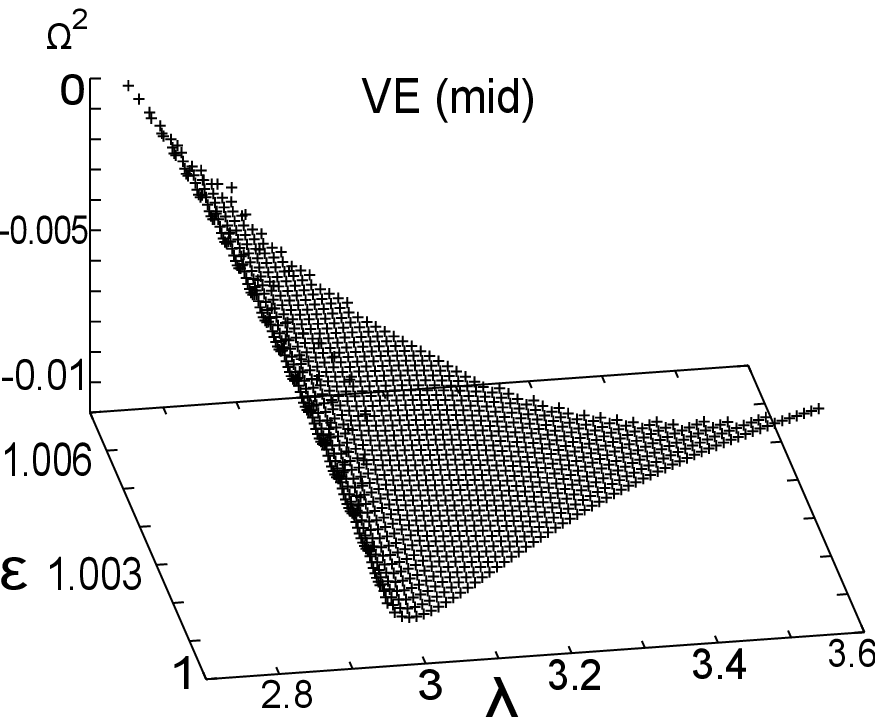} \\
\end{tabular}
\caption{Comparison of $\Omega^2$ vs. $\left[\mathcal{E},\lambda\right]$
of the middle critical point for constant height flow (CH), 
quasi-spherical flow (CF) and flow in vertical hydrostatic equilibrium (VE) 
($\gamma=1.35$, $a=0.1$).}
\label{fig4}
\end{figure}

\begin{figure}
\centering
\includegraphics[scale=0.7]{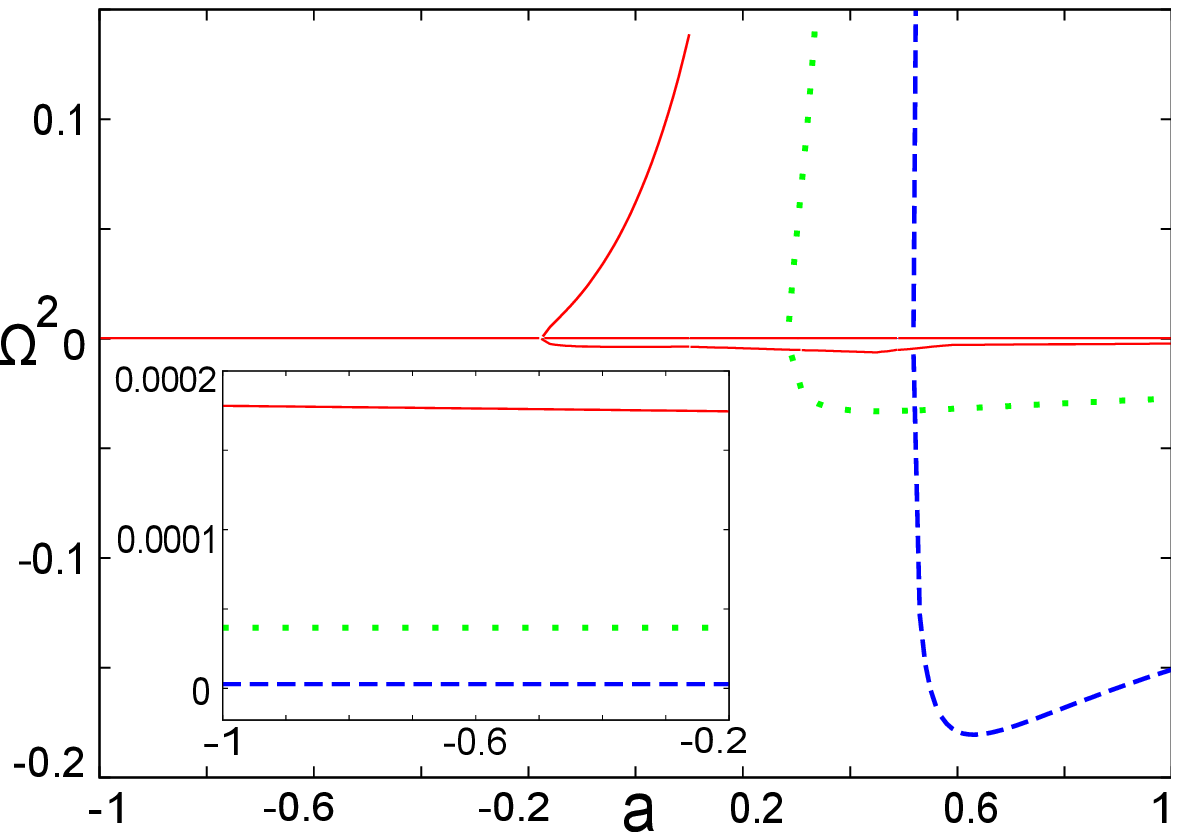}
\caption{Comparison of $\Omega^2$ vs. $a$ for constant height flow (blue dashed lines), 
quasi-spherical flow (green dotted lines) and flow in vertical hydrostatic equilibrium 
(red solid lines)($\gamma=1.35$, $\mathcal{E}=1.003$, $\lambda=3.0$). Inset shows a 
magnified view of the common monotransonic region for the three flow geometries.}
\label{fig5}
\end{figure}

In the previous section, we derived the explicit 
analytical expressions for calculating the numerical 
values of $\Omega^2$ once the locations of the critical 
points are known. We also argued that the solutions 
corresponding to multi-transonic accretion consist of 
three critical points- one of centre type and the 
other two of saddle type. In order to represent a real 
multi-transonic flow, the middle critical point 
is required to be of centre type such that the actual 
physical flow occurs through the inner and outer critical 
points, which are required to be of saddle type in nature. 
In terms of the present analytical 
formalism, $\Omega^2$ corresponding to the inner and outer 
critical points must assume a positive 
numerical value, whereas those corresponding to the middle 
critical points must be negative. 
Figs. \ref{fig3}, \ref{fig4} and \ref{fig5} establish the 
validity of this requirement.

In fig. \ref{fig3}, the variation of $\Omega^2$ for inner 
and outer critical points 
has been depicted over the entire physically accessible 
domain of $\left[\mathcal{E},
\lambda\right]$ for a given value of $\left[\gamma=1.35,a=0.1\right]$. As predicted, 
the numerical values are all positive, indicating a saddle 
nature. A similar observation is made 
in fig. \ref{fig4} where the values of $\Omega^2$ for the 
middle critical points over the 
entire domain of $\left[\mathcal{E},\lambda\right]$ with 
the same values of the other 
flow parameters, are negative, indicating a stable point 
of centre type in nature. 
An immediate comparision can be made between the absolute 
magnitudes of $\Omega^2$ for the inner 
and outer critical points. It is interesting to note that 
$\Omega^2_{inner}>>\Omega^2_{outer}$ indicating 
a correlation between numerical value of the quantity and 
the influence of gravity due to the 
central accretor, not only propagated through the value of 
metric components at the point, but also 
through the dynamical and thermodynamic variables 
pertaining to the flow. However, it is too 
far fetched to comment on any physical realization of the 
quantity at hand as we are dealing with 
a highly nonlinear system with a large number of 
parameters and variables with complicated 
implicit dependence on one another. It is only safe to 
state that the sign of the quantity is 
all that we are interested in at present, to understand 
the nature of the critical points in order to 
visualize the phase space orbits, without delving into 
actual numerics. A comparision of the three 
different flow geometries reveals that $|\Omega^2|_{CH}>|\Omega^2|_{CF}>|\Omega^2|_{VE}$.

Fig. \ref{fig5} provides an elegant pictorial method of 
realizing the nature of 
accretion over the entire range of black hole spin for a 
given value of specific energy and 
specific angular momentum of the flow at a particular 
polytropic index ($\mathcal{E}=1.003,
\lambda=0.3,\gamma=1.35$). The region with a single 
positive value of $\Omega^2$ (shown in 
inset) represents a saddle type critical point indicating at monotransonic flow for all three 
geometric configurations. The single positive value is 
then observed to split into 
one negative value and two positive values indicating the 
formation of one centre type middle 
critial point and two saddle type critical points. One of 
the two saddle points with its numerical 
value comparable with that of the single saddle type point 
in the monotransonic region represents 
the outer critical point, while the other with a higher 
value represents the inner critical point 
which is closer to the event horizon. It appears as if a 
saddle-centre pair is generated from the 
initial saddle at a particular value of spin and as one 
moves towards higher values of black hole 
spin, the new saddle, i.e. the inner critical point moving 
closer and closer to the horizon begins 
assuming higher values of $\Omega^2$ until it crosses the 
horizon and ultimately disappears from the 
physically accessible regime. And finally, one is left 
with the centre type middle critical point through 
which no physical flow can occur, and the previous saddle 
type outer critical point through which 
accretion continues as a purely monotransonic flow. The 
same universal trend can be observed in all 
three disc structures although splitting occurs at 
different values of $a$ and the relative 
magnitudes of $\Omega^2$ are distinct for each flow 
geometry. It is to be noted that for the 
same energy and angular momentum of the accreting fluid, 
flow in the hydrostatic equilibrium along the 
vertical direction allows for multitransonic solutions at 
the lowest values of black hole spin (even 
for counter-rotating black holes in the given case). It 
may also be observed that since the values 
of $\Omega^2$ represent critical points of a system, the 
splitting actually 
corresponds to a {\it{super-critical pitchfork 
bifurcation}} in the theory of dynamical systems 
where a stable critical point bifurcates into two stable 
critical points (inner and outer in this 
case through which actual flow occurs) and an unstable 
critical point (middle centre type point 
through which physical flow in not allowed).

\section{Integral flow solutions with shock for polytropic accretion}
In the previous section, one finds that it is possible to 
understand the nature of the critical points through 
some local stability analysis, i.e., the methodology is 
applicable in the close neighbourhood of the critical 
points. The global nature of the flow topology, however, 
is possible to know only through the stationary 
integral solutions of the corresponding flow equations. 
Such integral solutions are obtained through numerical 
techniques. For a particular set of values of 
$\left[\mathcal{E},\lambda,\gamma,a\right]$, one 
calculates the 
location of critical point(s). The values of 
$\left[u,c_s,\frac{du}{dr},\frac{dc_s}{dr}\right]$ on such 
critical points are then computed. Starting from the 
critical point, the expressions corresponding to 
$\frac{du}{dr}$ 
and $\frac{dc_s}{dr}$ are then numerically solved to 
obtain the radial Mach number vs. radial distance profile. 
For transonic flow with multiple critical points, a 
stationary shock may form. For such flow, integral 
stationary subsonic solutions pass through the outer sonic 
point (associated with the saddle type outer critical 
point) and becomes supersonic. The supersonic flow then 
encounters a discontinuous transition through shock and 
becomes subsonic once again. 
The location of the shock has to be determined by solving 
the corresponding shock conditions. The post-shock 
subsonic flow then passes through the inner sonic point 
(corresponding to the saddle type inner critical point) 
to become supersonic again and ultimately plunge into the 
event horizon.

\begin{figure}[h!]
\centering
\includegraphics[scale=0.7]{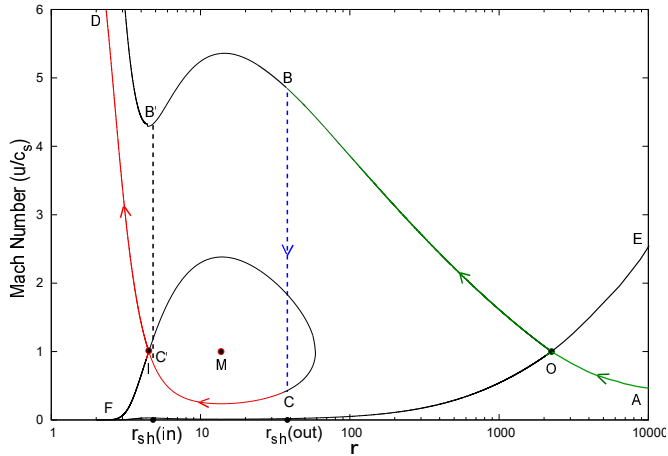}
\caption{Phase space portrait (Mach number vs. $r$ plot) for quasi-spherical disc 
($\mathcal{E}=1.0003$, $\lambda=3.5$, $\gamma=1.35$, $a=0.1$). $r_{\rm sh}^{\rm in}=4.805$, $r_{\rm sh}^{\rm out}=38.15$, 
$r^{\rm in}=4.5$ (inner sonic point $I$), $r^{\rm mid}=13.712$ (mid sonic point $M$), 
$r^{\rm out}=2244.313$ (outer sonic point $O$).}
\label{fig6}
\end{figure}

Figure \ref{fig6} shows the Mach number vs radial distance 
phase portrait of a shocked multi-transonic flow for 
accretion in quasi-spherical geometry. Branch $AOB$ 
(green curve) represents accretion through the outer 
sonic point $O$. The flow encounters a stable, standing, 
energy preserving shock at 
$r=r_{\rm sh}^{\rm out}$ whose location is obtained by 
using the numerical scheme of equating shock-invariant 
quantities (elaborated in the next subsection). It then 
jumps along the line of discontinuity $BC$ (blue dashed line). 
Thus being transformed into a subsonic, compressed and 
hotter flow, it then approaches the event horizon 
moving along the line $CC'ID$ (red curve) and becoming supersonic once 
again while passing through the inner sonic 
point $I$. $B'C'$ shows an unstable line of discontinuity 
which is inaccessible to physical flow. $FOE$ 
represents the corresponding wind solution, while $DIC'CF$ 
is a homoclinic orbit encompassing the middle 
critical point $M$. 

\subsection{Shock-invariant quantities ($S_{\rm h}$)}
The {\it{shock-invariant quantity}} ($S_{\rm h}$) is 
defined as a quantity whose numerical value remains the 
same on the integral solution branch passing through the 
outer sonic point as well as the branch passing through 
the inner sonic point, exclusively at the location(s) of 
physically allowed discontinuities obeying the general 
relativistic Rankine Hugoniot conditions. Thus, once 
expression for the shock-invariant quantities are 
obtained, the corresponding shock locations can be 
evaluated by numerically checking for the condition 
$S_{\rm h}^{\rm out}=S_{\rm h}^{\rm in}$, where 
$S_{\rm h}^{\rm out}$ and $S_{\rm h}^{\rm in}$ are the 
shock-invariant quantities defined on 
the integral flow solutions passing through the outer and 
the inner sonic points respectively.
The Rankine Hugoniot conditions applied to a fully general relativistic background flow are given by, \\
\begin{equation}
\left[\left[\rho u^\mu\right]\right]=0\text{, and }\left[\left[T^{\mu\nu}\right]\right]=0
\label{eqn54}
\end{equation}
where $\left[\left[V\right]\right]=V_--V_+$, $V_+$ and $V_-$ symbolically denote the values of some flow variable $V$ 
before and after the shock respectively. \\
Eqn.(\ref{eqn54}) can further be decomposed into the 
following three conditions, \\
\begin{equation}
\left[\left[\rho u^r\right]\right]=0,
\label{eqn55}
\end{equation}
\begin{equation}
\left[\left[(p+\epsilon)u_tu^r\right]\right]=0,
\label{eqn56}
\end{equation}
\begin{equation}
\left[\left[(p+\epsilon)u^ru^r+p\right]\right]=0,
\label{eqn57}
\end{equation}
where $u^r=\frac{u\Delta^\frac{1}{2}}{r\sqrt{1-u^2}}$. \\
Using the definition for specific enthalpy ($h$) of the fluid given by, \\
\begin{equation}
h=\frac{p+\epsilon}{\rho}
\label{eqn58}
\end{equation}
and using eqn.(\ref{eqn1a}) and eqn.(\ref{eqn3}) together 
with the polytropic equation of state $p=K\rho^\gamma$, 
one can express $\rho$, $p$ and $\epsilon$ in terms of the 
adiabatic sound speed $c_s^2$ as, \\
\begin{eqnarray}
\rho=\left[\frac{c_s^2(\gamma-1)}{K\gamma(\gamma-1-c_s^2)}\right]^\frac{1}{\gamma-1} \nonumber \\
p=K^\frac{-1}{\gamma-1}\left[\frac{c_s^2(\gamma-1)}{\gamma(\gamma-1-c_s^2)}\right]^\frac{\gamma}{\gamma-1} \nonumber \\
\epsilon=\left(\frac{c_s^2(\gamma-1)}{K\gamma(\gamma-1-c_s^2)}\right)^\frac{1}{\gamma-1}
\left[1+\frac{1}{\gamma}\left(\frac{c_s^2}{\gamma-1-c_s^2}\right)\right]
\label{eqn59}
\end{eqnarray}
Now, considering geometry of the flow eq.(\ref{eqn55}) can 
be re-written as \\
\begin{equation}
\left[\left[\rho u^r \mathcal{H}(r)\right]\right]=0
\label{eqn60}
\end{equation}
where, the accretion geometry dependent terms $\mathcal{H}(r)$ for three different flow structures are given by, \\
\begin{eqnarray}
\mathcal{H}_{CH}(r)=2\pi r H \nonumber \\ 
\mathcal{H}_{CF}(r)=\Theta r^2 \nonumber \\
\mathcal{H}_{VE}(r)=4\pi r H(r)
\label{eqn61}
\end{eqnarray}
$H$ being the thickness of the constant height disc, 
$\Theta$ being the solid angle subtended by 
the quasi-spherical disc at the horizon and $H(r)$ being 
the radius dependent thickness for flow in hydrostatic 
equilibrium along the vertical direction given by eq.(\ref{eqn22a}). \\
Substituting eqs.(\ref{eqn61}) and (\ref{eqn59}) in eqs.(\ref{eqn60}) and (\ref{eqn57}), and then solving 
simultaneously, we derive the shock-invariant quantitites 
($S_h$) for all three flow geometries as, \\
\begin{equation}
S_h\bigg\vert_{CH}=\frac{u^2(\gamma\frac{\Delta}{r^2}-c_s^2)+c_s^2}{u\sqrt{1-u^2}(\gamma-1-c_s^2)}
\label{eqn62}
\end{equation}
\begin{equation}
S_h\bigg\vert_{CF}=\frac{u^2(\gamma\frac{\Delta}{r^2}-c_s^2)+c_s^2}{u\sqrt{1-u^2}(\gamma-1-c_s^2)}
\label{eqn63}
\end{equation}
\begin{equation}
S_h\bigg\vert_{VE}=\frac{\sqrt{F}\lbrace u^2(\gamma\frac{\Delta}{r^2}-c_s^2)+c_s^2 \rbrace}{uc_s\sqrt{(1-u^2)(\gamma-1-c_s^2)}}
\label{eqn64}
\end{equation}

\section{Shock parameter space for polytropic accretion}
We now intend to see which region of the 
$\left[\mathcal{E}-\lambda\right]$ parameter space allows 
shock formation. For a 
fixed set of $\left[\gamma=1.35,a=0.57\right]$, we check 
the validity of the Rankine Hugoniot condition 
corresponding to every value of 
$\left[\mathcal{E},\lambda\right]$ for which the accretion 
flow possesses three critical points. This means the 
shock-invariant quantity is calculated for every 
$\left[\mathcal{E},\lambda\right]$ for which the 
multitransonic accretion is possible, and it is observed 
that the quantities calculated along the solution passing 
through the outer and the inner sonic points become equal 
at a particular radial distance, i.e. at the shock 
location, only for a subset of such 
$\left[\mathcal{E},\lambda\right]$. 
We then plot the corresponding 
$\left[\mathcal{E},\lambda\right]_{\rm{shock}}$ for various geometric configurations of matter. 

\begin{figure} [h!]
\centering
\includegraphics[scale=0.7]{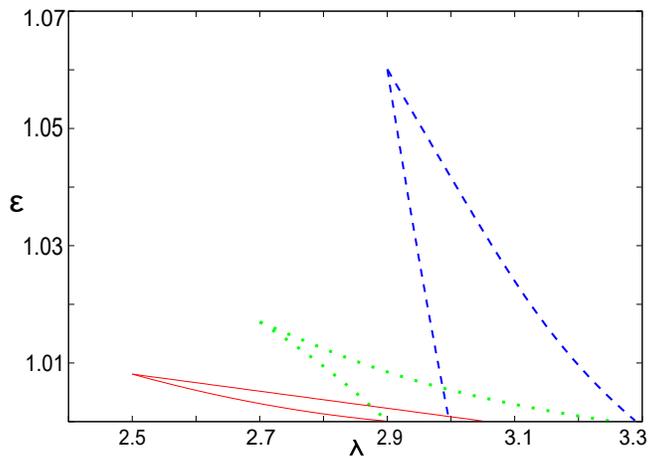}
\caption{Comparision of $\mathcal{E}$-$\lambda$ plot of allowed shocked multitransonic accretion
solutions for three different flow geometries ($\gamma=1.35$ and $a=0.57$). 
Constant height disc, quasi-spherical flow and flow in vertical hydrostatic equilibrium represented by blue dashed lines, 
green dotted lines and red solid lines respectively.}
\label{fig7}
\end{figure}

\begin{figure}[h!]
\centering
\includegraphics[scale=0.7]{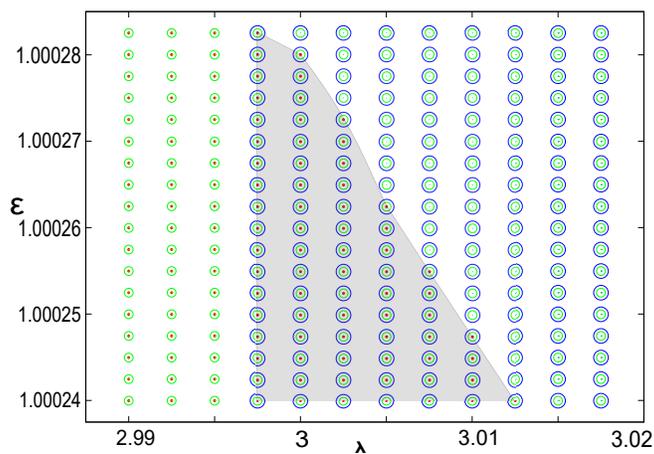}
\caption{Shaded region depicts the common domain of $\left[\mathcal{E},\lambda\right]$ 
($\gamma=1.35,a=0.57$) which allows shock formation in constant height disc (blue circles), 
quasi-spherical disc (green circles) and flow in hydrostatic equilibrium (red dots).}
\label{fig8}
\end{figure}

\begin{figure}[h!]
\centering
\includegraphics[scale=0.7]{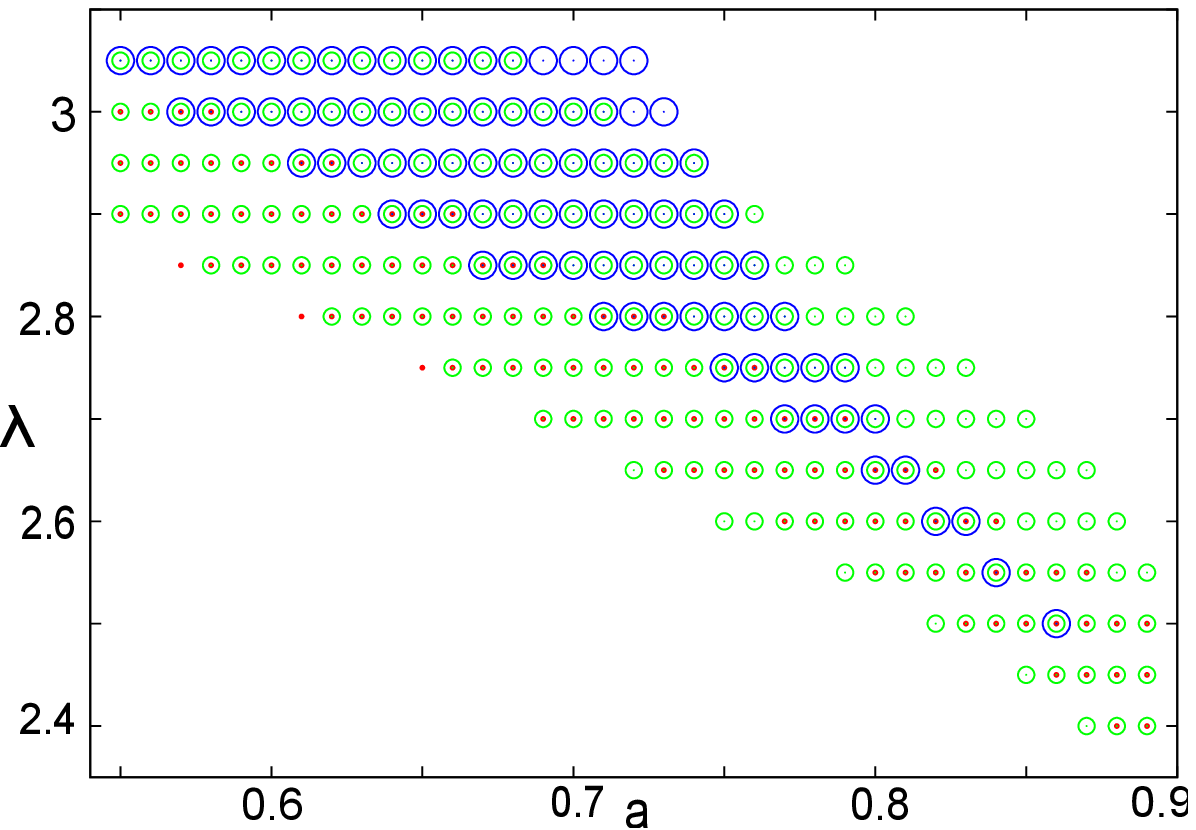}
\caption{Overlap region of $\left[a,\lambda\right]$ 
($\gamma=1.35,\mathcal{E}=1.00024$) which allows shock formation in 
constant height disc (blue circles), quasi-spherical disc (green circles) 
and flow in hydrostatic equilibrium (red dots).}
\label{fig9}
\end{figure}

In fig. \ref{fig7} we plot such shock forming parameter space for three different flow geometries. The 
shock forming region of $\left[\mathcal{E},\lambda\right]$ for a relevant combination of $a$ and $\gamma$, 
which is common to all three geometries is shown in fig. \ref{fig8}. 
Similarly, fig. \ref{fig9} shows the domain of $\left[a,\lambda\right]$ for a given value of 
$\mathcal{E}$ and $\gamma$ where shock forming regions of the three flow models overlap. 
This particular plot indicates at the requirement of an anti-correlation between the angular momentum 
of flow and spin of the central gravitating source for multitransonic accretion to occur. Moreover, it may 
be observed that a higher difference between these two values allows for a greater multitransonic shock 
forming region. The only probable reason behind this typical observation seems to be an 
increase in the effective centrifugal barrier experienced by the flow.
These overlapping parameter space domains are of extreme importance for our purpose. All the shock 
related flow properties for which the flow behaviour is to be compared for three different geometries, are 
to be characterized by $\left[\mathcal{E},\lambda,\gamma,a\right]$ 
corresponding to these common regions only. We will show this in greater detail in subsequent sections.

In what follows, we will study the dependence of the shock location ($r_{sh}$), shock strength (the ratio 
of pre to post shock values of the Mach number, $M_+/M_-$), shock compression ratio (the ratio of 
the post to pre shock matter density, $\rho_-/\rho_+$), the ratio of post to pre shock temperature 
($T_-/T_+$) and pressure ($P_-/P_+$), on the black hole spin parameter $a$. Subscripts `+' and `-' 
represent pre and post shock quantities respectively. One can study 
the dependence of such quantities on other accretion parameters, i.e. $\left[\mathcal{E},\lambda,\gamma
\right]$ as well. Such dependence, however, are not much relevant for our study in this work, since for a 
fixed value of the Kerr parameter, nature of such dependence should actually be equivalent to the 
corresponding nature of the dependence of $\left[r_{sh},M_+/M_-,\rho_-/\rho_+,P_-/P_+\right]$ on $
\left[\mathcal{E},\lambda,\gamma\right]$ as observed in the Schwarzschild metric which has already been 
investigated in P1. Hereafter, throughout this work, we will study the dependence of 
every physical quantity on the Kerr parameter only, for a fixed set of $\left[\mathcal{E},\lambda,\gamma
\right]$.

\begin{figure}[h!]
\centering
\includegraphics[scale=0.7]{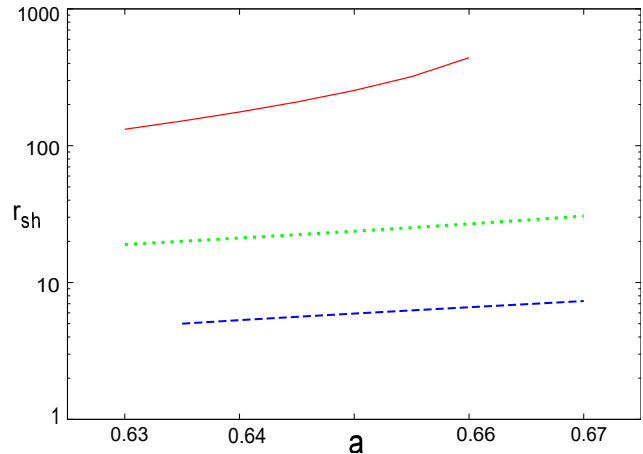}
\caption{Shock location ($r_{sh}$) vs. $a$ plot 
($\gamma=1.35,\mathcal{E}=1.00024,\lambda=2.9$) for 
constant height disc (dashed blue line), quasi-spherical disc (dotted green line) 
and flow in hydrostatic equilibrium (solid red line).}
\label{fig10}
\end{figure}

Fig. \ref{fig10} depicts variation of the shock location ($r_{sh}$) with spin parameter $a$. The value 
of $\lambda$ in this figure and all subsequent figures illustrating other shock related quantities 
has been chosen from the common region in fig. \ref{fig9} so as to ensure the maximum 
possible overlapping range of $a$ permissible for shocked accretion at the given value of $\mathcal{E}$ 
and $\gamma$ for all three flow models. The shock location is observed to shift further from the horizon 
as the black hole spin increases. This is what we may expect as increasing Kerr parameter for a fixed 
angular momentum of the flow implies growth in the difference of the two parameters, thus strengthening 
the effective centrifugal barrier. Thus transonicity and shock formation are speculated to occur in 
earlier phases of the flow at greater distances from the massive central source. A comparision of the 
models reveals the following trend at a given value of $a$, $r_{sh}(VE)>r_{sh}(CF)>r_{sh}(CH)$. This 
indicates at the fact that flow in hydrostatic equilibrium has to face much more opposition than the 
other two disc geometries for the same amount of impediment posed by rotation of the flow and that of 
the black hole. 
 
\begin{figure}[h!]
\centering
\begin{tabular}{cc}
\includegraphics[width=0.4\linewidth]{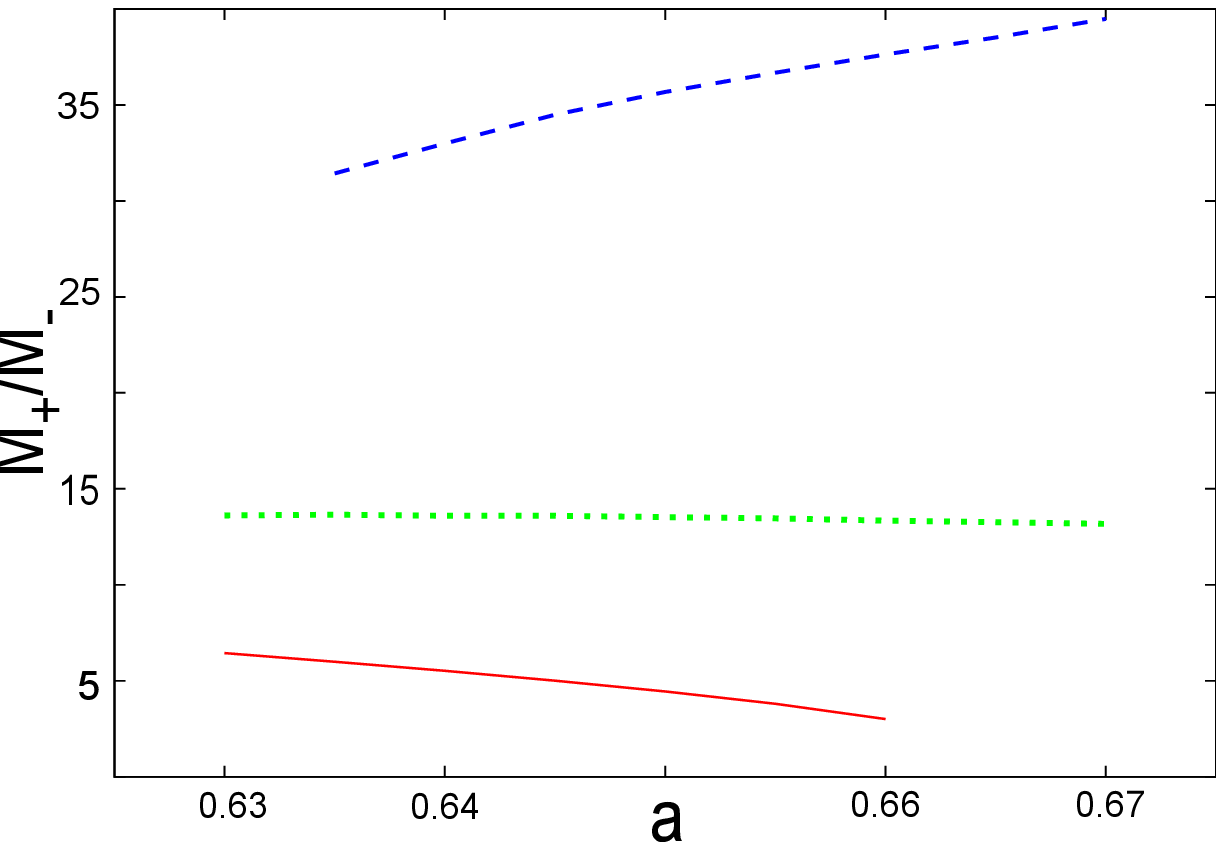} &
\includegraphics[width=0.4\linewidth]{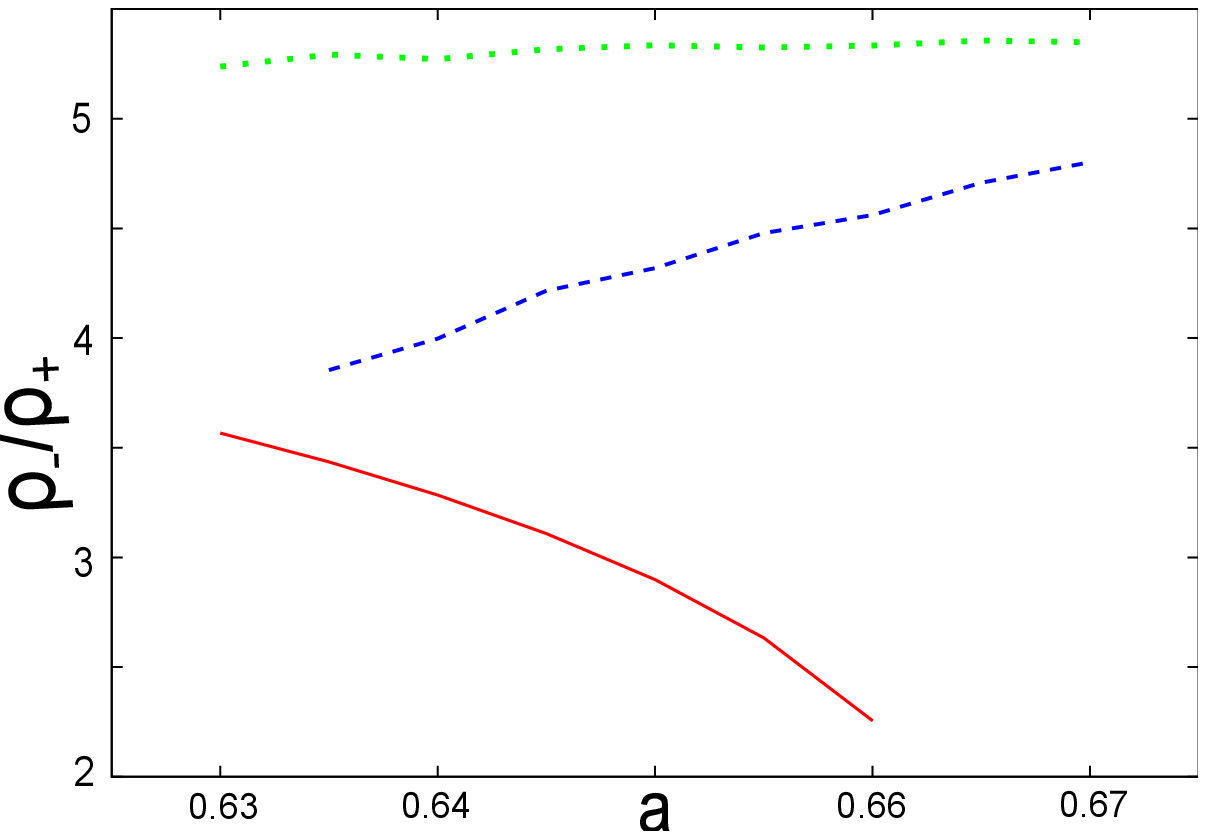} \\
\includegraphics[width=0.4\linewidth]{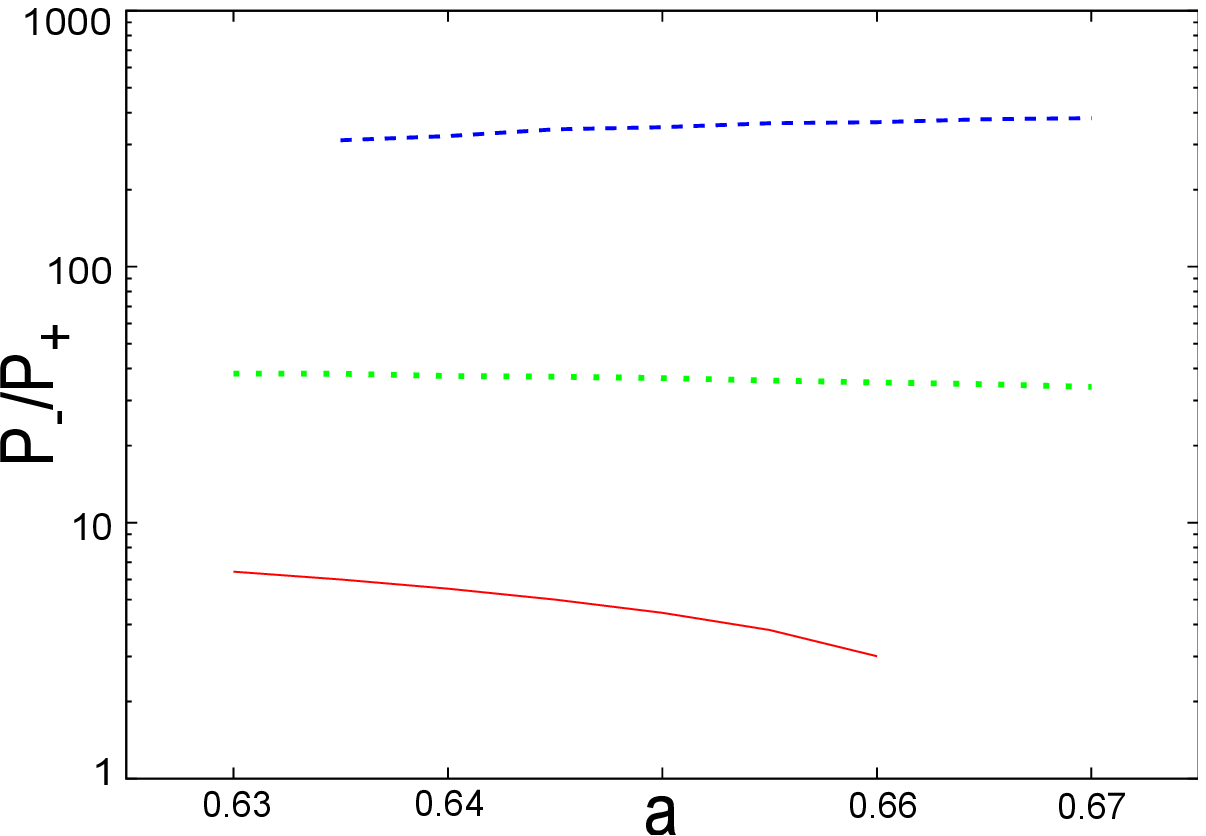} &
\includegraphics[width=0.4\linewidth]{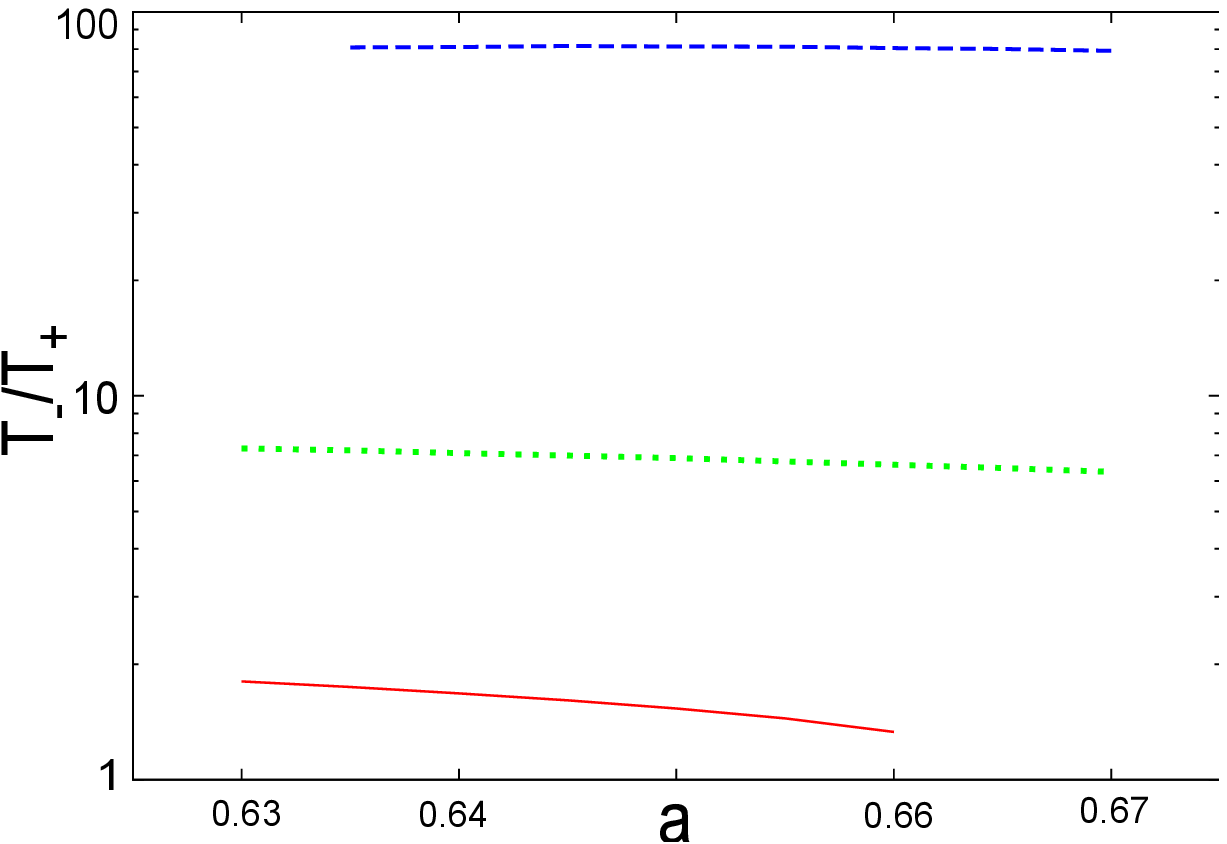} \\
\end{tabular}
\caption{Variation of shock strength ($M_+/M_-$), compression ratio ($\rho_-/\rho_+$), 
pressure ratio ($P_-/P_+$) and temperature ratio ($T_-/T_+$) with black hole 
spin parameter $a$ ($\gamma=1.35,\mathcal{E}=1.00024,\lambda=2.9$) for 
constant height flow (dashed blue lines), quasi-spherical flow (dotted green lines) 
and flow in hydrostatic equilibrium (solid red lines). Subscripts `+' and `-' 
represent pre and post shock quantities respectively.}
\label{fig11}
\end{figure}

It is also interesting to note from fig. \ref{fig11} that not only does the vertical equilibrium model 
experience maximum hindrance due to rotation, but it also exhibits the formation of shocks with the 
weakest strength, i.e. pre-shock to post-shock ratio of the Mach number ($M_+/M_-$), when compared with 
the other two models. The shocks are strongest in case of discs with a constant height and intermediate 
in the case of quasi-spherical flows. The strengths are observed to decrease with $a$. This might be 
explained by the dependence of shock location on the spin parameter. Greater values of $r_{sh}$ point 
at decreasing curvature of physical space-time leading to diminishing influence of gravity. Thus dropping 
of shock strength with increasing $a$, or in other words, higher values of $r_{sh}$, establishes that 
weaker gravity amounts to the formation of weaker discontinuities in the flow and vice versa. Naturally, 
waning shock strengths in turn lead to lower post to pre-shock compression ($\rho_-/\rho_+$), pressure 
($P_-/P_+$) and temperature ($T_-/T_+$) ratios, as observed in the figure. A seemingly 
anomalous behaviour is observed in this context for the constant height flow geometry, in which case, in 
spite of an outward shifting of the shock location, shock strength is seen to increase, although behaviour 
of the other related ratios fall in line with our previous arguments. We shall try to discuss the reason 
behind such an anomaly, in the next section.

\section{Quasi-terminal values}
Accreting matter manifests extreme behaviour before plunging through the event horizon because it experiences the 
strong curvature of space-time close to the black hole. The spectral signature of such matter corresponding to 
that length scale helps to understand the key features of the strong gravity space-time to the close 
proximity of the horizon. It may also help to study the spectral signature of black hole spin. The 
corresponding spectral profiles and the light curves may be used for constructing the relevant black hole 
shadow images (\cite{fma00apj},\cite{takahashi04iau},\cite{hcsy07mnras},\cite{hm09prd},\cite{zpin12nar},\cite{svagp12aa}). \\
For a very small positive value of $\delta$ ($\sim 0.0001$), any accretion variable $V_\delta$ measured at 
a radial distance $r_\delta=r_++\delta$ will be termed as 
`quasi-terminal value' of that corresponding accretion variable. In \cite{dnhbmcbwkn15na}, dependence of 
$V_\delta$ on the Kerr parameter was studied for polytropic accretion flow in hydrostatic equilibrium 
along the vertical direction. In the present work, we intend to generalize such work by computing the $V_
\delta$ for all three different matter geometries. This generalization will be of paramount 
importance in understanding the geometric configuration of matter flow close to the horizon through the 
imaging of the shadow. \\
In what follows, we will study the dependence of $\left[M,\rho,T,P\right]_{r_\delta}$ on the Kerr 
parameter for shocked multi-transonic accretion in three different flow geometries to understand how the 
nature of such dependence gets influenced by the flow structure. We will also study such dependence for 
monotransonic flows for the entire range of Kerr parameters, starting from $-1$ to $+1$, to study whether 
any general asymmetry exists between co-rotating and counter-rotating accretion in connection to 
values of the corresponding $V_\delta$.

\subsection{Dependence of $\left[M,T,\rho,P\right]_{r_\delta}$ on $a$ for shocked polytropic accretion}

\begin{figure}[h!]
\centering
\begin{tabular}{cc}
\includegraphics[width=0.4\linewidth]{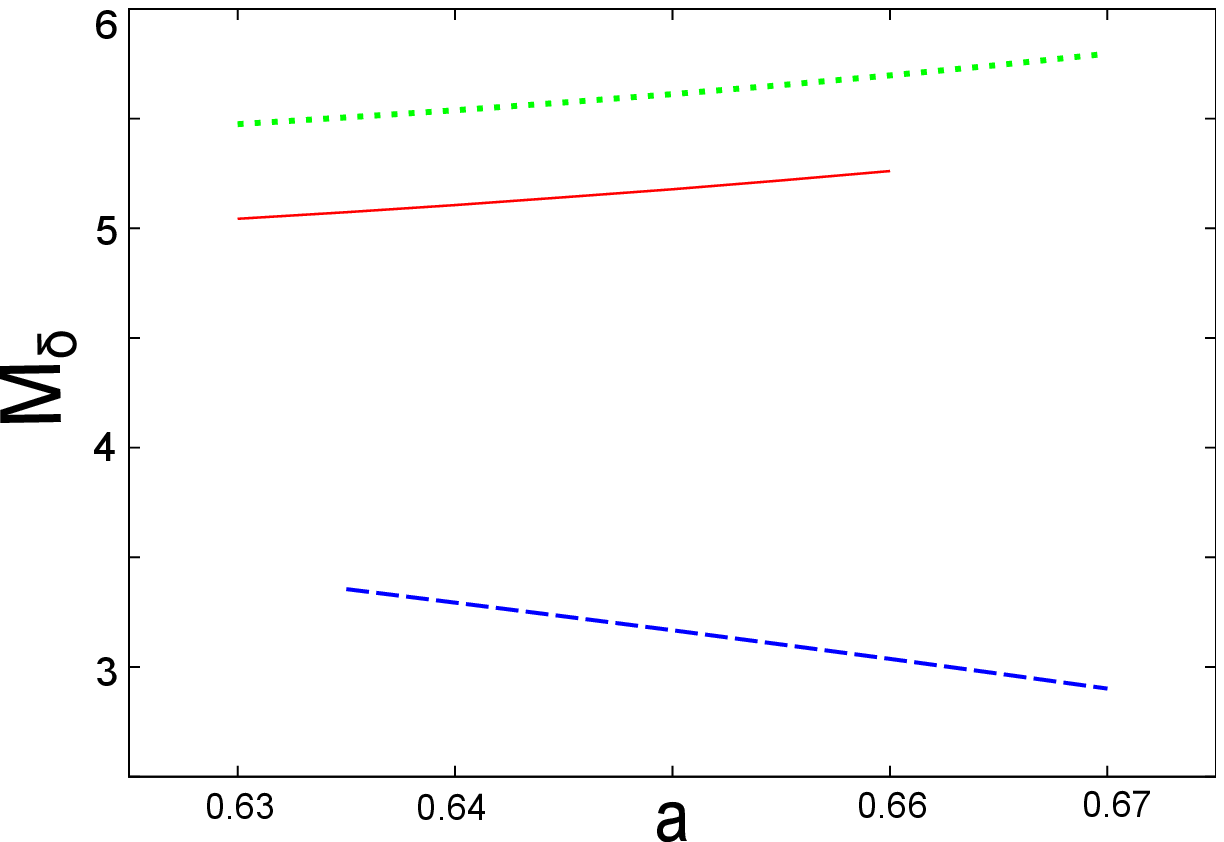} &
\includegraphics[width=0.4\linewidth]{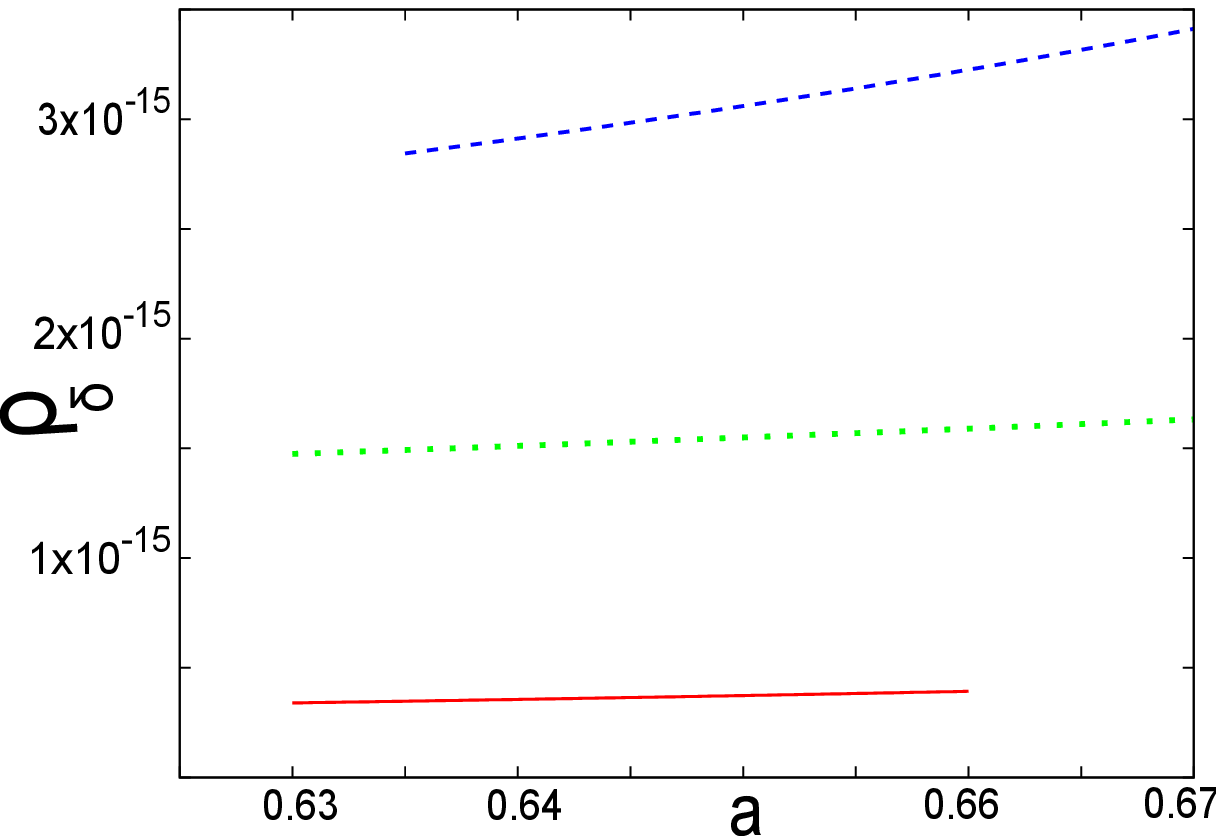} \\
\includegraphics[width=0.4\linewidth]{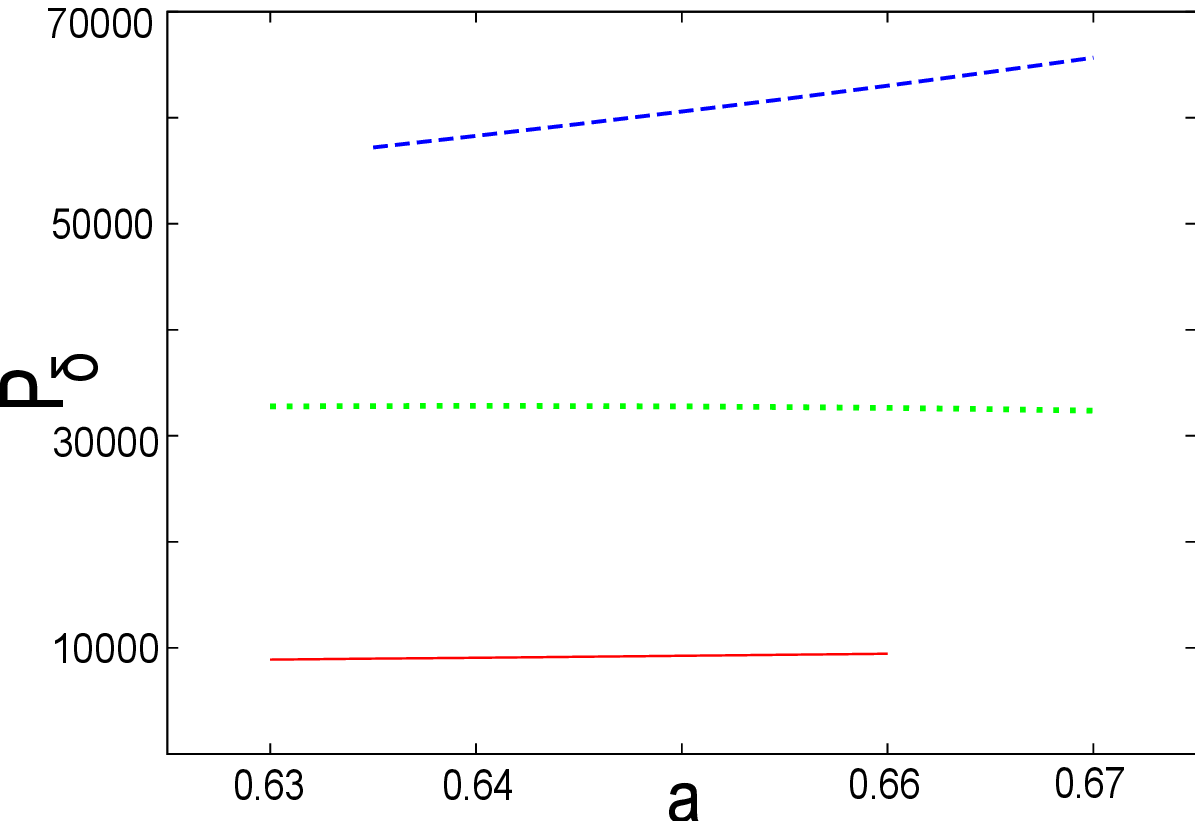} &
\includegraphics[width=0.4\linewidth]{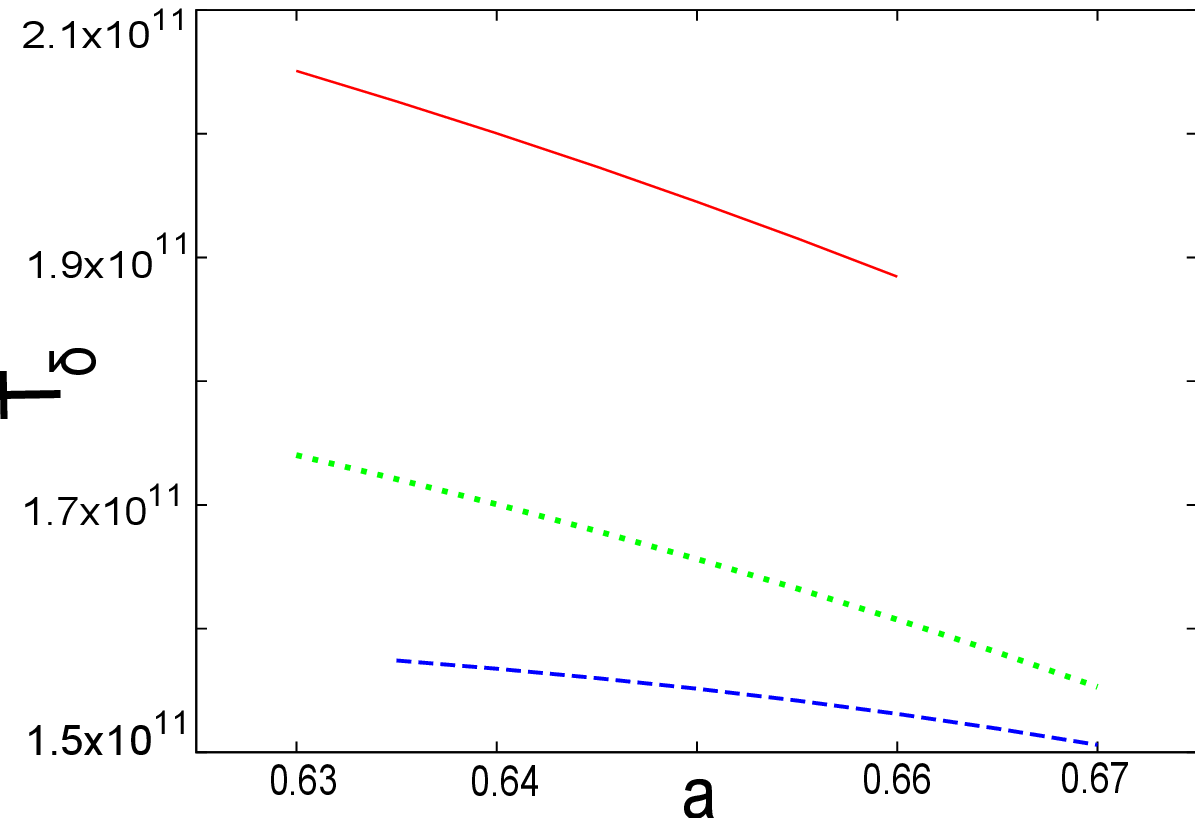} \\
\end{tabular}
\caption{Variation of quasi-terminal values of Mach number ($M_\delta$),
density ($\rho_\delta$), pressure ($P_\delta$) and temperature ($T_\delta$) with $a$ 
($\gamma=1.35,\mathcal{E}=1.00024,\lambda=2.9$) for 
constant height flow (dashed blue lines), quasi-spherical flow (dotted green lines) 
and flow in hydrostatic equilibrium (solid red lines). Density and pressure are in 
CGS units of $g$ $cm^{-3}$ and $dyne$ $cm^{-2}$ respectively and temperature is in 
absolute units of Kelvin.}
\label{fig12}
\end{figure}

Fig. \ref{fig12} demonstrates how the quasi-terminal values pertaining to Mach number ($M_\delta$), 
density ($\rho_\delta$), pressure ($P_\delta$) and the bulk ion temperature ($T_\delta$) vary with 
black hole spin $a$, at a given set of $\left[\mathcal{E},\lambda,\gamma\right]$ chosen such that 
a substantial range of $a$ is available for studying any observable trend of variation in the common 
shock regime for all three matter configurations. It might be noted that although a general course of 
dependence of the values may be observed within a local set of flow parameters for each geometry 
separately, however it is impossible to conclude upon any global trends of such sort. This is primarily 
due to the reason that each permissible set of $\left[\mathcal{E},\lambda,\gamma\right]$ offers an 
exclusively different domain of black hole spin for multitransonic accretion to occur and an even 
narrower common window for the viability of general relativistic Rankine Hugoniot type shocks in different 
geometric configurations of the flow. Hence, in spite of the fact that physical arguments may be able to 
specifically establish the observed results in certain cases, as it could be done with results obtained in 
the previous sections, however similar specific attempts made in all cases globally may not only turn out 
to be futile, but also dangerously misleading. The anomaly which was pointed out in the preceding section, 
is a stark example of such an instance. However, there is absolutely no reason for disbelief in the 
universality or the validity of previous physical arguments. It is only that nature offers a few 
selected cases to provide us the opportunity of peeking into its global behaviour. We present exactly 
such a case in the following subsection.

\subsection{Dependence of $\left[M,T,\rho,P\right]_{r_\delta}$ on $a$ for monotransonic accretion}

\begin{figure}[h!]
\centering
\begin{tabular}{cc}
\includegraphics[width=0.4\linewidth]{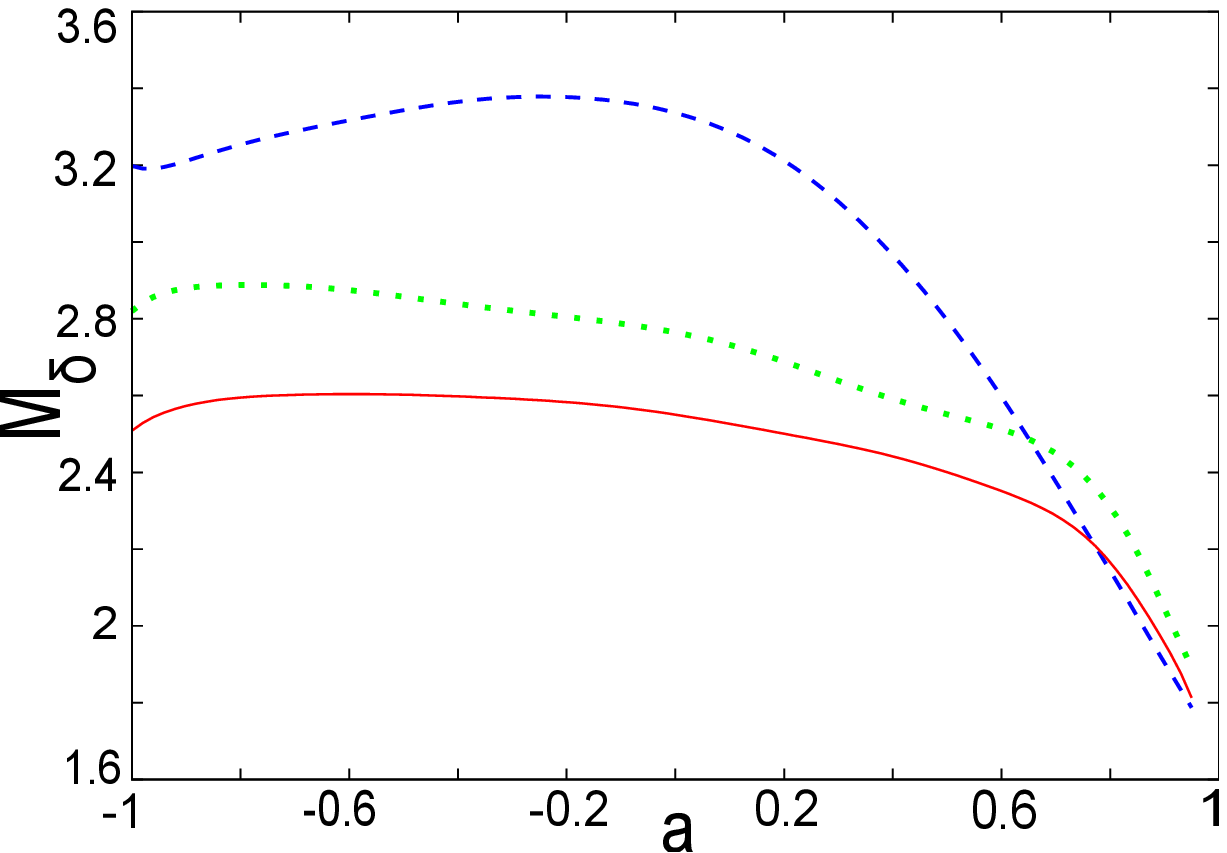} &
\includegraphics[width=0.4\linewidth]{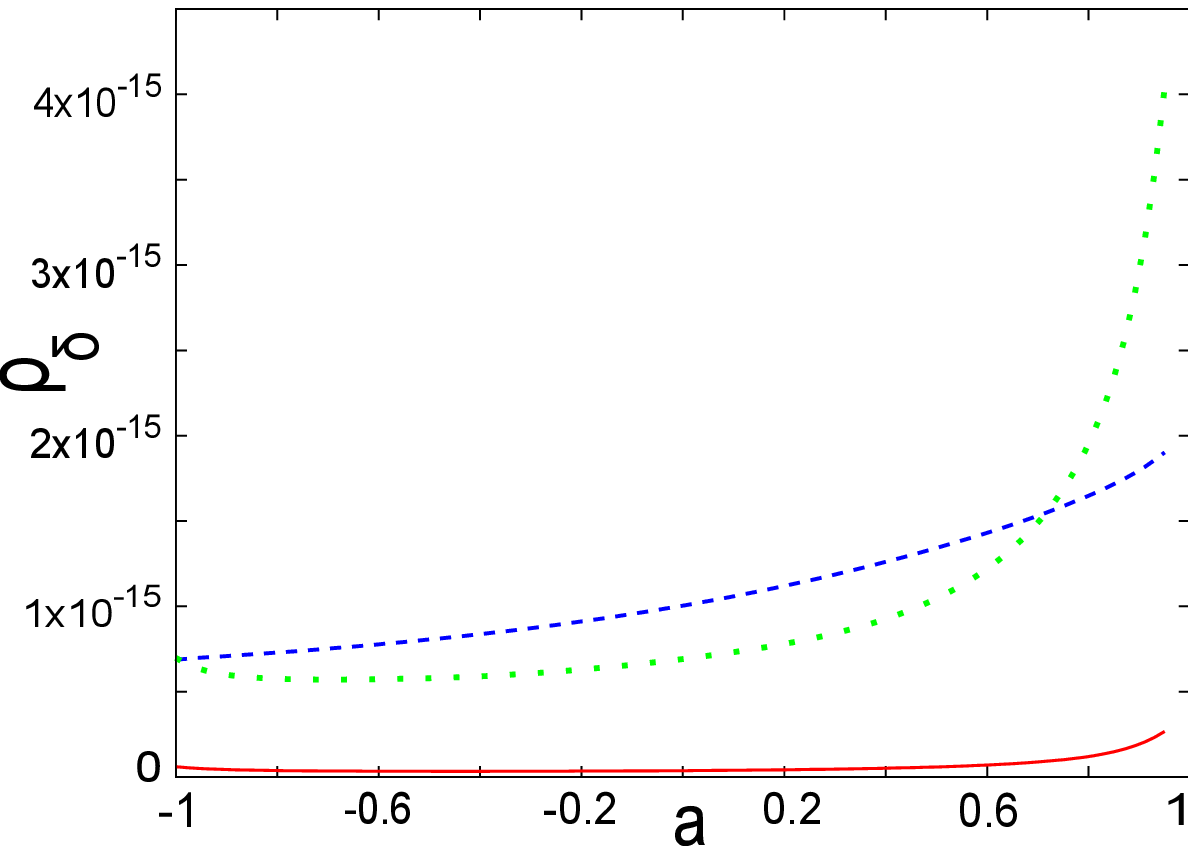} \\
\includegraphics[width=0.4\linewidth]{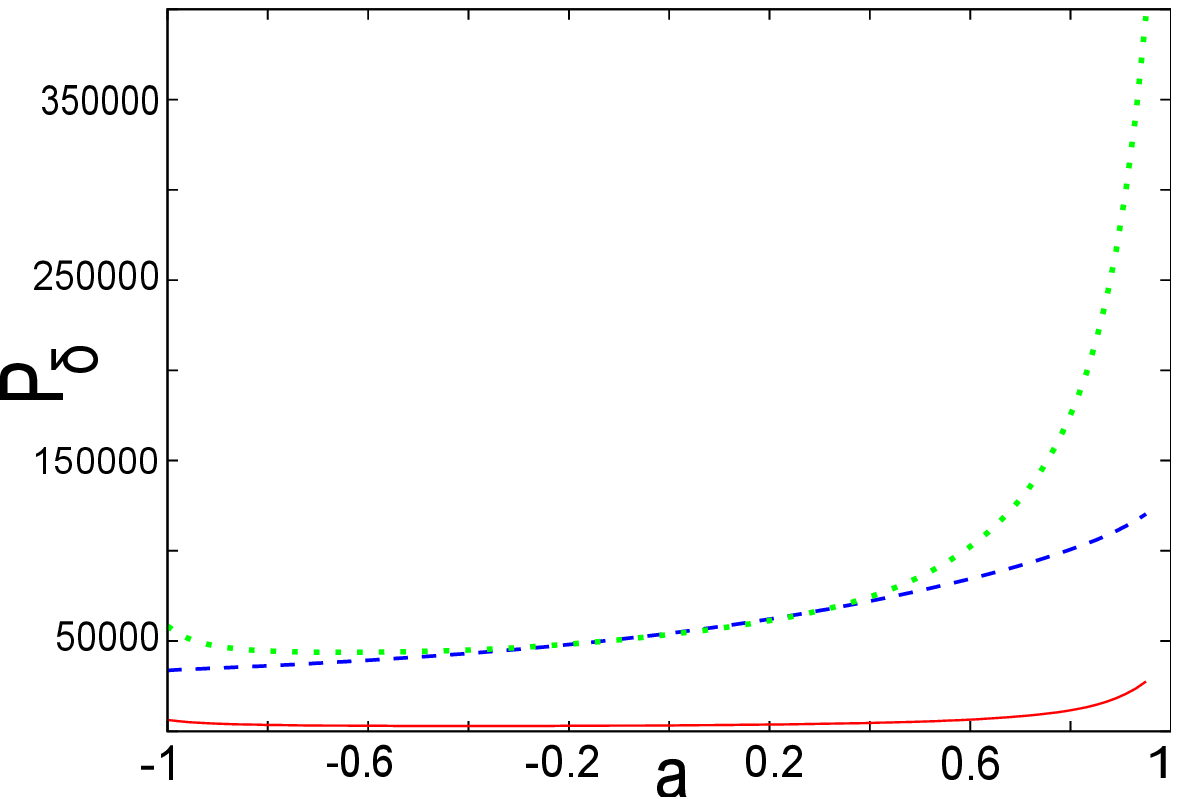} &
\includegraphics[width=0.4\linewidth]{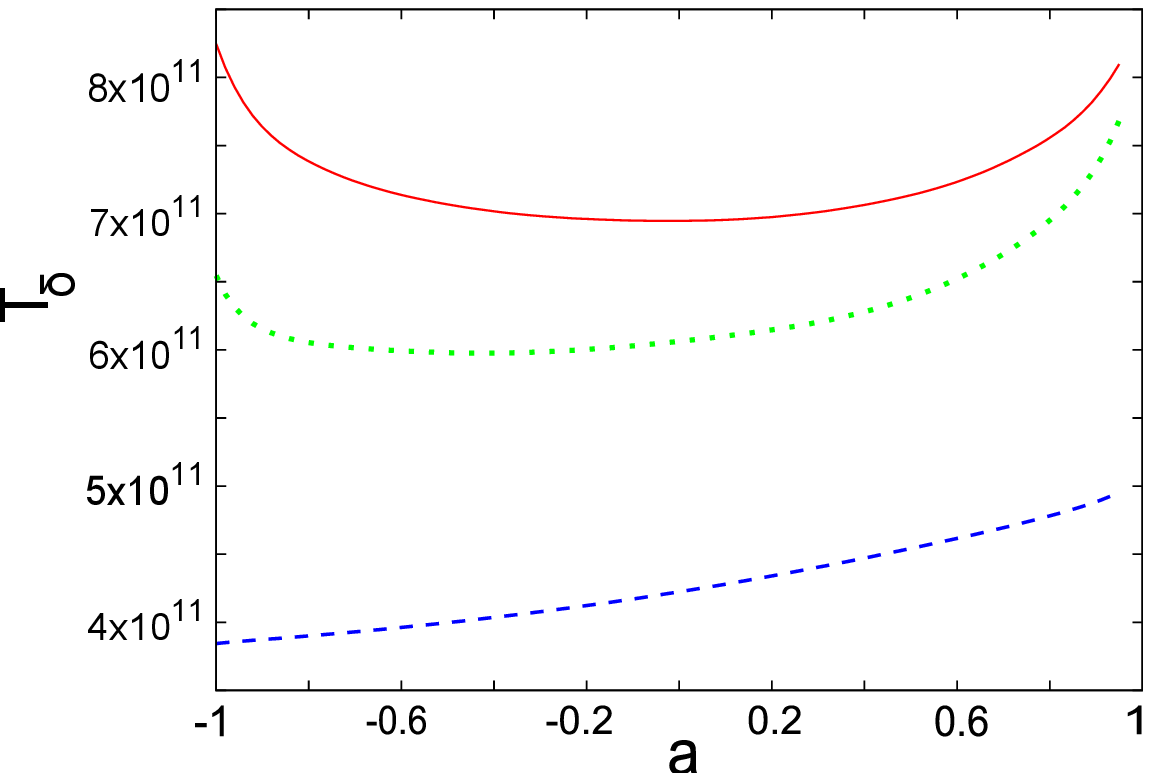} \\
\end{tabular}
\caption{Variation of quasi-terminal values of Mach number ($M_\delta$),
density ($\rho_\delta$), pressure ($P_\delta$) and temperature ($T_\delta$) with $a$ 
($\gamma=1.35,\mathcal{E}=1.2,\lambda=2.0$) for monotransonic accretion in 
constant height flow (dashed blue lines), quasi-spherical flow (dotted green lines) 
and flow in hydrostatic equilibrium (solid red lines). Density and pressure are in 
CGS units of $g$ $cm^{-3}$ and $dyne$ $cm^{-2}$ respectively and temperature is in 
absolute units of Kelvin.}
\label{fig13}
\end{figure}

In fig. \ref{fig13} we show the dependence of 
quasi-terminal values on black hole spin for monotransonic 
accretion. It is observed that weakly rotating and 
substantially hot flows allow for stationary 
monotransonic solutions over the entire range of Kerr 
parameters. From a careful glance at the results 
it becomes clear that reason behind the previously stated 
anomaly in general spin dependent behaviour of 
the corresponding physical quantities for three different 
flow geometries is essentially due to intrinsic 
limitations in the possibility of observing their variation 
over the complete range of spin. Since, for 
any given set of $\left[\mathcal{E},\lambda,\gamma\right]$, 
shocked stationary multitransonic accretion 
solutions for all matter configurations are allowed over a 
considerably small overlapping domain of $a$, 
one is able to look only through a narrow slit of the whole 
window. It is clearly evident from 
fig. \ref{fig13} that the quasi-terminal values indeed 
exhibit common global trends of variation over 
$a$ for all three geometric configurations. However, while 
concentrating upon a small portion of spin,  
asymmetry in the distribution of such trends leads to 
crossovers and apparently non-correlative or anti-
correlative mutual 
behaviours among the various flow models. Hence it is 
natural to question the utility of results 
with such constraints at the intrinsic level. But it is 
this very asymmetry, that turns out to be of 
supreme importance in pointing towards a prospective 
observational signature of the black hole spin.

\section{Isothermal flow structures for various matter geometries}

The equation of state characterising isothermal fluid flow 
is given by,
\begin{equation}
p=c_s^2\rho=\frac{\cal R}{\mu}\rho T=\frac{k_B\rho T}{\mu m_H}
\label{eqn65a}
\end{equation}
where $T$ is the bulk ion temperature, $\cal R$ is the 
universal gas constant, $k_B$ is Boltzmann constant, 
$m_H$ is mass of the Hydrogen atom and $\mu$ is the mean 
molecular mass of fully ionized hydrogen. 
The temperature $T$ as introduced in the above equation, 
and which has been used as one of the parameters to
describe the isothermal accretion, is the 
temperature-equivalent of the bulk ion flow velocity.
That is the reason why the value appears to be high 
($10^{10-11}$ K) in this work. 
The actual disc temperature is the corresponding electron 
temperature, 
which should be of the of the order of $10^{6-7}$ Kelvin.
The electron temperature may be computed from the 
corresponding ion temperature 
by incorporating various radiative processes, see, e.g. 
\cite{emn97apj}.
These calculations for our general relativistic model are, 
however, beyond the scope of this particular work 
and will be reported elsewhere. For low angular momentum 
shocked flow under the
influence of the Paczy{\'n}ski and Wiita 
pseudo-Schwarzschild black hole potential (\cite{pw80aa}), 
such computations have been performed, 
see, e.g. \cite{mdc06mnras}, as well as 
\cite{cmpds07RAGtime}. \\
The energy-momentum conservation equation obtained by 
setting the 4-divergence 
(covariant derivative w.r.t. $\nu$) 
of eqn.(\ref{eqn1}) to be zero is,
\begin{equation}
p_{,\nu}(g^{\mu\nu}+v^\mu v^\nu)+(p+\epsilon)v^\nu v^\mu_{;\nu}=0
\label{eqn65b}
\end{equation}
Using eqn.(\ref{eqn65a}), the general relativistic Euler equation for isothermal flow becomes,
\begin{equation}
\frac{c_s^2}{\rho}\rho_{,\nu}(g^{\mu\nu}+v^\mu v^\nu)+v^\nu v^\mu_{;\nu}=0
\label{eqn65c}
\end{equation}
Using the irrotationality condition $\omega_{\mu\nu}=0$, where 
$\omega_{\mu\nu}=l^\lambda_\mu l^\sigma_\nu v_{\left[\lambda ; \sigma\right]}$, \\
$\omega_{\mu\nu}$ being the vorticity of the fluid, $l^\lambda_\mu$ being the projection operator in the normal direction of $v^\mu$ \\
$l^\lambda_\mu=\delta^\lambda_\mu+v^\lambda v_\mu$, and 
$v_{\left[\lambda ; \sigma\right]}=\frac{1}{2}\left(v_{\sigma ;\lambda}-v_{\lambda ;\sigma}\right)$, we obtain,
\begin{equation}
\partial_\nu(v_\mu \rho^{c_s^2})-\partial_\mu(v_\nu \rho^{c_s^2})=0
\label{eqn65d}
\end{equation}
Taking the time component, we thus observe that for an irrotational isothermal flow, $v_t \rho^{c_s^2}$ turns out to be a conserved quantity. 
The square of this quantity is defined as the {\it{quasi-specific energy}} given by,
\begin{equation}
\xi=v_t^2 \rho^{2c_s^2}.
\label{eqn65e}
\end{equation}
$\xi$ is the first integral of motion for isothermal 
flows. The second 
integral of motion is $\dot{M}$, which is a function 
of the disc height $H$. 
The critical point conditions and expressions for the velocity gradients are computed using the same formalism as illustrated 
in case of polytropic flow for three different configurations of the disc geometry.

\subsection{Constant Height Flow}

Radial gradient of advective velocity: \\
\begin{equation}
\frac{du}{dr}|_{CH}=\frac{\frac{1-c_s^2}{2c_s^2}\frac{\Delta'}{\Delta}-\frac{1}{2c_s^2}\frac{B'}{B}}{\frac{1}{u}-\frac{u}{1-u^2}\frac{1-c_s^2}{c_s^2}}
\label{eqn65}
\end{equation}

Critical point conditions: \\
\begin{equation}
u_c^2|_{CH}={c_s}_c^2|_{CH}=1-\frac{B'}{B}\frac{\Delta}{\Delta'}
\label{eqn66}
\end{equation}

Velocity gradient at critical points: \\
\begin{equation}
\left(\frac{du}{dr}\right)_c|_{CH}=-\sqrt{\frac{\beta_{CH}}{\Gamma_{CH}}}
\label{eqn67}
\end{equation}

where, \\
$\Gamma_{CH}=\frac{2}{{c_s}_c^2(1-{c_s}_c^2)}$, \\
$\beta_{CH}=\beta^{(1)}_{CH}+\beta^{(2)}_{CH}+\beta^{(3)}_{CH}-\beta^{(4)}_{CH}-\beta^{(5)}_{CH}$, \\
$\beta^{(1)}_{CH}=\frac{2(1-{c_s}_c^2)(1-r_c)^2}{{c_s}_c^2({c_s}_c^2+r_c(r_c-2))^2}$, \\
$\beta^{(2)}_{CH}=\frac{{c_s}_c^2-1}{{c_s}_c^2({c_s}_c^2+r_c(r_c-2))}$, \\
$\beta^{(3)}_{CH}=\frac{\beta^{(31)}_{CH}}{r_c^4{c_s}_c^2({c_s}_c^2+\frac{2{c_s}_c^2}{r_c}+r_c^2-\frac{4{c_s}_c\lambda}{r_c}-\frac{(r_c-2)(r_c^3+{c_s}_c^2(r_c+2))\lambda^4}{r_c^3({c_s}_c^2(r_c+2)+r_c\lambda^2)})}$, \\
$\beta^{(31)}_{CH}=-2{c_s}_c^2r_c+5r_c^4+4{c_s}_cr_c\lambda$ \\
$+\frac{2{c_s}_c^2(r_c-2)(r_c^3+{c_s}_c^2(r_c+2))\lambda^4({c_s}_c^2+\lambda^2)}{({c_s}_c^2(r_c+2)+r_c\lambda^2)^3}-\frac{{c_s}_c^2(r_c-2)({c_s}_c^2+3r_c^2)\lambda^4}{({c_s}_c^2(r_c+2)+r_c\lambda^2)^2}-\frac{{c_s}_c^2(r_c^3+{c_s}_c^2(r_c+2))\lambda^4}{({c_s}_c^2(r_c+2)+r_c\lambda^2)^2}+\frac{(r_c^3-{c_s}_c^2(r_c^2-8))\lambda^4({c_s}_c^2+\lambda^2)}{({c_s}_c^2(r_c+2)+r_c\lambda^2)^2}+\frac{(2{c_s}_c^2-3r_c)r_c\lambda^4}{{c_s}_c^2(r_c+2)+r_c\lambda^2}$, \\
\resizebox{0.5\textwidth}{!}{$\beta^{(4)}_{CH}=\frac{4(-{c_s}_c^2r_c^2+r_c^5+2{c_s}_cr_c^2\lambda-\frac{{c_s}_c^2(r_c-2)(r_c^3+{c_s}_c^2(r_c+2))\lambda^4}{({c_s}_c^2(r_c+2)+r_c\lambda^2)^2}+\frac{(-r_c^3+{c_s}_c^2(r_c^2-8))\lambda^4}{{c_s}_c^2(r_c+2)+r_c\lambda^2})}{{c_s}_c^2r_c^5({c_s}_c^2+\frac{2{c_s}_c^2}{r_c}+r_c^2-\frac{4{c_s}_c\lambda}{r_c}-\frac{(r_c-2)(r_c^3+{c_s}_c^2(r_c+2))\lambda^4}{r_c^3({c_s}_c^2(r_c+2)+r_c\lambda^2)})}$}, \\
$\beta^{(5)}_{CH}=\frac{\beta^{(51)}_{CH}}{\beta^{(52)}_{CH}}$, \\
$\beta^{(51)}_{CH}=2\left[-{c_s}_c^6r_c^2(r_c+2)^2+2{c_s}_cr_c^4\lambda^5+2{c_s}_c^5r_c^2(r_c+2)^2\lambda\right.$ \\
$\left.+4{c_s}_c^3r_c^3(r_c+2)\lambda^3+r_c^4\lambda^4(r_c^3-\lambda^2)\right.$ \\
$\left.+{c_s}_c^2r_c\lambda^2(4r_c^5+2r_c^6-3r_c^3\lambda^2-8\lambda^4+r_c^2\lambda^4)\right.$ \\
$\left.+{c_s}_c^4(r_c+2)(2r_c^5+r_c^6-2r_c^3\lambda^2-6\lambda^4-r_c\lambda^4+r_c^2\lambda^4)\right]^2$, \\
$\beta^{(52)}_{CH}=\left[{c_s}_c^2r_c^2({c_s}_c^2(r_c+2)+r_c\lambda^2)^2\right]$ \\
$\left[{c_s}_c^4r_c^2(r_c+2)^2-4{c_s}_c^3r_c^2(r_c+2)\lambda-4{c_s}_cr_c^3\lambda^3\right.$ \\
$\left.+r_c^3\lambda^2(r_c^3-(r_c-2)\lambda^2)+{c_s}_c^2(r_c+2)(r_c^5+r_c^3\lambda^2-(r_c-2)\lambda^4)\right]^2$ \\

Eqn.(\ref{eqn65}) can be integrated numerically using eqns.(\ref{eqn66}) and (\ref{eqn67}) to obtain exact topology of the flow in the phase space.

\subsection{Conical Flow}

Radial gradient of advective velocity: \\
\begin{equation}
\frac{du}{dr}|_{CF}=\frac{\frac{1-c_s^2}{2c_s^2}\frac{\Delta'}{\Delta}-\frac{1}{2c_s^2}\frac{B'}{B}-\frac{1}{r}}{\frac{1}{u}-\frac{u}{1-u^2}\frac{1-c_s^2}{c_s^2}}
\label{eqn68}
\end{equation}

Critical point conditions: \\
\begin{equation}
u_c^2|_{CF}={c_s}_c^2|_{CF}=\frac{\frac{\Delta'}{\Delta}-\frac{B'}{B}}{\frac{2}{r}+\frac{\Delta'}{\Delta}}
\label{eqn69}
\end{equation}

Velocity gradient at critical points: \\
\begin{equation}
\left(\frac{du}{dr}\right)_c|_{CF}=-\sqrt{\frac{\beta_{CF}}{\Gamma_{CF}}}
\label{eqn70}
\end{equation}

where, \\
$\Gamma_{CF}=\frac{2}{{c_s}_c^2(1-{c_s}_c^2)}$, \\
$\beta_{CF}=\beta^{(0)}_{CF}+\beta^{(1)}_{CF}+\beta^{(2)}_{CF}+\beta^{(3)}_{CF}-\beta^{(4)}_{CF}-\beta^{(5)}_{CF}$, \\
$\beta^{(0)}_{CF}=-\frac{1}{r_c^2}$, \\
$\beta^{(1)}_{CF}=\frac{2(1-{c_s}_c^2)(1-r_c)^2}{{c_s}_c^2({c_s}_c^2+r_c(r_c-2))^2}$, \\
$\beta^{(2)}_{CF}=\frac{{c_s}_c^2-1}{{c_s}_c^2({c_s}_c^2+r_c(r_c-2))}$, \\
$\beta^{(3)}_{CF}=\frac{\beta^{(31)}_{CF}}{r_c^4{c_s}_c^2({c_s}_c^2+\frac{2{c_s}_c^2}{r_c}+r_c^2-\frac{4{c_s}_c\lambda}{r_c}-\frac{(r_c-2)(r_c^3+{c_s}_c^2(r_c+2))\lambda^4}{r_c^3({c_s}_c^2(r_c+2)+r_c\lambda^2)})}$, \\
$\beta^{(31)}_{CF}=-2{c_s}_c^2r_c+5r_c^4+4{c_s}_cr_c\lambda$ \\
$+\frac{2{c_s}_c^2(r_c-2)(r_c^3+{c_s}_c^2(r_c+2))\lambda^4({c_s}_c^2+\lambda^2)}{({c_s}_c^2(r_c+2)+r_c\lambda^2)^3}-\frac{{c_s}_c^2(r_c-2)({c_s}_c^2+3r_c^2)\lambda^4}{({c_s}_c^2(r_c+2)+r_c\lambda^2)^2}-\frac{{c_s}_c^2(r_c^3+{c_s}_c^2(r_c+2))\lambda^4}{({c_s}_c^2(r_c+2)+r_c\lambda^2)^2}+\frac{(r_c^3-{c_s}_c^2(r_c^2-8))\lambda^4({c_s}_c^2+\lambda^2)}{({c_s}_c^2(r_c+2)+r_c\lambda^2)^2}+\frac{(2{c_s}_c^2-3r_c)r_c\lambda^4}{{c_s}_c^2(r_c+2)+r_c\lambda^2}$, \\
\resizebox{0.5\textwidth}{!}{$\beta^{(4)}_{CF}=\frac{4(-{c_s}_c^2r_c^2+r_c^5+2{c_s}_cr_c^2\lambda-\frac{{c_s}_c^2(r_c-2)(r_c^3+{c_s}_c^2(r_c+2))\lambda^4}{({c_s}_c^2(r_c+2)+r_c\lambda^2)^2}+\frac{(-r_c^3+{c_s}_c^2(r_c^2-8))\lambda^4}{{c_s}_c^2(r_c+2)+r_c\lambda^2})}{{c_s}_c^2r_c^5({c_s}_c^2+\frac{2{c_s}_c^2}{r_c}+r_c^2-\frac{4{c_s}_c\lambda}{r_c}-\frac{(r_c-2)(r_c^3+{c_s}_c^2(r_c+2))\lambda^4}{r_c^3({c_s}_c^2(r_c+2)+r_c\lambda^2)})}$}, \\
$\beta^{(5)}_{CF}=\frac{\beta^{(51)}_{CF}}{\beta^{(52)}_{CF}}$, \\
$\beta^{(51)}_{CF}=2\left[-{c_s}_c^6r_c^2(r_c+2)^2+2{c_s}_cr_c^4\lambda^5+2{c_s}_c^5r_c^2(r_c+2)^2\lambda\right.$ \\
$\left.+4{c_s}_c^3r_c^3(r_c+2)\lambda^3+r_c^4\lambda^4(r_c^3-\lambda^2)\right.$ \\
$\left.+{c_s}_c^2r_c\lambda^2(4r_c^5+2r_c^6-3r_c^3\lambda^2-8\lambda^4+r_c^2\lambda^4)\right.$ \\
$\left.+{c_s}_c^4(r_c+2)(2r_c^5+r_c^6-2r_c^3\lambda^2-6\lambda^4-r_c\lambda^4+r_c^2\lambda^4)\right]^2$, \\
$\beta^{(52)}_{CF}=\left[{c_s}_c^2r_c^2({c_s}_c^2(r_c+2)+r_c\lambda^2)^2\right]$ \\
\resizebox{0.5\textwidth}{!}{$\left[{c_s}_c^4r_c^2(r_c+2)^2-4{c_s}_c^3r_c^2(r_c+2)\lambda-4{c_s}_cr_c^3\lambda^3+r_c^3\lambda^2(r_c^3-(r_c-2)\lambda^2)\right.$} \\
$\left.+{c_s}_c^2(r_c+2)(r_c^5+r_c^3\lambda^2-(r_c-2)\lambda^4)\right]^2$ \\

The flow profile is then obtained by integrating the velocity gradient using critical point conditions and values of velocity gradients 
evaluated at the critical points.

\subsection{Flow in vertical hydrostatic equilibrium}

The general equation for the height of an accretion 
disc held by hydrostatic equilibrium in the vertical 
direction is given by (\cite{alp97apj}),
\begin{equation}
-\frac{2p}{\rho}+\left(\frac{H}{r}\right)^2\frac{F}{r^2}=0,
\end{equation}
where $F=\lambda^2u_t^2-a^2(u_t-1)$.
The equation had been derived for flow in the Kerr 
metric and holds for any general equation of state of 
the infalling matter. Hence, for isothermal flows, 
the disc height can be calculated as, 
\begin{equation}
H=\left[\frac{2c_s^2r^4}{F}\right]^\frac{1}{2},
\end{equation}
leading to the following results. \\
Radial gradient of advective velocity: \\
\begin{equation}
\frac{du}{dr}|_{VE}=\frac{{c_s}_c^2(\frac{\Delta'}{2\Delta}+\frac{2}{r}-(2\lambda^2 v_t-a^2)\frac{v_t P1}{4F})-\frac{P1}{2}}{\frac{u}{1-u^2}-\frac{{c_s}_c^2}{u(1-u^2)}(1-(2\lambda^2 v_t-a^2)\frac{u^2 v_t}{2F})}
\label{eqn71}
\end{equation}

Critical point conditions: \\
\begin{eqnarray}
u_c^2|_{VE}=\frac{P1}{\frac{\Delta'}{\Delta}+\frac{4}{r}} \\
{c_s}_c^2|_{VE}=\frac{u_c^2}{1-\frac{u_c^2v_t(2\lambda^2 v_t-a^2)}{2F}}
\label{eqn72}
\end{eqnarray}

Velocity gradient at critical points: \\
\begin{equation}
\left(\frac{du}{dr}\right)_c|_{VE}=-\frac{\beta_{VE}}{2\alpha_{VE}} \pm \frac{1}{2\alpha_{VE}}\sqrt{\beta_{VE}^2-4\alpha_{VE}\Gamma_{VE}}
\label{eqn73}
\end{equation}
where, \\
$\alpha_{VE}=\frac{1+u_c^2}{\left(1-u_c^2\right)^2}-D_2D_6$, $\beta_{VE}=D_2D_7+\tau_4$, $\Gamma_{VE}=-\tau_3$, \\
$D_2=\frac{c_s^2}{u\left(1-u^2\right)}\left(1-D_3\right)$, $D_6=\frac{3u^2-1}{u\left(1-u^2\right)}-\frac{D_5}{1-D_3}$, \\
$D_7=\frac{D_3D_4v_tP1}{2\left(1-D_3\right)}$, $\tau _3=\left(c_s^2\tau_2-c_s^2v_5v_t\frac{P1}{2}\right)-\frac{P1'}{2}$, \\
$\tau _4=\frac{c_s^2v_5v_tu}{1-u^2}$, $v_1=\frac{\Delta'}{2\Delta
}+\frac{2}{r}-\left(2\lambda ^2v_t-a^2\right)v_t\frac{\text{P1}}{4F}$, \\
$D_3=\frac{u^2v_t\left(2\lambda^2v_t-a^2\right)}{2F}$, $D_4=\frac{1}{v_t}+\frac{2\lambda^2}{2\lambda^2v_t-a^2}-\frac{2\lambda^2v_t-a^2}{F}$,
$D_5=D_3\left(\frac{2}{u}+\frac{D_4v_tu}{1-u^2}\right)$, $\tau_2=\tau _1-\frac{v_t\left(2\lambda^2v_t-a^2\right)}{4F}P1'$, $v_5=\left(2\lambda^2v_t-a^2\right)\frac{P1}{4F}v_4$, \\
$\tau_1=\frac{1}{2}\left(\frac{\Delta''}{\Delta}-\frac{(\Delta')^2}{\Delta ^2}\right)-\frac{2}{r^2}$, 
$v_4=\frac{v_3}{\left(2\lambda^2v_t-a^2\right)F}$, $v_3=\left(4\lambda^2v_t-a^2\right)F-\left(2\lambda^2v_t-a^2\right)^2v_t$. \\

Eqn.(\ref{eqn71}) is integrated numerically using eqns.(\ref{eqn72}) and (\ref{eqn73}) to obtain the flow profile and also to locate the sonic points 
corresponding to the respective critical points. 

\section{Parameter Space for Isothermal Accretion}

Since, the parameter space is three dimensional in case of isothermal accretion in the Kerr metric, for convenience, 
we deal with a two dimensional parameter space among $^3C_2$ such possible combinations. 
The limits for two of the parameters governing the flow are 
$\left[0 < \lambda < 4, -1 < a < 1 \right]$. For the time being, we concentrate 
on $\left[T-\lambda\right]$ parameter space for a fixed value of $a$. A general $\left[T-\lambda\right]$ 
diagram for a given accretion disc geometry would look similar to the generic diagram for polytopic 
accretion shown in fig.\ref{fig1}. \\


\begin{figure}[h!]
\centering
\includegraphics[scale=0.7]{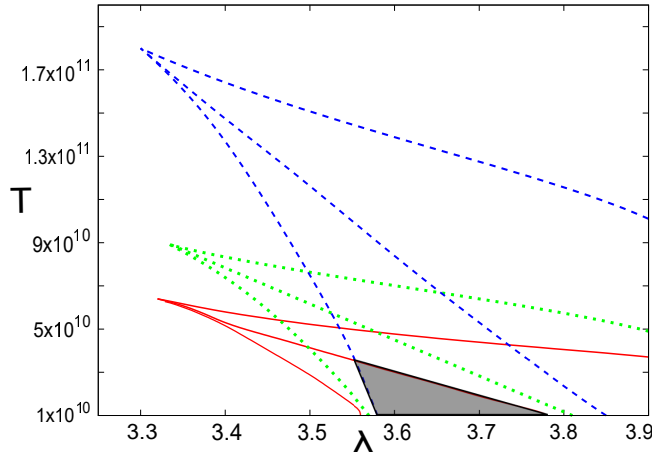}
\caption{Comparision of $T$-$\lambda$ plot for three different flow geometries ($a=0.1$, $T$ in Kelvin). 
Constant height disc, quasi-spherical flow and flow in vertical hydrostatic equilibrium represented by blue dashed lines, 
green dotted lines and red solid lines respectively. The shaded region allows for multicritical solution
in all flow configurations.}
\label{fig15}
\end{figure}

In figure \ref{fig15}, for $a=0.1$, we compare the parameter spaces for three different 
flow geometries. The common domain for which multiple critical points are formed for all three flow geometries is shown as a shaded region.

\section{Classification of critical points for isothermal accretion}
Using the same technique elaborated in section $4$, eigenvalues of the stability matrices for isothermal accretion can be computed 
for the three disc geometries.
\subsection{Constant Height Flow}
\begin{equation}
{\Omega^{iso}_{CH}}^2 = {\mathcal{B}_{CH}^{iso}}{\mathcal{C}_{CH}^{iso}}
\label{eqn74}
\end{equation}
where, \\
\begin{eqnarray}
{{\cal B}_{CH}^{iso}} &=& \frac{{f_c^{'}}}{f_c}\frac{a^2-1-(r_c-1)^2}{\Delta(r_c-1)}-\frac{{f_c^{''}}}{f_c}+\left(\frac{{f_c^{'}}}{f_c}\right)^2\\
{{\cal C}_{CH}^{iso}} &=& \frac{1}{u_c^2(1-u_c^2)}
\label{eqn75}
\end{eqnarray}

\subsection{Conical Flow}
\begin{equation}
{\Omega^{iso}_{CF}}^2 = {\mathcal{B}_{CF}^{iso}}{\mathcal{C}_{CF}^{iso}}
\label{eqn76}
\end{equation}
where, \\
\begin{eqnarray}
{{\cal B}_{CF}^{iso}} &=& \frac{{f_c^{'}}}{f_c(2r_c^2-3r_c+a^2)}(-\frac{\Delta_c}{r_c}+\frac{r_c}{\Delta_c}(a^2-1-(r_c-1)^2)) \nonumber \\
& &-\frac{{f_c^{''}}}{f_c}
+\left(\frac{{f_c^{'}}}{f_c}\right)^2\\
{{\cal C}_{CF}^{iso}} &=& \frac{1}{u_c^2(1-u_c^2)}
\label{eqn77}
\end{eqnarray}

\subsection{Flow in hydrostatic equilibrium along the vertical direction}
\begin{eqnarray}
{\Omega^{iso}_{VE}}^2 = {\mathcal{B}_{VE}^{iso}}{\mathcal{C}_{VE}^{iso}}-{\mathcal{A}_{VE}^{iso}}{\mathcal{D}_{VE}^{iso}}
\label{eqn78}
\end{eqnarray}
\begin{eqnarray}
{{\cal A}_{VE}^{iso}} &=& \frac{{{c_s}_c}^2}{g_2}\left(\frac{(2\lambda^2 v_t-a^2){f_c}^{'}{g_2}^{'}}{2g_2\sqrt{(1-u_c^2)f_c}}-\delta_3\right) \\
{{\cal B}_{VE}^{iso}} &=& {c_s}_c^2\left(\frac{2}{\Delta_c}-\frac{4}{r_c^2}-\frac{4(r_c-1)^2}{\Delta_c^2}-\frac{\delta_4}{g_2}\right. \nonumber \\
& &\left.+\left(\frac{(2\lambda^2v_t-a^2){f_c}^{'}}{2g_2\sqrt{(1-u_c^2)f_c}}\right)^2\right) \nonumber \\
& &-\frac{f_c^{''}}{f_c}+\left(\frac{f_c^{'}}{f_c}\right)^2 \\
{{\cal C}_{VE}^{iso}} &=& \frac{u_c^4-2{c_s}_c^2u_c^2+{c_s}_c^2}{u_c^4(1-u_c^2)^2}+\frac{{c_s}_c^2\delta_1}{g_2}-\frac{{c_s}_c^2{g_2^{'}}^2}{g_2^2} \\
{{\cal D}_{VE}^{iso}} &=& -{{\cal A}_{VE}^{iso}}
\label{eqn79}
\end{eqnarray}
where, \\
$g_2=(\lambda v_t)^2-v_ta^2+a^2$, \\
$\delta_1=\frac{2\lambda^2f}{(1 - u_c^2)^3}-\frac{3a^2}{4}\sqrt{\frac{f}{(1 - u_c^2)^5}}$, \\
$\delta_3=\frac{\lambda^2f'}{(1 - u_c^2)^2}-\frac{a^2f'}{4\sqrt{f(1-u_c^2)^3}}$, \\
$\delta_4=\frac{\lambda^2f^{''}}{1 - u_c^2}-\frac{a^2}{4\sqrt{1-u_c^2}}\frac{2ff^{''} - f'^2}{f^{\frac{3}{2}}}$ \\

\section{Dependence of $\Omega^2$ on flow and spin parameters for isothermal accretion in various matter geometries}

\begin{figure}[h!]
\centering
\begin{tabular}{cc}
\includegraphics[width=0.4\linewidth]{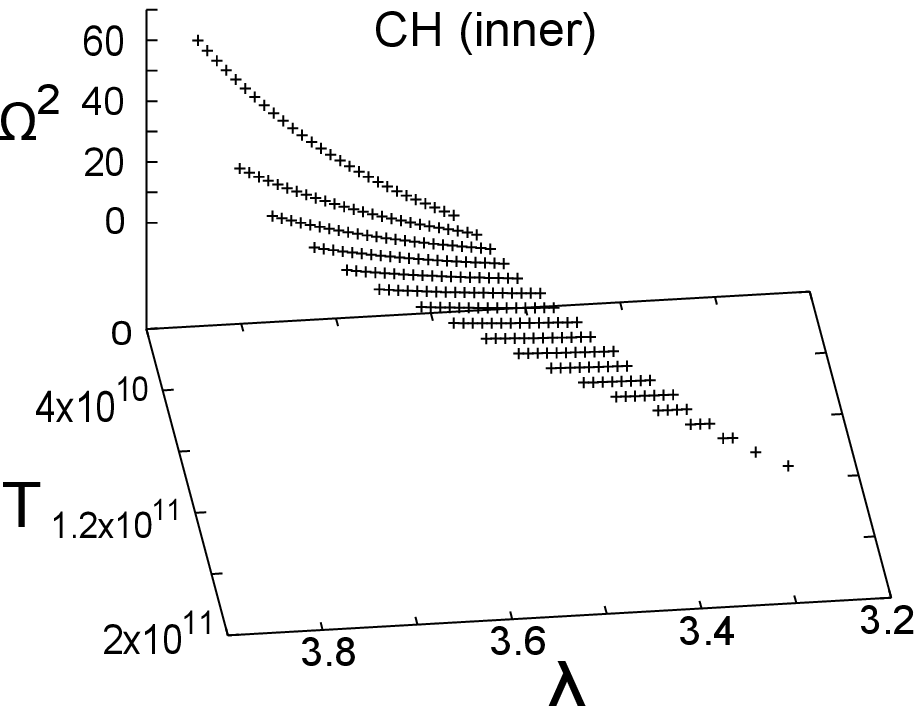} &
\includegraphics[width=0.4\linewidth]{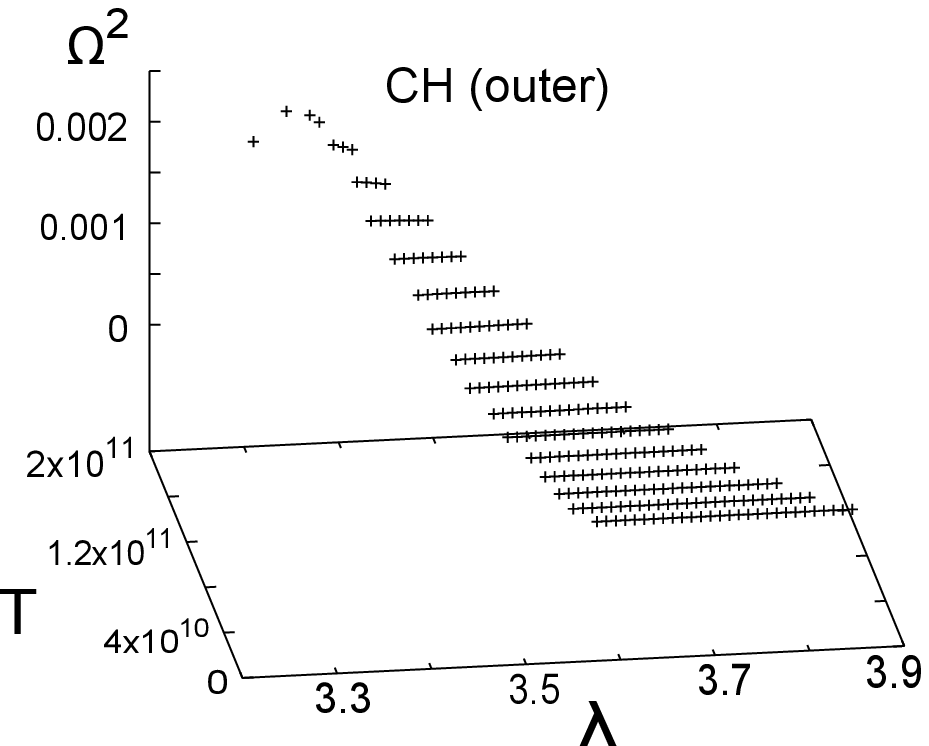} \\
\includegraphics[width=0.4\linewidth]{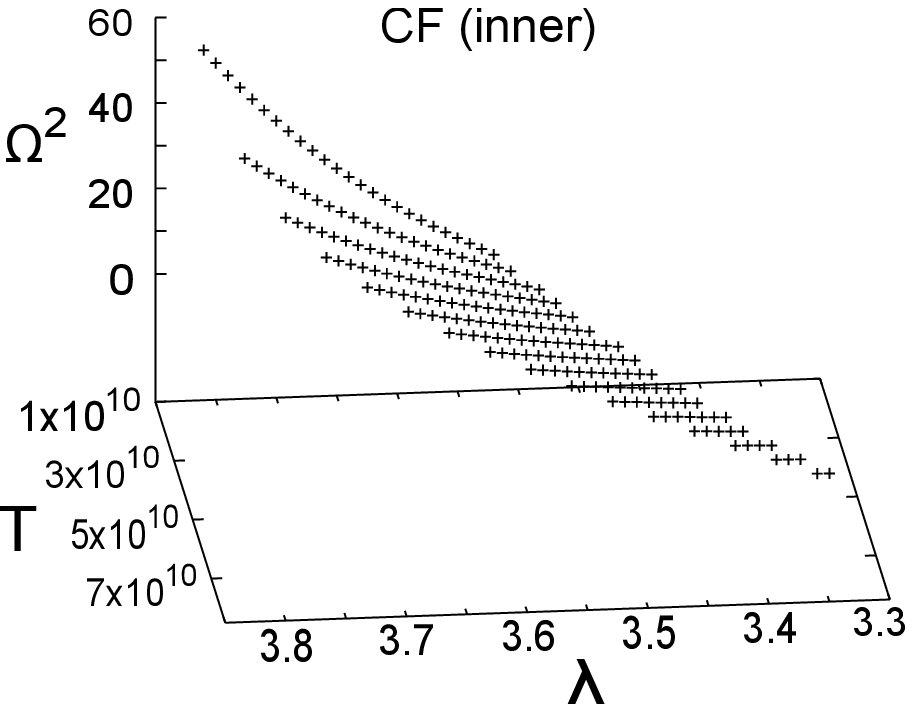} &
\includegraphics[width=0.4\linewidth]{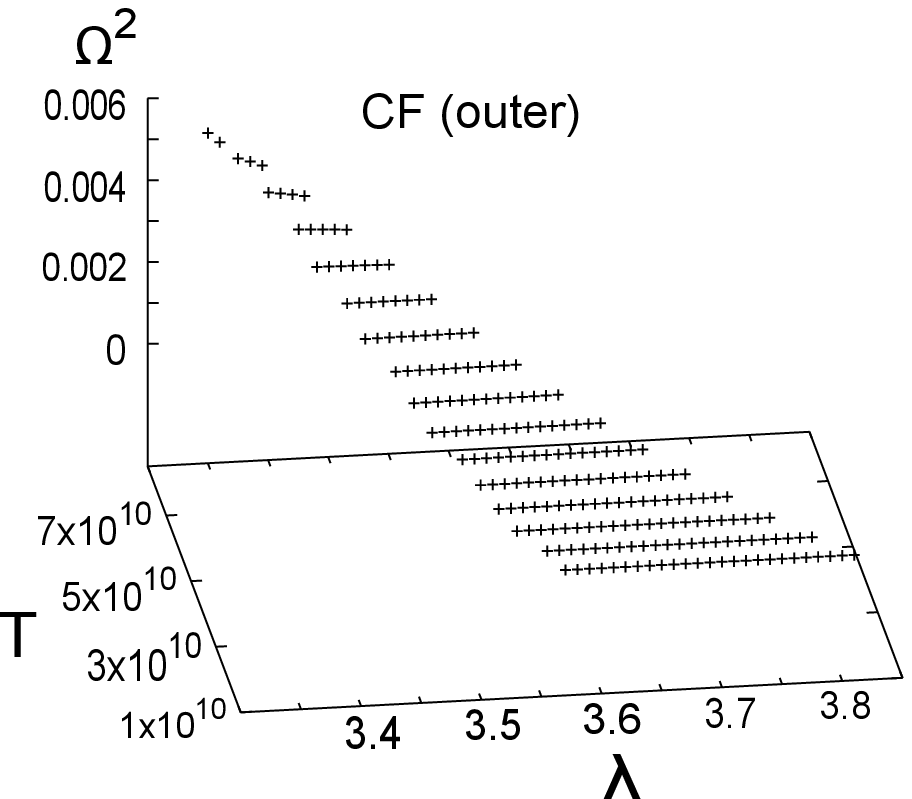} \\
\includegraphics[width=0.4\linewidth]{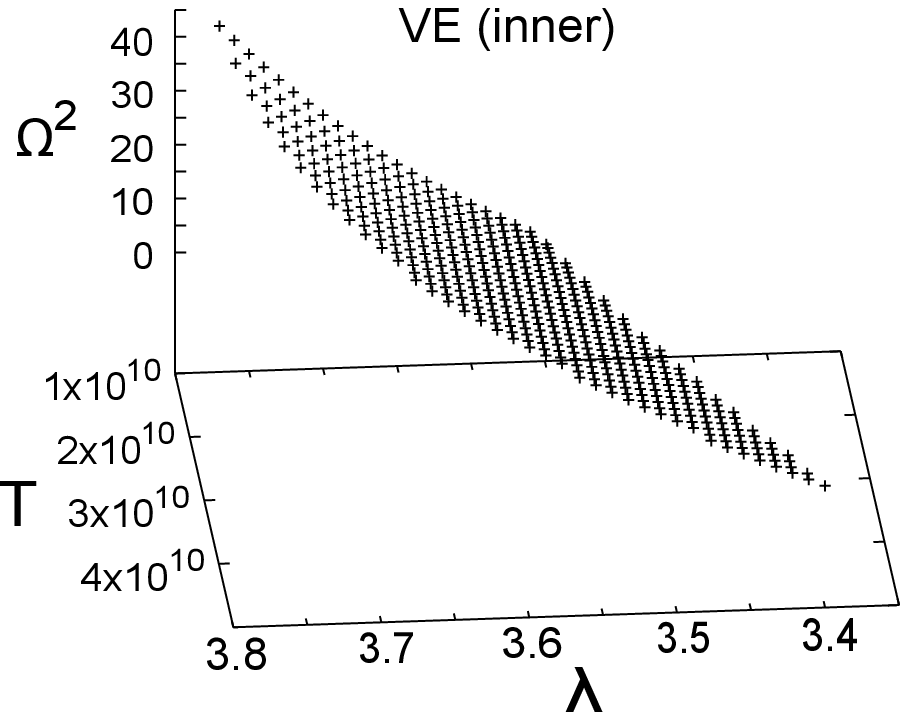} &
\includegraphics[width=0.4\linewidth]{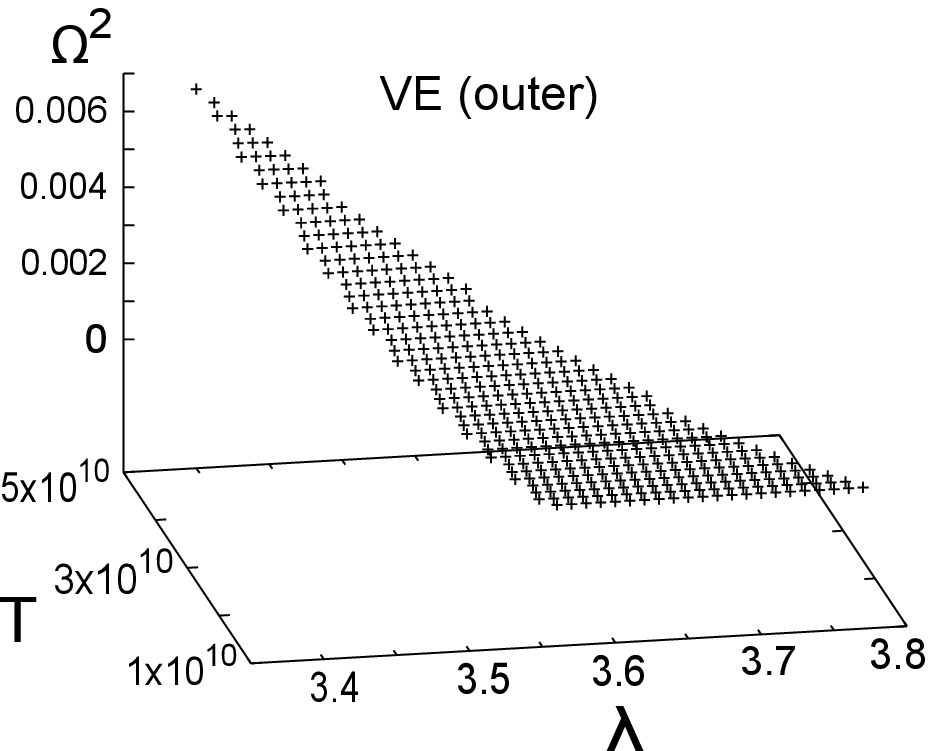} \\
\end{tabular}
\caption{Comparison of $\Omega^2$ vs. $\left[T,\lambda\right]$
of the inner and outer critical points for constant height flow (CH), 
quasi-spherical flow (CF) and flow in vertical hydrostatic equilibrium (VE) 
($a=0.1$). Left and right panels depict $\Omega^2$ for inner and outer 
critical points respectively for CH, CF and VE from top to bottom in the 
respective order.}
\label{fig16}
\end{figure}

\begin{figure}[h!]
\centering
\begin{tabular}{cc}
\includegraphics[width=0.4\linewidth]{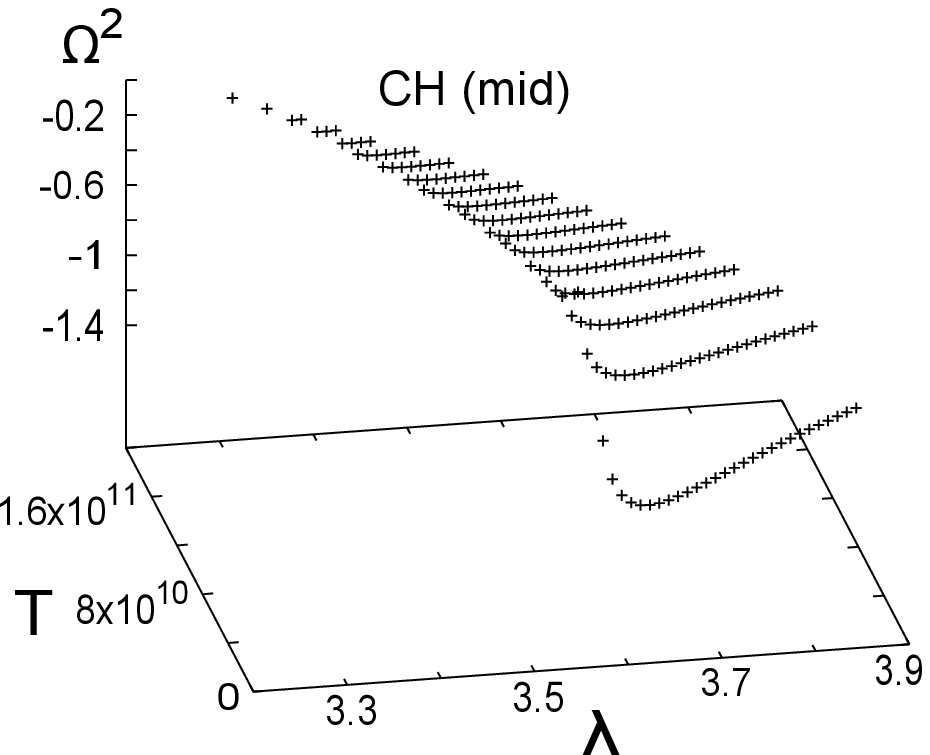} &
\includegraphics[width=0.4\linewidth]{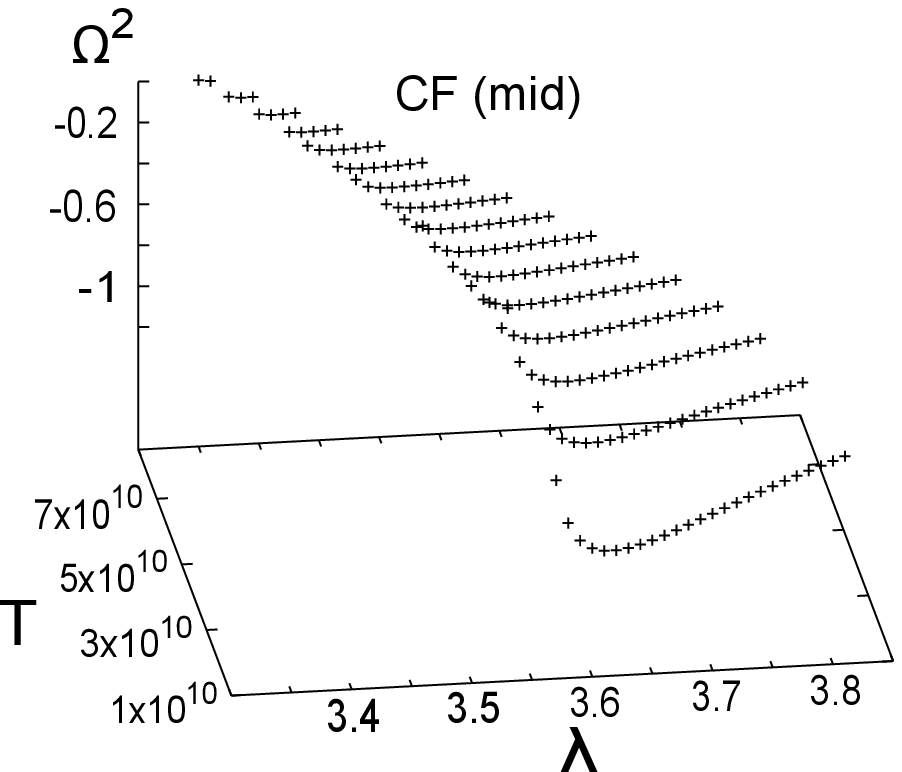} \\
\includegraphics[width=0.4\linewidth]{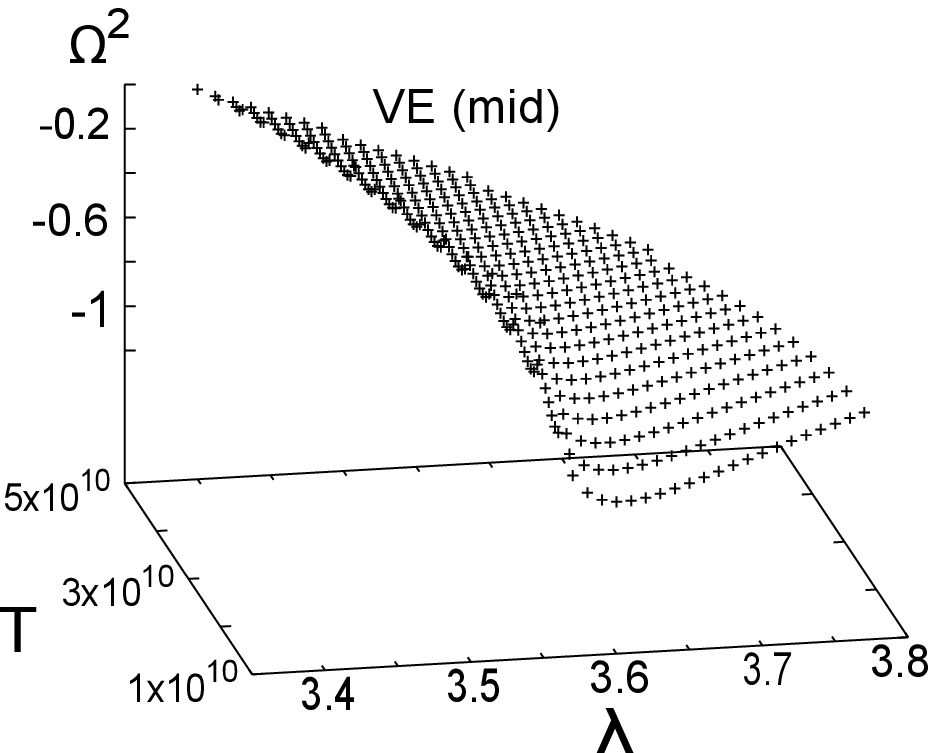} \\
\end{tabular}
\caption{Comparison of $\Omega^2$ vs. $\left[T,\lambda\right]$
of the middle critical point for constant height flow (CH) (top left), 
quasi-spherical flow (CF) (top right) and flow in vertical hydrostatic equilibrium (VE) 
(bottom left)($a=0.1$).}
\label{fig17}
\end{figure}

\begin{figure}[h!]
\centering
\includegraphics[scale=0.7]{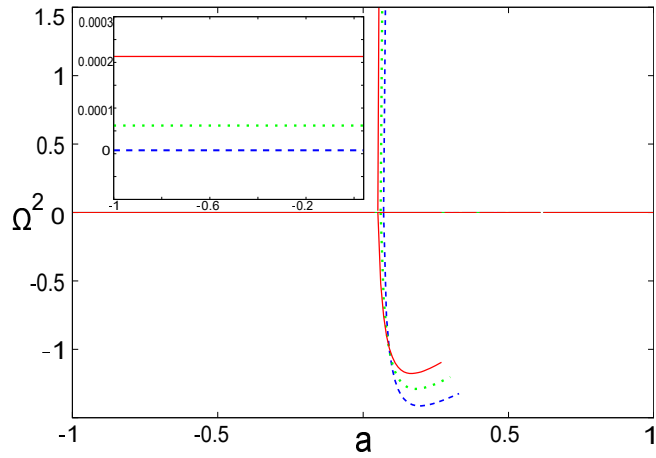}
\caption{Comparison of $\Omega^2$ vs. $a$ for constant height flow (blue dashed lines), 
quasi-spherical flow (green dotted lines) and flow in vertical hydrostatic equilibrium 
(red solid lines)($T=10^{10}$ K, $\lambda=3.6$). Inset shows a 
magnified view of the common monotransonic region for the three flow geometries.}
\label{fig18}
\end{figure}

Figs. \ref{fig16}, \ref{fig17} and \ref{fig18} obtained by evaluating the analytical expressions for isothermal flow derived in the 
previous section at $r_c$ establish the same argument about multi-transonicity and nature of the critical points as presented 
in the corresponding section for polytropic flows. The numerical value of $\Omega^2$ assumes positive sign for the inner and 
outer saddle-type critical points and negative sign for the middle centre-type critical points.

In fig. \ref{fig16}, the variation of $\Omega^2$ for inner and outer critical points 
has been depicted over the entire physical domain of $\left[T,\lambda\right]$ 
for a given value of $a=0.1$. Positive values indicate critical points of saddle nature. 
Fig. \ref{fig17} depicts the values of $\Omega^2$ for the middle critical points over the 
full accessible domain of $\left[T,\lambda\right]$ for the same value of $a$. 
Negative values indicate critical points which are centre-type. 
The same trend of comparision is observed between the absolute magnitudes of $\Omega^2$ for the saddle type 
critical points in the case of isothermal flow as well. $\Omega^2_{inner}>>\Omega^2_{outer}$ once again points 
towards a correlation between the absolute value of $\Omega^2$ and space-time curvature at the critical points. 
A comparision of the three different flow geometries reveals that for inner and middle critical points, 
$|\Omega^2|_{CH}>|\Omega^2|_{CF}>|\Omega^2|_{VE}$, whereas for the outer critical point 
$|\Omega^2|_{VE}>|\Omega^2|_{CF}>|\Omega^2|_{CH}$.

Fig. \ref{fig18} is a similar plot as was obtained for polytropic flow depicting the bifurcation of 
$\Omega^2$ along black hole spin parameter $a$ for a given value of temperature and 
specific angular momentum ($T=10^{10} K,\lambda=3.6$). 
Monotransonic flow through a saddle type critical point is shown in the inset where 
$\Omega^2$ assumes a single positive value for all three flow geometries.
The monotransonic flow then bifurcates into multi-transonic flow with a centre-type middle critical point 
(with negative value), a saddle-type inner critical point (with a larger positive value) and a 
saddle-type outer critical point (with a smaller positive value). Thus, a saddle-centre pair is generated at a definite value of 
$a$. The inner saddle gradually shifts closer to the event horizon acquiring higher values of $\Omega^2$ and finally one is left with a 
single saddle point through which physical monotransonic flow can occur.   
As in the case of polytropic flow, the value of parameter $a$ at which the bifurcation occurs is different for different flow geometries 
and that value is found to be minimum for discs in vertical hydrostatic equilibirum.

\section{Shock-invariant quantities ($S_h$)}

Applying the technique described in section 6.1, {\it{shock-invariant quantities}} ($S_h$) for all three 
isothermal flow geometries are obtained as under- \\
\subsection{Constant height flow}
\begin{equation}
S_h{|_{CH}^{iso}} = \left(\frac{u}{\sqrt{1-u^2}}\right)^{2{c_s^2}-1}(u^2\Delta+r^2{c_s^2}(1-u^2))
\label{eqn80}
\end{equation}

\subsection{Conical flow}
\begin{equation}
S_h{|_{CF}^{iso}} = \left(\frac{u}{\sqrt{1-u^2}}\right)^{2{c_s^2}-1}(u^2\Delta+r^2{c_s^2}(1-u^2))
\label{eqn81}
\end{equation}

\subsection{Flow in hydrostatic equilibrium in the vertical direction}
\begin{equation}
S_h{|_{VE}^{iso}} = u^{2{c_s^2}-1}(u^2\Delta+r^2{c_s^2}(1-u^2))
\label{eqn82}
\end{equation}

\section{Shock parameter space for isothermal accretion}
We now intend to see which region of the $T-\lambda$ parameter space allows shock formation. For a fixed 
of $a=0.1$, we check validity of the Rankine Hugoniot condition for every value of 
$\left[T,\lambda\right]$ for which the accretion flow possesses three critical points. This means the 
shock-invariant quantity is calculated for every $\left[T,\lambda\right]$ for which the multitransonic 
accretion is possible, and it is observed that only for some subset of such $\left[T,\lambda\right]$, 
the shock-invariant quantities calculated along the solution passing through the outer and the inner sonic points 
become equal at a particular radial distance, i.e., at the shock location. We then plot the corresponding 
$\left[T,\lambda\right]_{\rm{shock}}$ for various geometric configurations of matter. 

\begin{figure}[h!]
\centering
\includegraphics[scale=0.7]{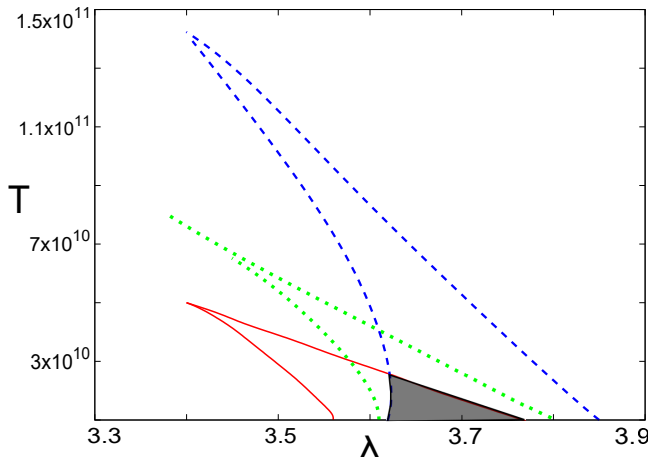}
\caption{Comparision of $T$-$\lambda$ plot of allowed shocked multitransonic accretion
solutions for three different flow geometries ($a=0.1$, $T$ in Kelvin). 
Constant height disc, quasi-spherical flow and flow in vertical hydrostatic equilibrium represented by blue dashed lines, 
green dotted lines and red solid lines respectively. Shaded region depicts the overlapping domain of 
shock formation in all the three geometries.}
\label{fig19}
\end{figure}

In fig. \ref{fig19} we plot the subsets of the $T$-$\lambda$ spaces for three different flow geometries at a fixed 
$a$ ($=0.1$), for which, value of the shock-invariant quantities $S_h$, when evaluated along the flow branches through inner 
and outer critical points, become equal at particular value(s) of $r$. This value of radial distance $r_{sh}$ is the location 
of shock. The shaded region depicts overlap of shock-forming $\left[T,\lambda\right]$ parameter set of the three disc configurations 
for a given $a$.

Once the common region for shock formation is obtained, we investigate the variation of shock location ($r_{sh}$), shock strength ($M_+/M_-$), 
compression ratio ($\rho_-/\rho_+$), pressure ratio ($P_-/P_+$) and quasi-specific energy dissipation ratio ($\xi_+/\xi_-$)
($+$ and $-$ have the same meanings as defined for polytropic accretion) on the black hole spin parameter $a$, 
also comparing the trends of variation for various disc geometries.

\begin{figure}[h!]
\centering
\includegraphics[scale=0.7]{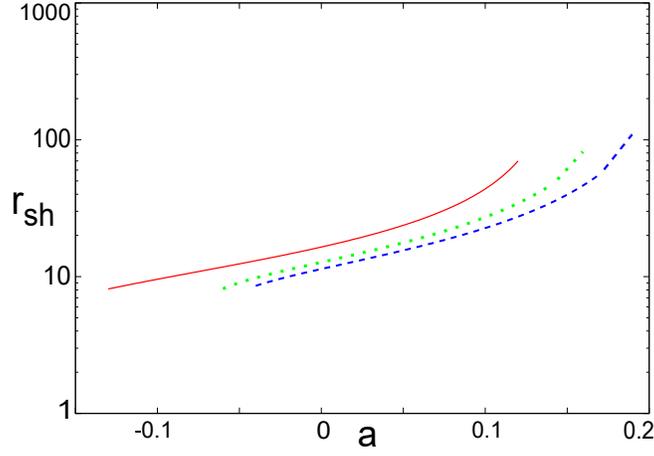}
\caption{Shock location ($r_{sh}$) vs. $a$ plot 
($T=10^{10} K,\lambda=3.75$) for 
constant height disc (dashed blue line), quasi-spherical disc (dotted green line) 
and flow in hydrostatic equilibrium (solid red line).}
\label{fig20}
\end{figure}

Fig. \ref{fig20} shows how the shock location ($r_{sh}$) varies with spin parameter $a$. The bulk ion temperature 
has been fixed at $10^{10}$ K and the value of $\lambda$ has been selected accordingly ($=3.75$) from the region of shock overlap
observed in fig. \ref{fig19} so that the available range of $a$ is maximum. The same set of $\left[T,\lambda\right]$ 
has been used in all subsequent shock related plots.
As already argued in the corresponding section for polytropic flow, growth in strength of the effective centrifugal barrier 
due to increase in the difference between $\lambda$ and $a$ explains the formation of shock farther away from the gravitating 
source as the value of black hole spin is increased while keeping the value of specific angular momentum fixed. 
Again, for a particular value of $a$, it is observed that $r_{sh}(VE)>r_{sh}(CF)>r_{sh}(CH)$, which indicates that even in the case of isothermal flow, 
an accretion disc in vertical hydrostatic equilibrium is exposed to maximum resistance for a given centrifugal barrier.
 
\begin{figure}[h!]
\centering
\begin{tabular}{cc}
\includegraphics[width=0.4\linewidth]{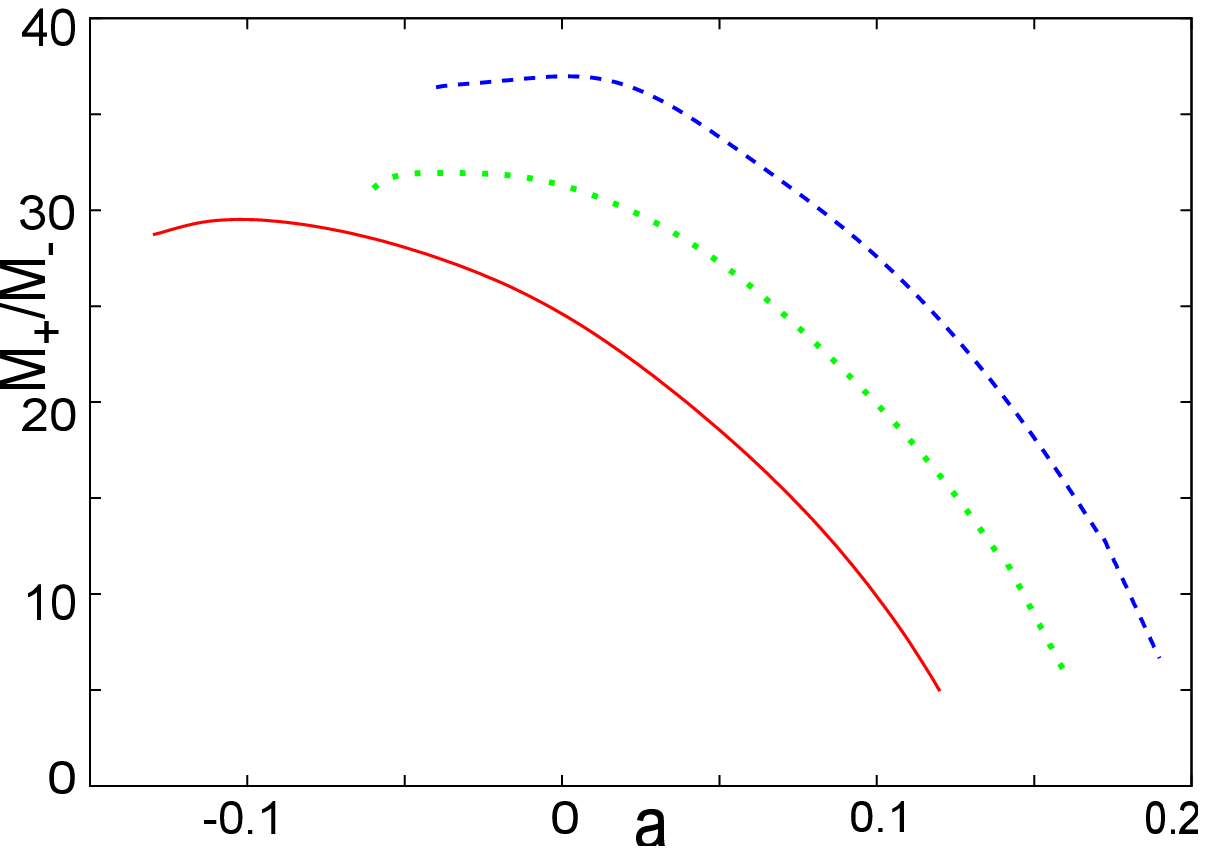} &
\includegraphics[width=0.4\linewidth]{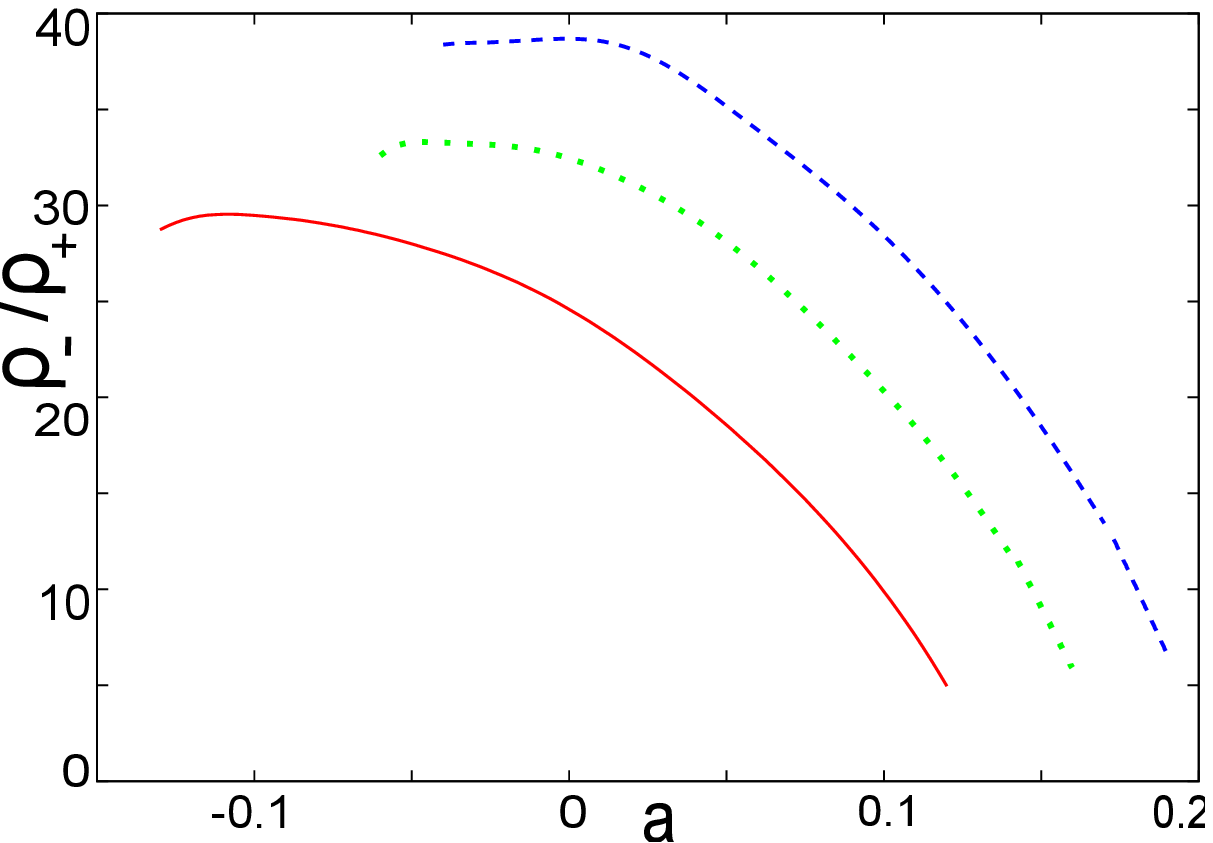} \\
\includegraphics[width=0.4\linewidth]{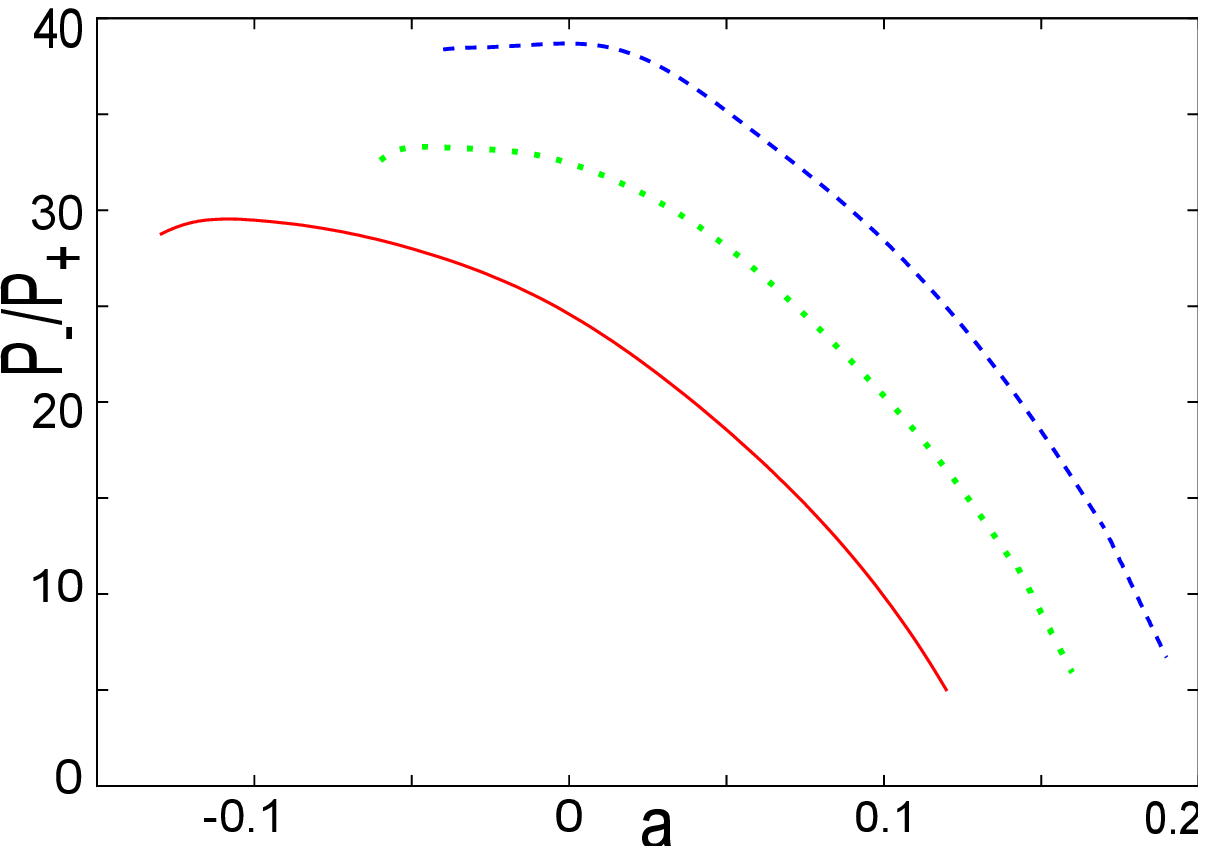} \\
\end{tabular}
\caption{Variation of shock strength ($M_+/M_-$), compression ratio ($\rho_-/\rho_+$) 
and pressure ratio ($P_-/P_+$) with black hole 
spin parameter $a$ ($E=10^{10} K,\lambda=3.75$) for 
constant height flow (dashed blue lines), quasi-spherical flow (dotted green lines) 
and flow in hydrostatic equilibrium (solid red lines). Subscripts `+' and `-' 
represent pre and post shock quantities respectively.}
\label{fig21}
\end{figure}

Figure (\ref{fig21}) on comparing with 
fig.(\ref{fig11}) establishes the fact that irrespective of 
whether the flow is polytropic or isothermal, a gradual 
increase in black hole spin for a specific flow angular 
momentum shifts the shock location outwards by boosting the 
effective centrifugal barrier. A shock formed far away from 
the event horizon is weaker in strength owing to the 
eventual flattening of space-time. Moreover in both 
polytropic and isothermal cases, 
for a given $\left[a,\lambda\right]$, the strongest shocks 
are formed in constant height discs whereas discs in 
hydrostatic vertical equilibrium exhibit the weakest 
shocks. The same trend is consistently observed for all the 
relevant ratios across 
the discontinuity.

\section{Powering the flares through the energy dissipated at the shock}

\begin{figure}[h!]
\centering
\begin{tabular}{cc}
\includegraphics[width=0.4\linewidth]{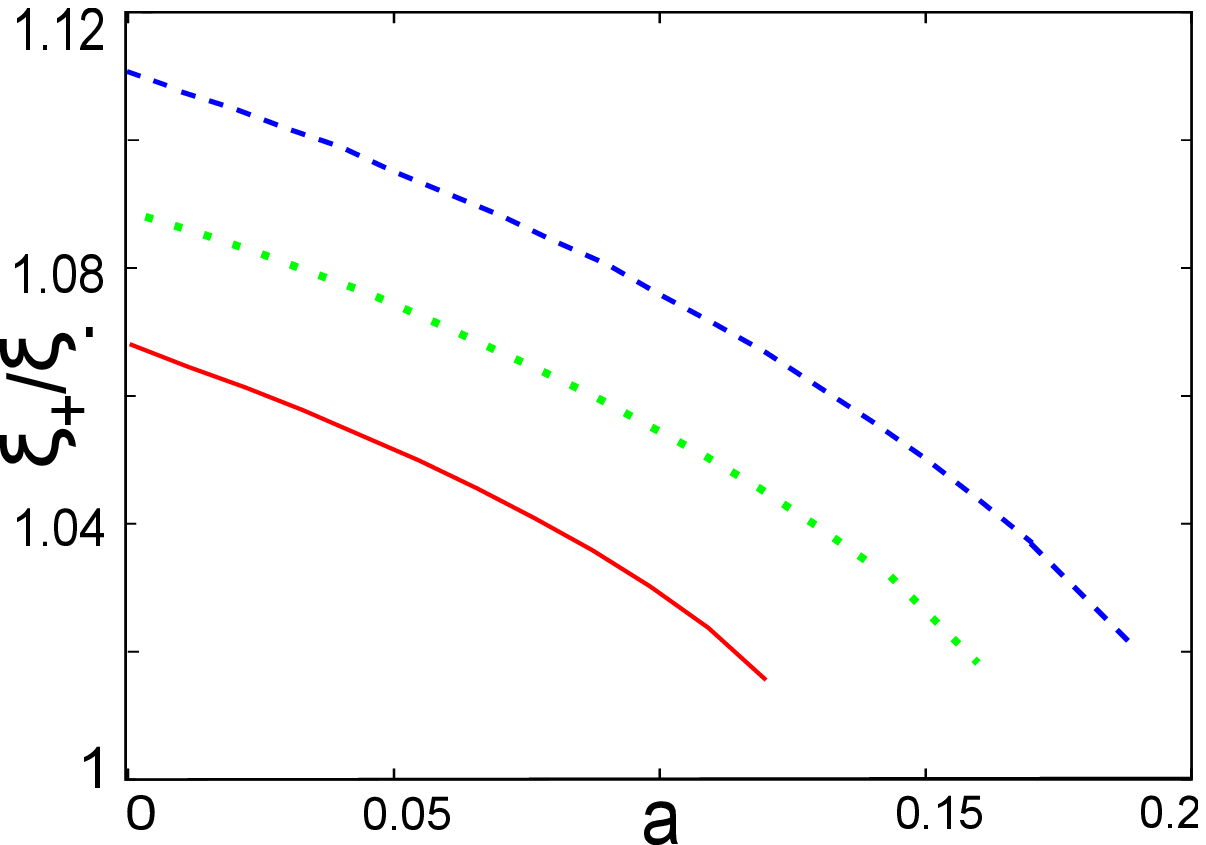} &
\includegraphics[width=0.4\linewidth]{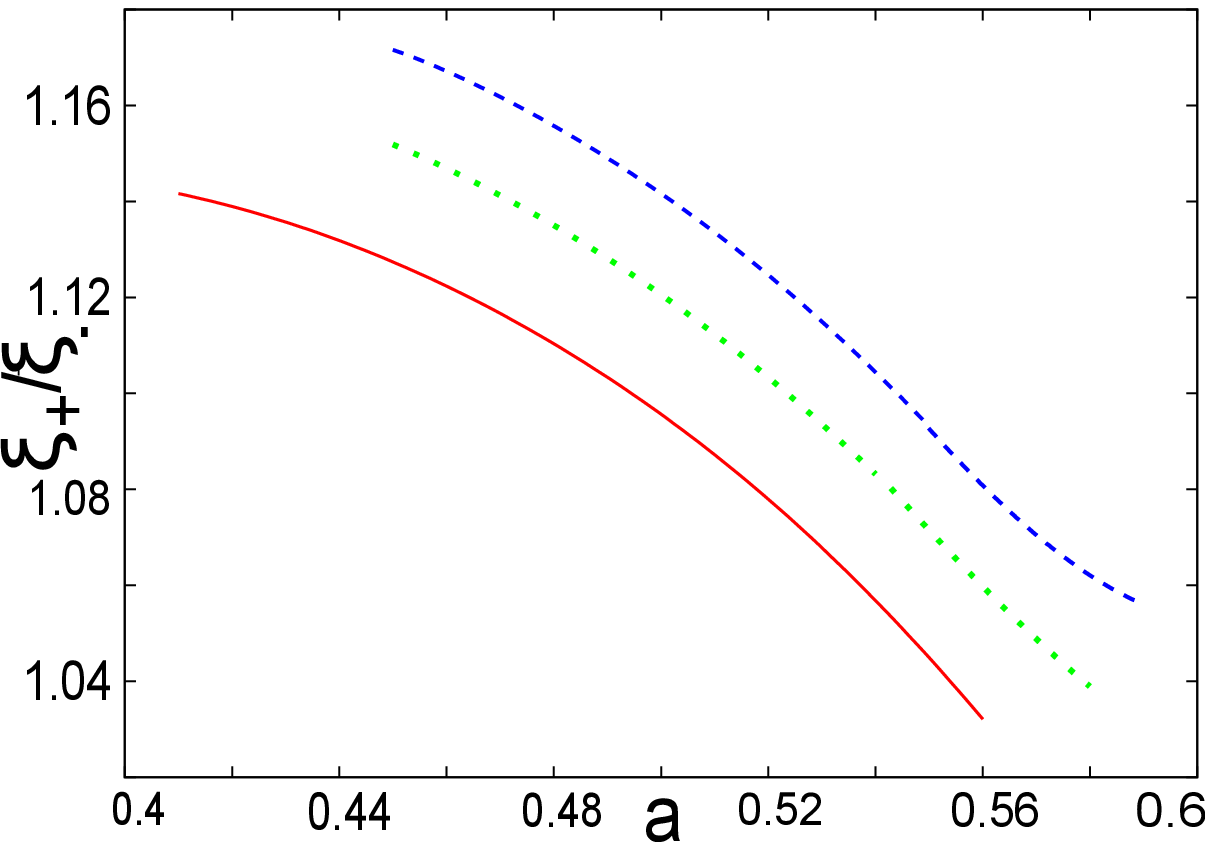} \\
\includegraphics[width=0.4\linewidth]{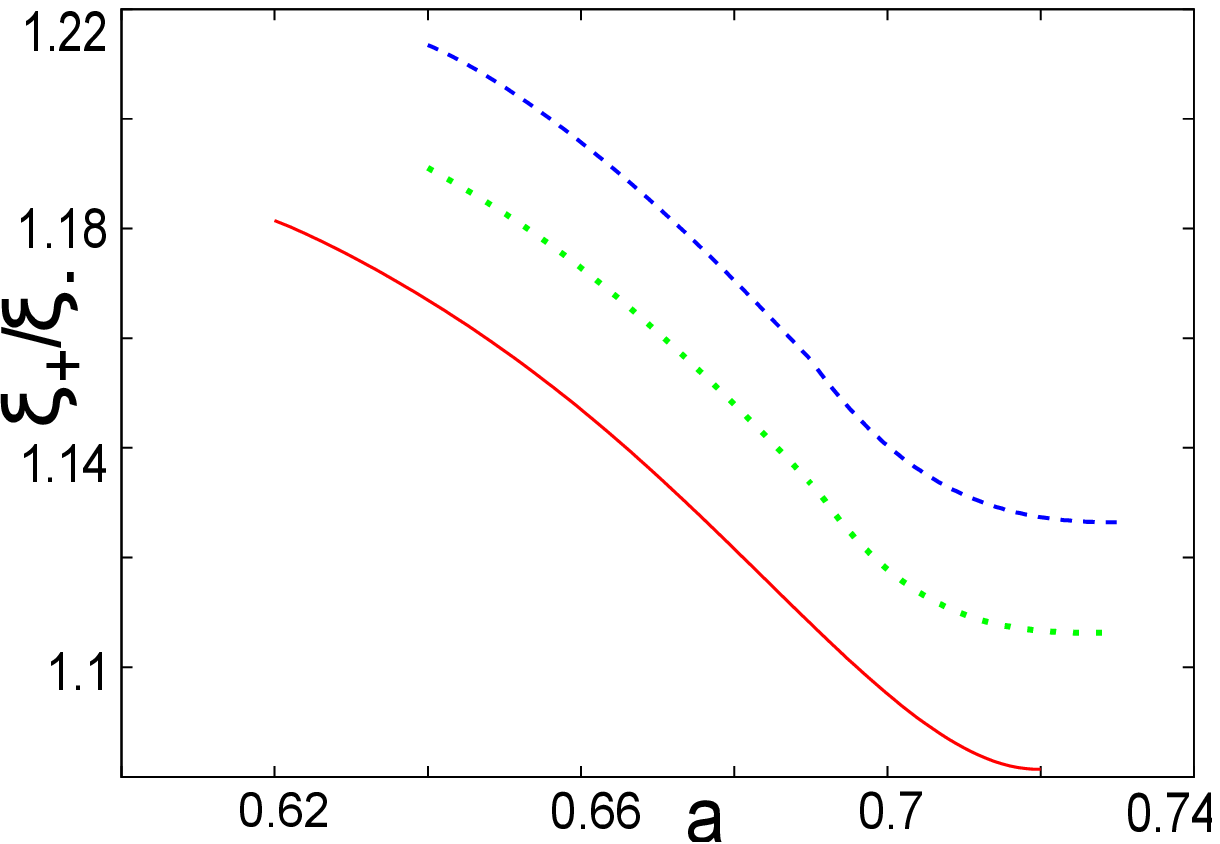} &
\includegraphics[width=0.4\linewidth]{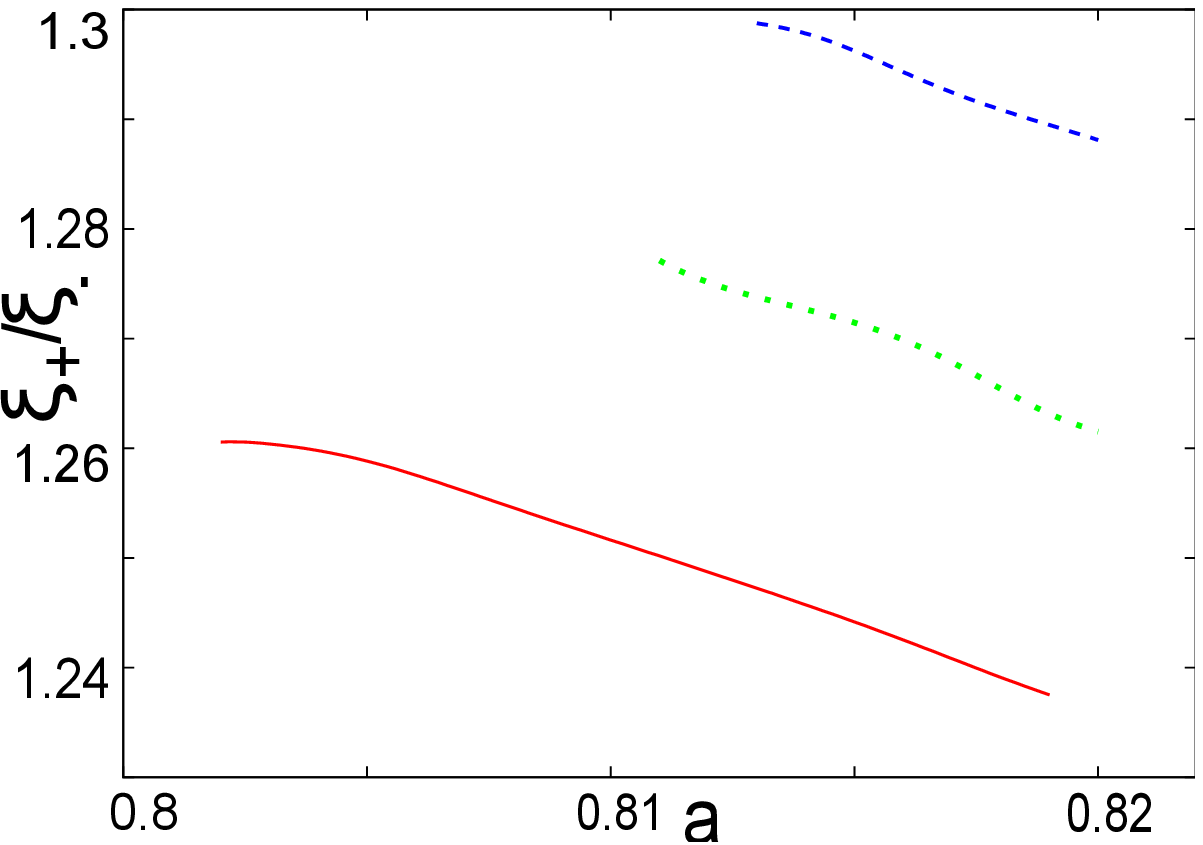} \\
\end{tabular}
\caption{Variation of quasi-specific energy ratio ($\xi_+/\xi_-$) with black hole 
spin parameter $a$ ($T=10^{10}$ K). In order to obtain multicritical domains with shock 
over four different ranges of $a$, four different values of $\lambda$ have been set at the given 
$T$, i.e. $\lambda=3.75$ (upper left), $\lambda=3.25$ (upper right), 
$\lambda=3.0$ (lower left), $\lambda=2.7$ (lower right). 
Constant height flow, quasi-spherical flow 
and flow in hydrostatic equilibrium have been represented by 
dashed blue lines, dotted green lines and solid red lines. Subscripts `+' and `-' 
represent pre and post shock quantities respectively.}
\label{fig22}
\end{figure}

For the isothermal accretion onto a rotating black hole 
considered in the present work, we concentrate on dissipative 
shocks. Unlike the standing Ranking-Hugoniot type energy-preserving shocks studied for the polytropic flow, a substantial 
amount of energy is dissipated at the shock 
location to maintain 
the temperature invariance of the isothermal flow. As a 
consequence, the flow thickness does not change abruptly at the 
shock location, and handling the pressure balance 
equation across the shock becomes more convenient as compared to 
that for the polytropic accretion. The amount of energy 
dissipated at the shock might make an isothermal shock to appear 
`bright', since for inviscid, dissipationless flow as considered 
in our work, accretion remains grossly radiatively inefficient 
throughout. \\
The type of low angular momentum inviscid flow we consider in 
the present work, is believed to be ideal to mimic the accretion 
environment of our galactic centre black hole 
(\cite{mdc06mnras}). Sudden substantial energy dissipation from 
the shock surface may thus be conjectured to feed the X-ray and 
IR flares emanating from our galactic centre black hole 
(\cite{baganoff01nature}, \cite{gsoealra03nature}, 
\cite{mbmmghdmbrbglmzrb08apj}, \cite{cldkpd10aa}, 
\cite{wnmbnycddfghhjlnprs13science}, 
\cite{pdmmmchzngmnrdtg15mnras}, \cite{kbevdkh17mnras}, 
\cite{mg17aa}, \cite{ywlw18mnras}). \\
In our formalism, the ratios of the quasi-specific energies 
corresponding to the pre-shock and post-shock flows is assumed 
to be a measure of the amount of the dissipated energy at the 
shock surface. In fig. \ref{fig22}, we plot such ratios for 
various ranges of the black hole spins (the Kerr parameter 
`$a$') for three different types of the geometrical thicknesses 
of the flow as considered in our present work. There are four 
panels in the figure, each plane corresponding to a certain 
range of values of the black hole spin angular momentum. As 
already discussed, for a fixed value of $\left[\mathcal{E},
\lambda,\gamma\right]$ or $\left[T,\lambda\right]$, shock 
formation over a continuous range of `$a$' spanning the entire 
domain of the Kerr parameter $-1>a>1$, is allowed neither for 
polytropic nor isothermal accretion. Four different panels in 
the figure are thus characterised by four different sets of $
\left[T,\lambda\right]$ as mentioned in the figure caption. The 
following interesting features are observed: \\

Depending on the initial conditions, substantial amount 
of energy gets liberated from the shock surface. Sometimes even 
as high as $30\%$ of the rest mass may be converted into 
radiated energy, which is a huge amount. Hence the shock-generated dissipated energy can, in principle, be considered as 
a good candidate to explain the source of energy dumped into the 
flare. \\
The length scale on the disc from which the flare may be 
generated actually matches with the shock location. Such `flare 
generating' length scales obtained in our theoretical 
calculations are thus, in good agreement with the observational 
works (\cite{kbevdkh17mnras}). \\
What we actually observe is that the amount of dissipated energy 
anti-correlates with the shock location, which is perhaps 
intuitively obvious because closer the shock forms to the 
horizon, greater is the available gravitational energy to be 
converted into dissipated radiation. \\
Following the same line of argument, the amount of dissipated 
energy anti-correlates with the flow angular momentum. Lower is 
the angular momentum of the flow, the centrifugal pressure 
supported region forms closer to the boundary. Such regions slow 
down the flow and break the flow behind it, and hence the shock 
forms. The locations of such region are thus markers 
anticipating from which region of the disc, the flare may be 
generated. \\
It is imperative to study the influence of the black hole spin 
in determining the amount of energy liberated at the shock. What 
we find here is that for the prograde flow, such amount anti-
correlates with the black hole spin. Thus, for a given flow 
angular momentum, slowly rotating black holes produce the 
strongest flares. Hence, for a given value of $\left[T,\lambda
\right]$, if shocked multitransonic accretion solutions 
exist over a positive span of $a$ including $a=0$, then flares 
originating from 
the vicinity of the Schwarzschild hole would consequently contain 
the maximum amount of energy. Hence, unlike the Blandford-Znajek 
mechanism (\cite{bz77mnras}, \cite{dc12mnras}, \cite{opm16apj}, 
\cite{cy16an}, \cite{bambi17revmodphys}), the amount of energy 
transferred to a flare is not extracted at the expense of black 
hole spin. \\
Certain works based on the observational results argue that 
there is no obvious correlation between the black hole spin and 
the jet power (\cite{fgr10mnras}, \cite{bf11mnras} and references 
therein). Our present finding is in accordance with such 
arguments. \\
In this connection, however, it is to be noted that BZ 
mechanism is usually associated with the electromagnetic energy 
extractions, whereas energy liberation at the shock is 
associated with the hydrodynamic flow. Hence no direct 
comparision can 
perhaps be made between the Blandford Znajek process and the 
process considered in our work. \\

\begin{figure}[h!]
\centering
\includegraphics[scale=0.7]{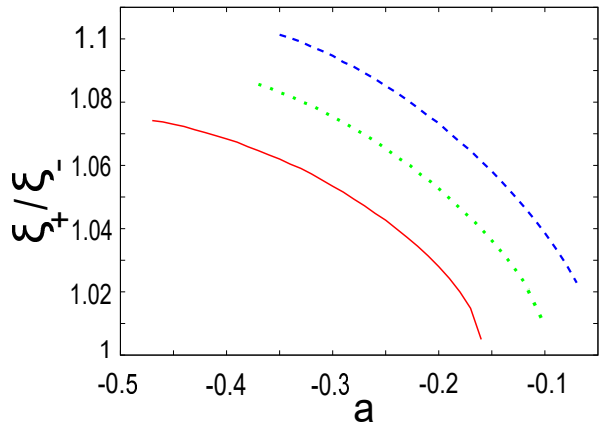}
\caption{Variation of quasi-specific energy ratio ($\xi_+/\xi_-$) 
with black hole spin parameter $a$ ($T=10^{10}$ K, $\lambda=4.0$) 
for retrograde flow. Constant height flow, quasi-spherical flow 
and flow in hydrostatic equilibrium have been represented by 
dashed blue lines, dotted green lines and solid red lines. 
Subscripts `+' and `-' represent pre and post shock quantities 
respectively.}
\label{fig25}
\end{figure}

In recent years, the study of retrograde flow close to the Kerr 
holes are also of profound interest (\cite{garofalo13aa}, 
\cite{mpgnb18mnras} 
and references therein). We thus study the 
spin dependence of the amount of energy dissipation at the 
shock. The result is shown in fig. \ref{fig25}. Here we observe 
that the amount of dissipated energy is more for faster 
{\it{counter-rotating}} holes. For retrograde flow, the negative 
Kerr parameter essentially reduces the overall measure of the 
angular momentum of the flow and the effective angular momentum 
may probably be thought of as $\lambda_{eff}=\lambda-a$, which 
explains such finding. \\
It is also observed that the amount of shock dissipated energy 
is also influenced by the geometric configuration of the flow. 
We find that axially symmetric flow with constant thickness 
produces largest amount of liberated energy at the shock, 
whereas the flow in hydrostatic equilibrium along the vertical 
direction produces the smallest amount. The conical wedge shaped 
flow contributes at a rate which is intermediate to the rates 
for the constant height disc and the disc in vertical 
equilibrium. This feature remains unaltered for the prograde as 
well as the retrograde flows. \\

\section{Quasi-terminal values}

\subsection{Dependence of $\left[M,\rho,P\right]_{r_\delta}$ on $a$ for shocked isothermal accretion}

\begin{figure}[h!]
\centering
\begin{tabular}{cc}
\includegraphics[width=0.4\linewidth]{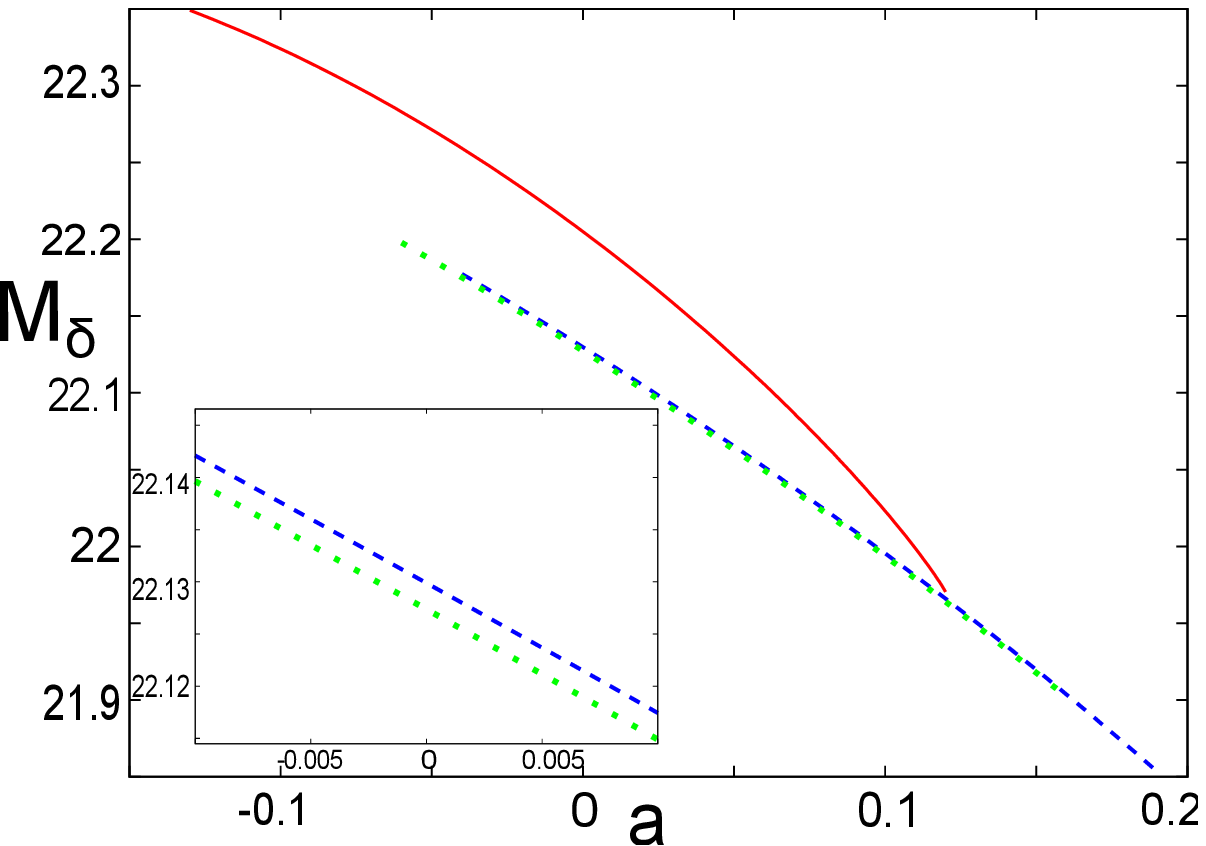} &
\includegraphics[width=0.4\linewidth]{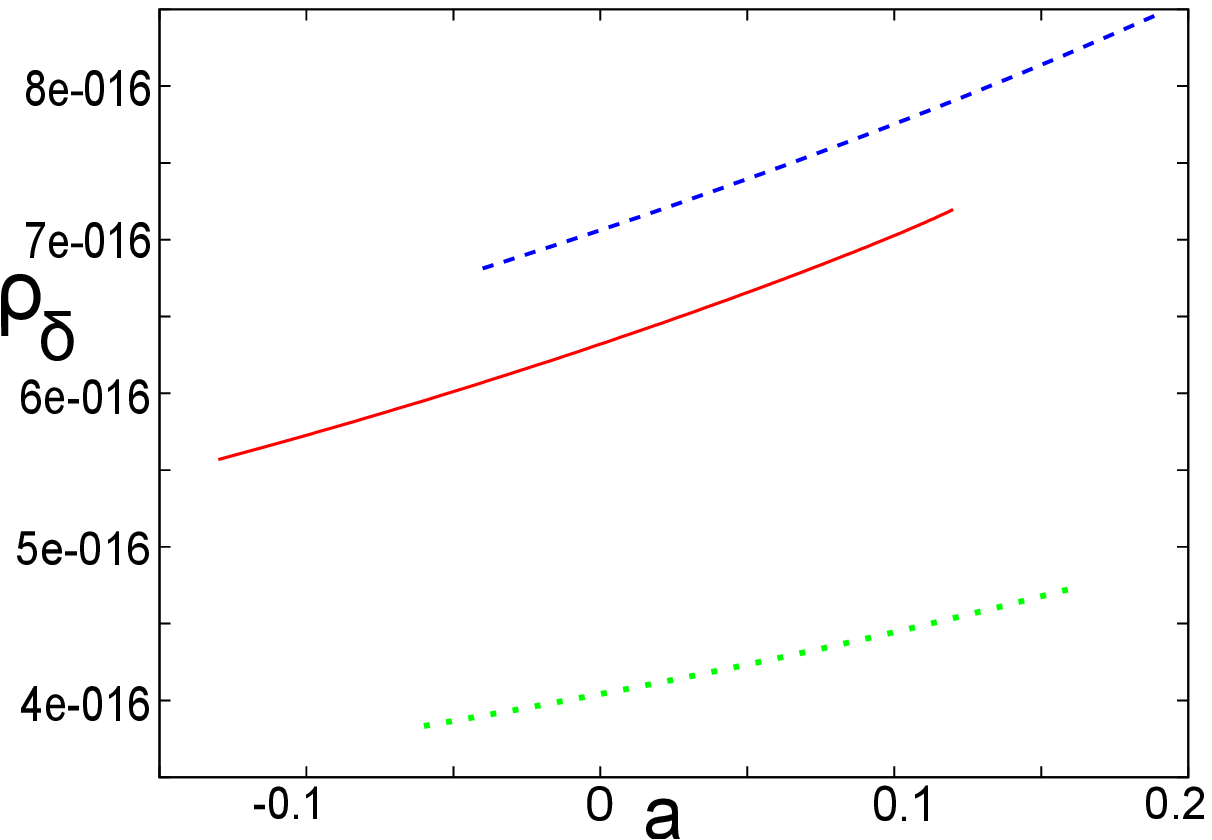} \\
\includegraphics[width=0.4\linewidth]{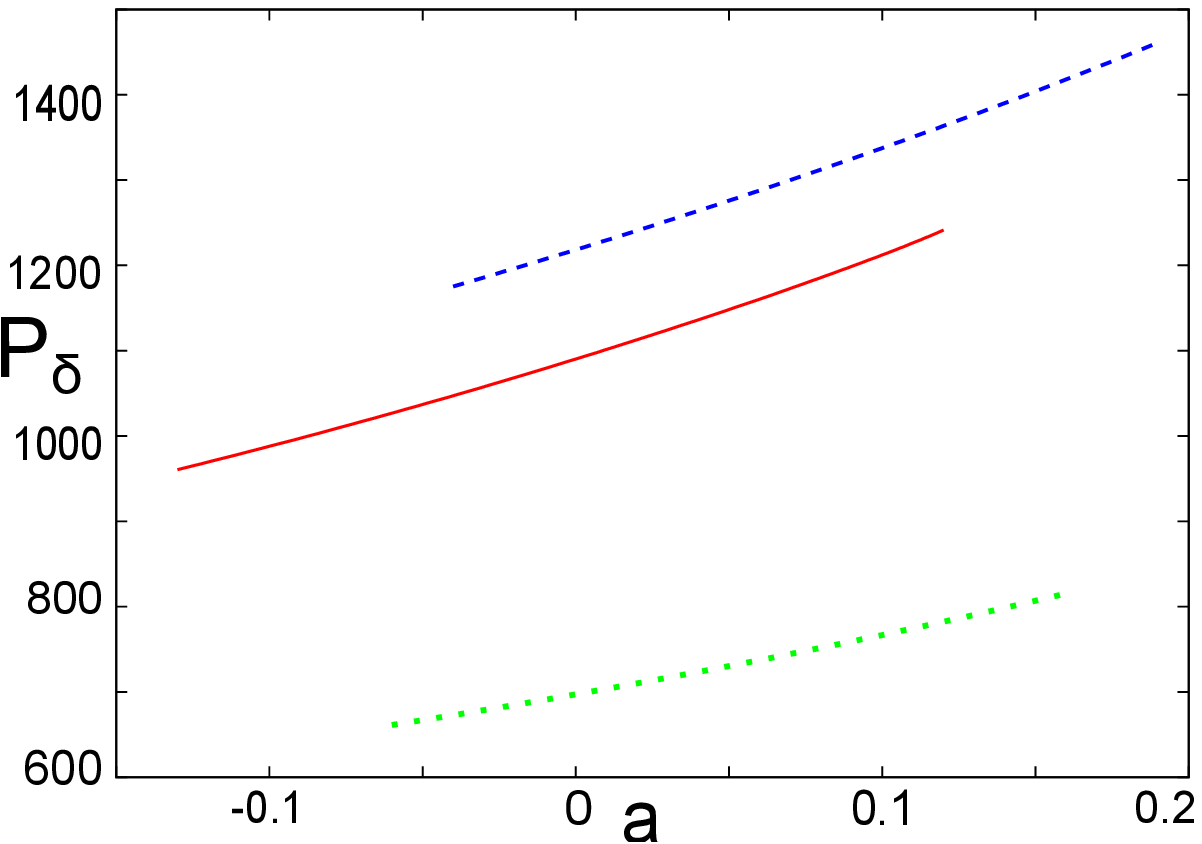} \\
\end{tabular}
\caption{Variation of quasi-terminal values of Mach number ($M_\delta$),
density ($\rho_\delta$) and pressure ($P_\delta$) with $a$ 
($T=10^{10} K,\lambda=3.75$) for 
constant height flow (dashed blue lines), quasi-spherical flow (dotted green lines) 
and flow in hydrostatic equilibrium (solid red lines). Density and pressure are in 
CGS units of $g$ $cm^{-3}$ and $dyne$ $cm^{-2}$ respectively and temperature is in 
absolute units of Kelvin.}
\label{fig23}
\end{figure}

Variation of the quasi-terminal values of Mach number ($M_\delta$), 
density ($\rho_\delta$) and pressure ($P_\delta$) with spin parameter $a$ has been shown in fig.(\ref{fig23}) for a 
given $T$ ($=10^{10}$) and $\lambda$ ($=3.75$). It is observed that although the variations are similar in nature to those 
for polytropic flow, but even in this case, limitations in the availibility and overlap of a broad range of spin for 
shocked multitransonic accretion for all geometric models, make it impossible to comment on the global trend 
with which such quantities vary in accordance to black hole spin or the disc configuration. Hence, we try to resolve this issue 
in the next subsection by looking at the case of monotransonic isothermal accretion.

\subsection{Dependence of $\left[M,\rho,P\right]_{r_\delta}$ on $a$ for monotransonic isothermal accretion}

\begin{figure}[h!]
\centering
\begin{tabular}{cc}
\includegraphics[width=0.4\linewidth]{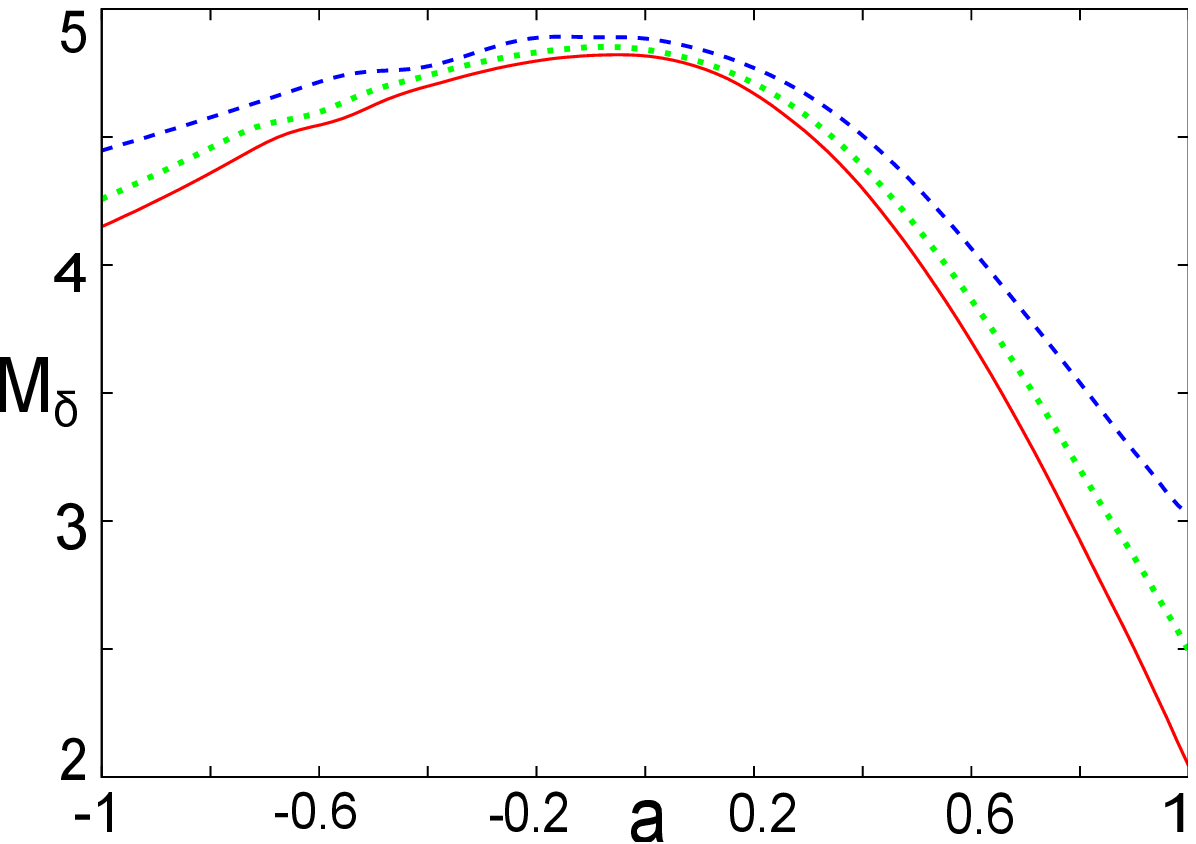} &
\includegraphics[width=0.4\linewidth]{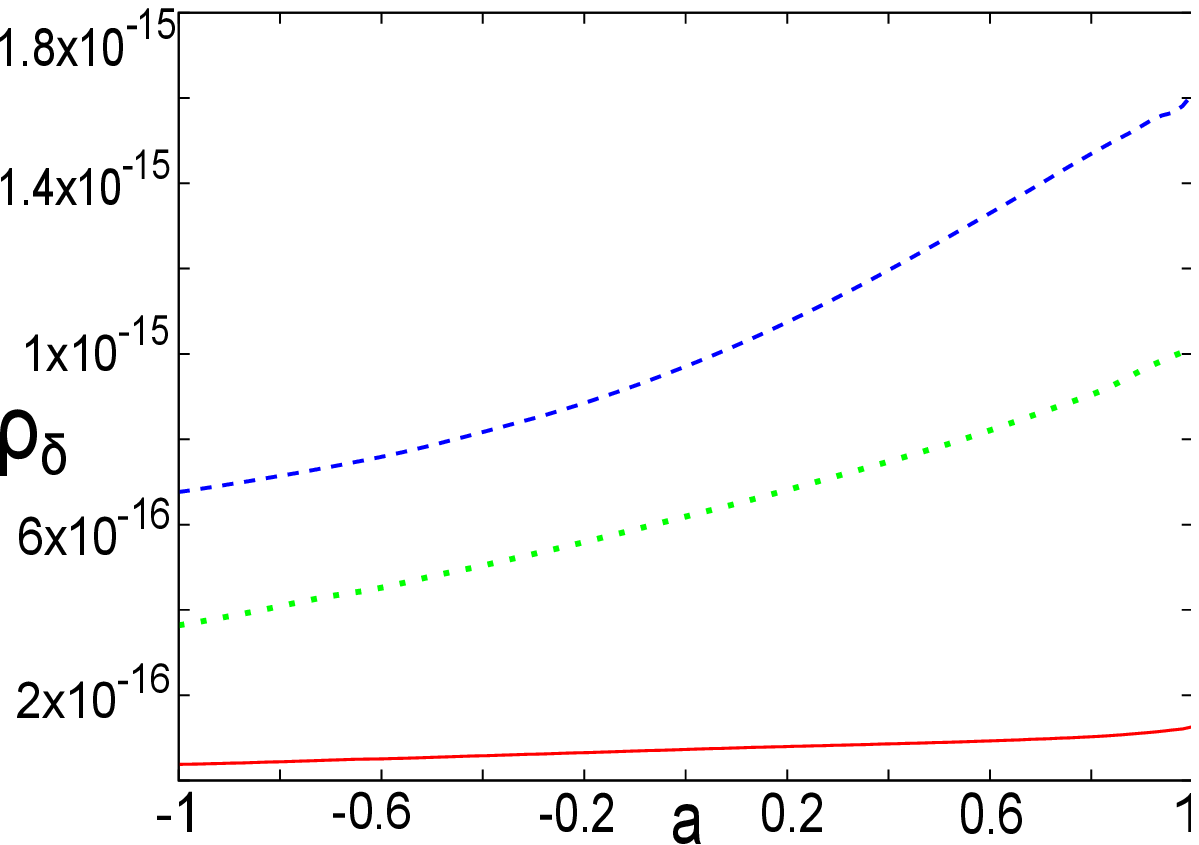} \\
\includegraphics[width=0.4\linewidth]{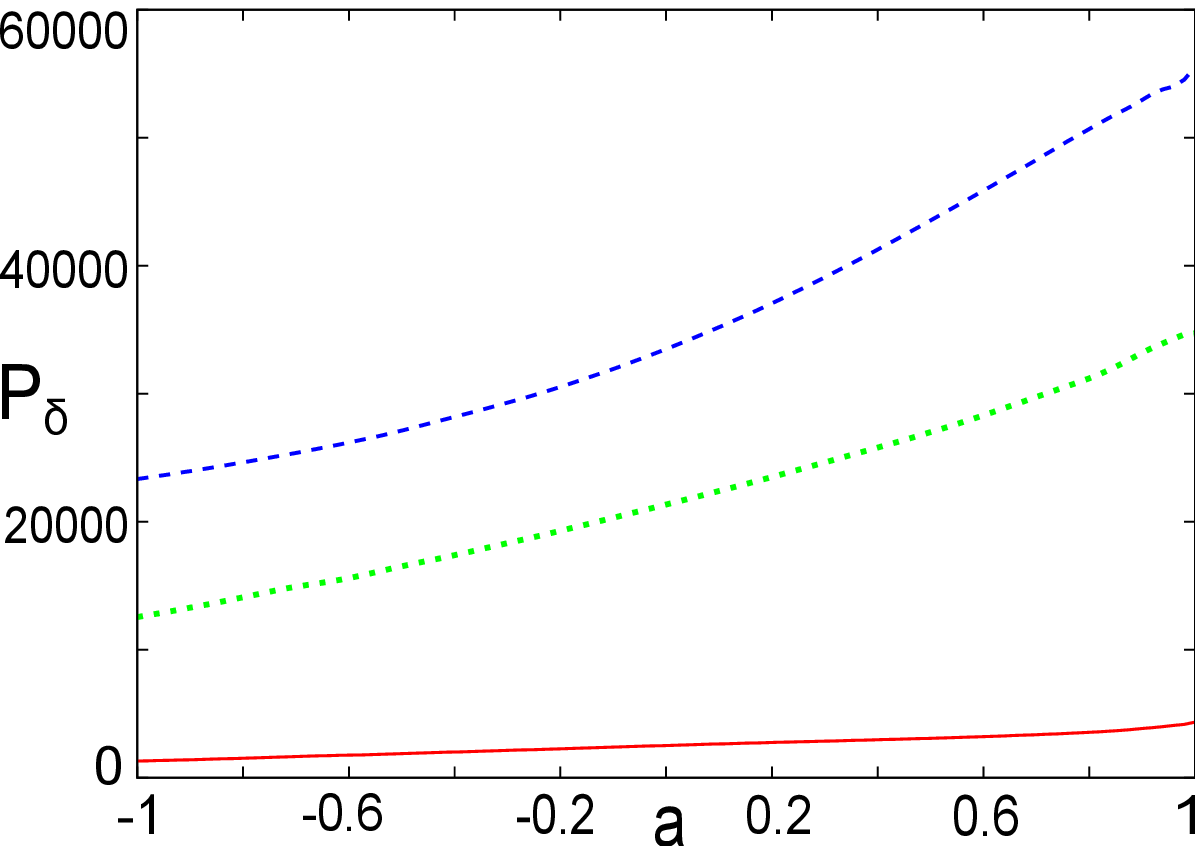} \\
\end{tabular}
\caption{Variation of quasi-terminal values of Mach number ($M_\delta$),
density ($\rho_\delta$) and pressure ($P_\delta$) with $a$ 
($T=2 \times 10^{11} K,\lambda=2.0$) for monotransonic accretion in 
constant height flow (dashed blue lines), quasi-spherical flow (dotted green lines) 
and flow in hydrostatic equilibrium (solid red lines). Density and pressure are in 
CGS units of $g$ $cm^{-3}$ and $dyne$ $cm^{-2}$ respectively and temperature is in 
absolute units of Kelvin.}
\label{fig24}
\end{figure}

Fig.(\ref{fig24}) depicts how quasi-terminal values of Mach number, density and pressure of monotransonic isothermal 
flow depend on the Kerr parameter. Hot flows with low angular momentum exhibit stationary accretion solutions 
spanning the full domain of black hole spin. It is observed that the general spin dependent behaviour of 
the corresponding physical quantities for three different flow geometries is quite well behaved in case of 
isothermal accretion, as opposed to the polytropic case. However, the previously stated intrinsic 
limitations in the possibility of observing variations over complete range of spin still exist for 
multitransonic flows. It is clear from 
fig.(\ref{fig24}) that quasi-terminal values possess common global trends of variation over 
$a$ for all three geometric configurations. The important observation in this context is the existence 
of asymmetry in variation of the related quantities between prograde and retrograde spin of the 
black hole. As mentioned in the case of polytropic accretion, such asymmetry is an extremely 
significant finding for the observation of black hole spin related effects.

\section{Concluding Remarks}

Computation of the quasi-terminal values helps us to 
understand the nature of spectra for which photons 
have emanated from a close proximity of the horizon. Hence, 
variation of the quasi-terminal values may be 
useful to understand how the black hole spin influences 
the configuration of the image of the black hole 
shadow. The present work puts forth two important findings 
which are worth mentioning in this context. 
Firstly, the spin dependence of quasi-terminal values has 
been studied for different geometrical 
configurations of matter. Secondly, we have found that 
(see, e.g. figs.(\ref{fig13}) and (\ref{fig24})) the 
prograde and the 
retrograde flows are distinctly marked by asymmetric 
distributions of relevant quasi-terminal values over 
the entire theoretical range of Kerr parameter. This 
indicates that the constructed image of shadow will 
be different for the co- and counter rotating flows. We 
also observe that the physical quantities 
responsible to construct the black hole spectra (velocity, 
density, pressure, temperature (for polytropic accretion) 
and quasi-specific energy (for isothermal accretion) of the 
flow) change abruptly at the shock location. This indicates 
that the discontinuous changes in the physical 
quantities should be manifested as a break in the 
corresponding spectral index, and will also show up during 
the procedure of black 
hole shadow imaging. Our work is thus expected to predict 
how the shape of the image of the shadow might 
be governed by the dynamical and thermodynamic properties 
of the accretion flow along with the spin of 
black hole. Through the construction of such image (work in 
progress), we will not only be able to provide 
a possible methodology (atleast at a qualitative level) for 
the observational signature of the black hole 
spin, but such images will also possibly shed light on the 
difference between the prograde and retrograde 
flows from an observational point of view. We have analysed 
general relativistic accretion of both polytropic and 
isothermal fluids in the Kerr metric to study the effects 
of matter geometry and black hole spin parameter 
on multitransonic shocked accretion flow.

\section*{Acknowledgments}
PT and SN would like to acknowledge the kind hospitality 
provided by HRI, Allahabad, India, for several 
visits through the $\rm{XII^{th}}$ plan budget of Cosmology 
and High Energy Astrophysics grant. The long term
visiting student position of DBA at HRI was supported 
by the aforementioned grant as well. The authors 
would like to thank Sonali Saha Nag for her kind help in 
formulating the initial version of a numerical 
code which has partly been used in this work. 
TKD acknowledges the support from the Physics and Applied 
Mathematics Unit, Indian Statistical Institute, Kolkata, India 
(in the form of a long term visiting scientist), 
where parts of the present work have been carried out. 

\bibliographystyle{plainnat}
\bibliography{paper}

\end{document}